%% file: rgbagb.tex
\documentclass{emulateapj}
\usepackage{natbib}
\newcommand{\dmo}{\mbox{$(m\!-\!M)_{0}$}}

\newcommand{\av}{\mbox{$A_V$}}
\newcommand{\feh}{\mbox{\rm [{\rm Fe}/{\rm H}]}}

\newcommand{\Msun}{\mbox{$M_{\odot}$}}
\newcommand{\Teff}{\mbox{$T_{\rm eff}$}}

\newcommand{\beq}{\begin{equation}}
\newcommand{\eeq}{\end{equation}}
\newcommand{\beqa}{\begin{eqnarray}}
\newcommand{\eeqa}{\end{eqnarray}}

\slugcomment{To appear in ApJ \date}

\shorttitle{AGB lifetimes from ANGST galaxies}
\shortauthors{Girardi et al.}

\begin{document}


\title{The ACS Nearby Galaxy Survey Treasury IX. Constraining  
asymptotic giant branch evolution with old metal-poor galaxies}


\author{
L\'eo Girardi\altaffilmark{1}, 
Benjamin F. Williams\altaffilmark{2}, 
Karoline M. Gilbert\altaffilmark{2}, 
Philip Rosenfield\altaffilmark{2}, 
Julianne J. Dalcanton\altaffilmark{2}, 
Paola Marigo\altaffilmark{3},
Martha L. Boyer\altaffilmark{7}, 
Andrew Dolphin\altaffilmark{4}, 
Daniel R. Weisz\altaffilmark{6}, 
Jason Melbourne\altaffilmark{9}, 
Knut A.G. Olsen\altaffilmark{8},
Anil C. Seth\altaffilmark{5}, 
Evan Skillman\altaffilmark{6}
}
\altaffiltext{1}{Osservatorio Astronomico di Padova - INAF, Vicolo dell'Osservatorio 5, I-35122 Padova, Italy}
\altaffiltext{2}{Department of Astronomy, University of Washington, Box 351580, Seattle, WA 98195, USA}
\altaffiltext{3}{Dipartimento di Astronomia, Università di Padova, Vicolo dell'Osservatorio 2, I-35122 Padova, Italy}
\altaffiltext{4}{Raytheon Company, 1151 East Hermans Road, Tucson, AZ 85756, USA}
\altaffiltext{5}{Harvard-Smithsonian Center for Astrophysics, 60 Garden Street Cambridge, MA 02138, USA}
\altaffiltext{6}{Department of Astronomy, University of Minnesota, 116 Church Street SE, Minneapolis, MN 55455, USA}
\altaffiltext{7}{Space Telescope Science Institute, 3700 San Martin Drive, Baltimore, MD 21218, USA}
\altaffiltext{8}{National Optical Astronomy Observatory, 950 North Cherry Avenue, Tucson, AZ 85719, USA}
\altaffiltext{9}{Caltech Optical Observatories, Division of Physics, Mathematics and Astronomy, Mail Stop 301-17, California Institute of Technology, Pasadena, CA 91125, USA}

\newpage
\begin{abstract}
In an attempt to constrain evolutionary models of the asymptotic giant
branch (AGB) phase at the limit of low masses and low metallicities,
we have examined the luminosity functions and number ratio between AGB
and red giant branch (RGB) stars from a sample of resolved galaxies
from the ACS Nearby Galaxy Survey Treasury (ANGST). This database
provides HST optical photometry together with maps of completeness,
photometric errors, and star formation histories for dozens of
galaxies within 4~Mpc. We select 12 galaxies characterized by
predominantly metal-poor populations as indicated by a very steep and
blue RGB, and which do not present any indication of recent star
formation in their color--magnitude diagrams. Thousands of AGB stars
brighter than the tip of the RGB (TRGB) are present in the sample
(between 60 and 400 per galaxy), hence the Poisson noise has little
impact in our measurements of the AGB/RGB ratio. We model the
photometric data with a few sets of thermally pulsing AGB (TP-AGB)
evolutionary models with different prescriptions for the mass
loss. This technique allows us to set stringent constraints to the
TP-AGB models of low-mass metal-poor stars (with $M<1.5$~\Msun,
$\feh\la-1.0$). Indeed, those which satisfactorily reproduce the
observed AGB/RGB ratios have TP-AGB lifetimes between 1.2 and 1.8~Myr,
and finish their nuclear burning lives with masses between 0.51 and
0.55~\Msun. This is also in good agreement with recent observations of
white dwarf masses in the M\,4 old globular cluster. These constraints
can be added to those already derived from Magellanic Cloud star
clusters as important mileposts in the arduous process of calibrating
AGB evolutionary models.

\end{abstract}





\section{Introduction}

The thermally pulsing asymptotic giant branch phase (TP-AGB) is both
one of the most important and one of the more uncertain phases of
stellar evolution. Its importance resides mainly in its sizeable
contribution to the integrated light and chemical yields of stellar
populations, which are essential for the understanding of galaxy
evolution and the interpretation of the light from distant galaxies
\citep[e.g.][]{Maraston_etal06, Eminian_etal08, ConroyGunn09, 
ConroyGunn10}. Its uncertainties derive from a series of
circumstances including their complex internal structure, the critical
role of difficult-to-model processes like convective dredge-up, mass
loss, circumstellar dust formation and long-period variability, and
the scarcity of clear-cut and unequivocal observational constraints on
their evolution in the most immediate universe.

Gigantic steps are being made in all of these subjects, but the
present situation is that the evolutionary timescales of stars in the
TP-AGB phase are far from being settled. Uncertainties by a factor of
{\em a few} still exist at the extremes of the age--metallicity region
allowed for AGB stars. This contrasts sharply with other evolutionary
phases like the main sequence and red giant branch (RGB), for which
the evolutionary times are known with errors smaller than a few
tenths, as demonstrated by a large variety of observations
\citep[see][and references therein]{Gallart_etal05}.

Present day constraints on the TP-AGB evolution are largely based on
observations of stars in the Magellanic Clouds and of the Milky Way
(MW). The lifetimes as a function of stellar mass, for slightly
subsolar metallicities, can be derived for star counts in Magellanic
Cloud clusters \citep{Frogel_etal90, vanLoon_etal05,
GirardiMarigo07a}. The same can be done for low-mass stars in MW
globular clusters.  Even for the most populous star clusters, however,
AGB stars are few and their counts are affected by large Poisson
fluctuations; in the case of MW globular clusters, AGB stars brighter
than the tip of the red giant branch (TRGB), and long period
variables, practically disappear from observed samples at
$\feh\la-1.0$ \citep{FrogelElias88, FrogelWhitelock98}.

Thus, to obtain useful constraints on the AGB lifetime from star
clusters, it is necessary to either sum the star counts in many
clusters into age and metallicity bins
\citep[e.g.][]{GirardiMarigo07a}, or embark on a more detailed study 
of the dust, chemical and pulsational properties of individual cluster
stars (see e.g. \citealt{LebzelterWood07, Lebzelter_etal08,
Kamath_etal10} for Magellanic Cloud clusters, and
\citealt{Lebzelter_etal06, vanLoon_etal06, McDonald_etal09,
McDonald_etal10, Boyer_etal09b, Boyer_etal10} for MW globulars). Less
direct constraints on AGB evolution come from integrated cluster
properties, like their colors and surface brightness fluctuations
\citep[see e.g.][]{Maraston05, Pessev_etal08, Raimondo09,
ConroyGunn09, ConroyGunn10}, which by their very nature cannot
disentangle different evolutionary properties like stellar
luminosities, lifetimes, and chemical types, and hence often provide
somewhat ambiguous constraints on the numbers of stars at different
evolutionary stages. It may happen, for instance, that TP-AGB models
which do not consider the third dredge-up and the formation of carbon
stars \citep[such as the BaSTI ones;][]{Cordier_etal07}, or models in
which carbon stars are shifted to the high effective temperature,
($\Teff$) range of 4000 to 3100~K \citep[][with their $\log\Teff$
shift of $\sim+0.1$~dex for TP-AGB stars at Magellanic Cloud
metallicities]{ConroyGunn09, ConroyGunn10}, provide acceptable (but
spurious) fits to integrated colors of Magellanic Cloud clusters.
Unfortunately, the properties of these models are inconsistent with
the most basic observations of over 10\,000 C-type AGB stars in the
Magellanic Clouds \citep[see][]{NikolaevWeinberg00, CioniHabing03}
with a significant fraction being at $3100>\Teff/{\rm K}>2650$
\citep{Groenewegen_etal09}. Such high discrepancies and gaps in the
models remain hidden when only the integrated properties are examined.

The problem of low number counts in star clusters could, in principle,
be circumvented by using AGB star counts in entire galaxies, but, in
practice, this implies meeting a series of difficulties. For galaxies
with moderate-to-high metallicities, the AGB develops at low effective
temperatures, and consequently near-infrared photometry is required to
unveil them, as dramatically demonstrated by the Magellanic Clouds
\citep{Frogel_etal90, Cioni_etal00, WeinbergNikolaev01}. Moreover,
since a non-negligible fraction of the AGB population becomes
dust-enshrouded, mid-infrared data are necessary for a complete census
of the AGB population \citep[e.g.,][]{Blum_etal06, Bolatto_etal07,
Boyer_etal09}.

At the limit of very low metallicities, the nearest low surface
brightness dwarf spheroidal galaxies (dSph) seem to be particularly
useful for the study of their AGB populations because they appear
uncrowded at magnitudes accessible with present day 4-m class
telescopes equipped with near-infrared cameras. Indeed,
\citet{Gullieuszik_etal08} and \citet{Held_etal10} are able to derive
important indications from ground-based data of the Leo~II and Leo~I
dSphs \citep[see also][for Fornax and
Sagittarius]{Lagadec_etal08}. The main uncertainties in these works
derive from the small numbers of AGB stars, from the high MW
foreground contamination, and the uncertainties in the dSphs star
formation histories (SFH). Also, \citet{Boyer_etal09} call attention
to the large fraction (30 to 40\%) of AGB stars in dwarf irregulars
(dIrr) which are bright in mid-infrared light, but are either missing
or misclassified in optical studies because they are enshrouded in
thick dust shells. It is still not clear whether this fraction is also
representative of dSphs.

The possibilities for a better calibration of AGB luminosities and
lifetimes become much wider when milli-arcsec resolution imaging is
available. A good example is provided by \citet{Melbourne_etal10}, who
use images taken with the Advanced Camera for Surveys (HST/ACS) in the
$I$ band (F814W) together with Keck Adaptive Optics (AO) in $K$ to
study the AGB population of the dIrr KKH~98 at a distance of 2.5
Mpc. This work clearly demonstrates the utility of moving towards more
distant resolved dwarf galaxies, for which (1) more stars can be
observed in a single pointing, even considering the small
field-of-view available for ACS and AO, and (2) the MW foreground
becomes dramatically smaller. On the other hand, some potential
disadvantages of these targets are also evident: (a) they present SFHs
that are somewhat more uncertain than the nearest dSphs; (b) there is
no spectroscopic information to allow a clearcut separation between
the hottest C- and O-rich stars (cooler stars, instead, are well
separated by their infrared colors), which limits the analysis of
aspects related to the third dredge-up events in the AGB; (c) the
mid-infrared photometry is also not available, due to the limited
resolution of Spitzer and AKARI, hence limiting the analysis to
non-obscured AGB stars; (d) the chance of crowding/blending of AGB and
RGB stars also increases with distance.

A result common to the low-metallicity dwarf galaxies studied by
\citet{Gullieuszik_etal08}, \citet{Held_etal10} and 
\citet{Melbourne_etal10} is that the TP-AGB lifetimes from 
\citet[][hereafter MG07]{MarigoGirardi07} and \citet{Marigo_etal08} 
models are, apparently, largely overestimated. The reason for this is
still not clear, but is possibly due to the underestimation of the
mass-loss rates at low metallicities. As a concurring factor, the
circumstellar extinction presently assumed in the models may be
somewhat underestimated, as suggested by the Spitzer observations of
nearby dwarf irregulars by \citet{Jackson_etal07a, Jackson_etal07b}
and \citet{Boyer_etal09}. Regardless, it is likely that the problem is
limited to low metallicities, since the MG07 TP-AGB models have been
calibrated in the intermediate-metallicity Magellanic Clouds. Their
evolutionary behavior at very low metallicities can be considered as
either educated guesses, or extrapolations of the behavior met at
intermediate metallicities, since they are a result of the straight
application of theoretically-uncertain metallicity dependences.
 
In this paper, we use HST optical data from the ANGST survey to derive
constraints on the AGB evolution for a set of metal-poor galaxies in
which the recent star formation activity is generally very low, if not
absent. Our target is to derive constraints on the optically-visible
TP-AGB lifetimes of low-mass, metal-poor stars. The advantage of using
a set of Mpc-distant galaxies is clear: we will have thousands of AGB
stars in our samples, reducing Poisson noise to a minimum. The price
to pay, as in the case of \citet{Melbourne_etal10}, is that our
analysis can be affected by uncertainties in the SFHs of individual
galaxies. The hope however is that these SFH errors become less
relevant when averaged over a large sample of galaxies.

In Sect.~\ref{sec_data} we describe the data used in this work.  In
Sect.~\ref{sec_models} we model the data using two different sets of
TP-AGB models, deriving clear quantitative constraints on them. We
then provide a revised set of TP-AGB tracks that brings models and
observations into better agreement.

\section{The data}
\label{sec_data}

\subsection{The ANGST/ANGRRR photometry database}

We used data from the ACS Nearby Galaxy Survey Treasury
(ANGST)\footnote{http://www.nearbygalaxies.org, 
http://archive.stsci.edu/prepds/angst/ }
\citep{Dalcanton_etal09} and from the ``Archive of Nearby Galaxies:
Reduce, Reuse, Recycle''
(ANGRRR)\footnote{http://archive.stsci.edu/prepds/angrrr} databases.
These two programs have archived stellar photometry for tens of
millions of stars in nearby galaxies outside the Local Group, based on
images taken with the ACS and the Wide Field Planetary Camera 2
(WFPC2) on the HST.  The imaging primarily used blue observations in
the F475W, F555W, or F606W filters, and F814W for the red filter.  The
photometry was performed using
DOLPHOT\footnote{http://purcell.as.arizona.edu/dolphot} and
HSTPHOT\footnote{http://purcell.as.arizona.edu/hstphot} \citep[for ACS
and WFPC2 imaging, respectively;][]{Dolphin00}, as described fully in
\citet{Dalcanton_etal09} and \citet{Williams_etal09}.  The depth of 
the resulting color--magnitude diagrams (CMD) varies with distance,
crowding, and exposure time, but reaches several magnitudes below the
TRGB in all cases, and below the red clump ($M_I\sim-0.5$~mag) in all
but 2 of the cases we analyze here.

For this paper, the photometry from the first data release for ANGST
and ANGRRR has been supplemented with extensive artificial star
tests. For each of the fields studied, a minimum of 100,000 artificial
star experiments were performed to determine the completeness and
errors as a function of color and magnitude.  In each test, one star
of known color and magnitude was added to the data, and the photometry
routine was re-executed to determine the difference between the input
and output magnitudes if the star was recovered. 

SFHs were determined by fitting the observed CMDs with the MATCH
fitting package \citep{Dolphin02}.  The parameters of MATCH were set
to match those used by \citet{Williams_etal09}.  In brief, a
\citet{Salpeter55} IMF was used to generate a base set of Hess
diagrams using the isochrones of \citet[][with updates in
\citealp{Marigo_etal08}]{Girardi_etal02} and the error and
completeness from our artificial star tests.  The best-fit combination
of this base set of diagrams was then determined using the statistics
described in \citet{Dolphin02}.  First, the fit was performed on the
full data set brightward of the 50~\% completeness limit as determined
by our artificial star tests.  Then, the fit was repeated excluding
all data brightward of the TRGB.  This second fit allowed us to
quantify the effects of the bright AGB stars on the resulting SFH (as
discussed in Sect.~\ref{sec_resmg07}). 

We further analyzed the CMDs to estimate extinctions \av, distance
moduli \dmo, and the magnitude of the TRGB.  The total extinction and
distance moduli are determined automatically by MATCH, based on the
optimal reddening and distance needed to reproduce the observed CMD
\citep[see the full discussion in][]{Williams_etal09}, assuming
$R_V=3.1$ at optical wavelengths.  Uncertainties are determined by
identifying the range over which \av\ and \dmo\ can vary without
producing a statistically significant reduction in the quality of the
fit to the observed CMD.

\subsection{The galaxy sample and their SFHs}

\begin{figure}
\plotone{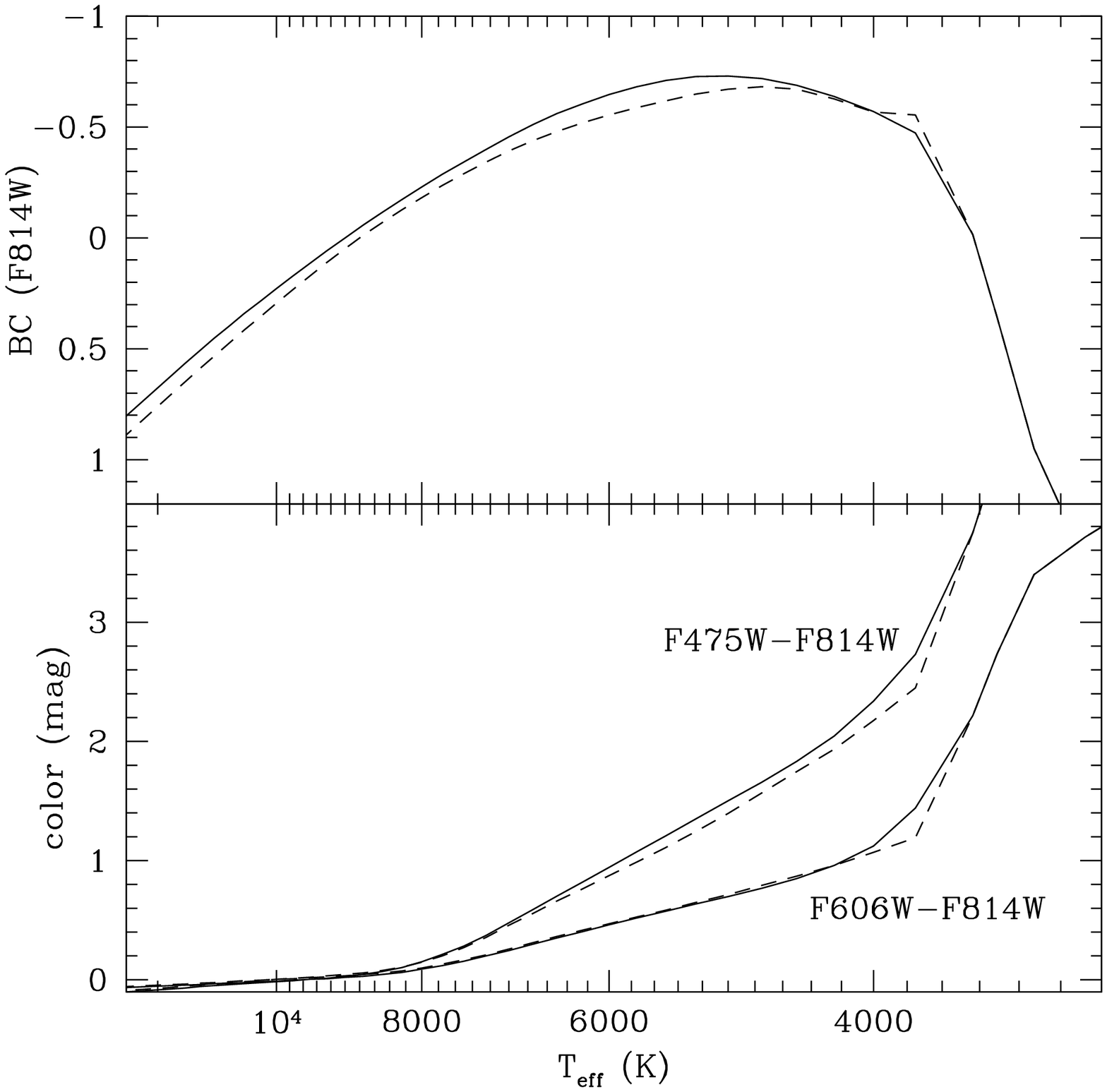}
\caption{{\bf Top panel:} F814W-band bolometric corrections as a
function of \Teff, for both $\feh=0$ (continuous line) and $\feh=-2$
(dashed line). {\bf Bottom panel:} The \Teff-color relations for both
F475W$-$F814W and F606$-$F814W, for the same metallicities.  These
relations are described in \citet{Girardi_etal08}. The small kinks at
$\Teff\sim3800$~K correspond to the transition between
\citet[][for $\log g=2$]{CastelliKurucz03} and \citet{Fluks_etal94}
model atmospheres. }
\label{fig_bc}
\end{figure}

\begin{figure*}
\includegraphics[width=0.33\textwidth]{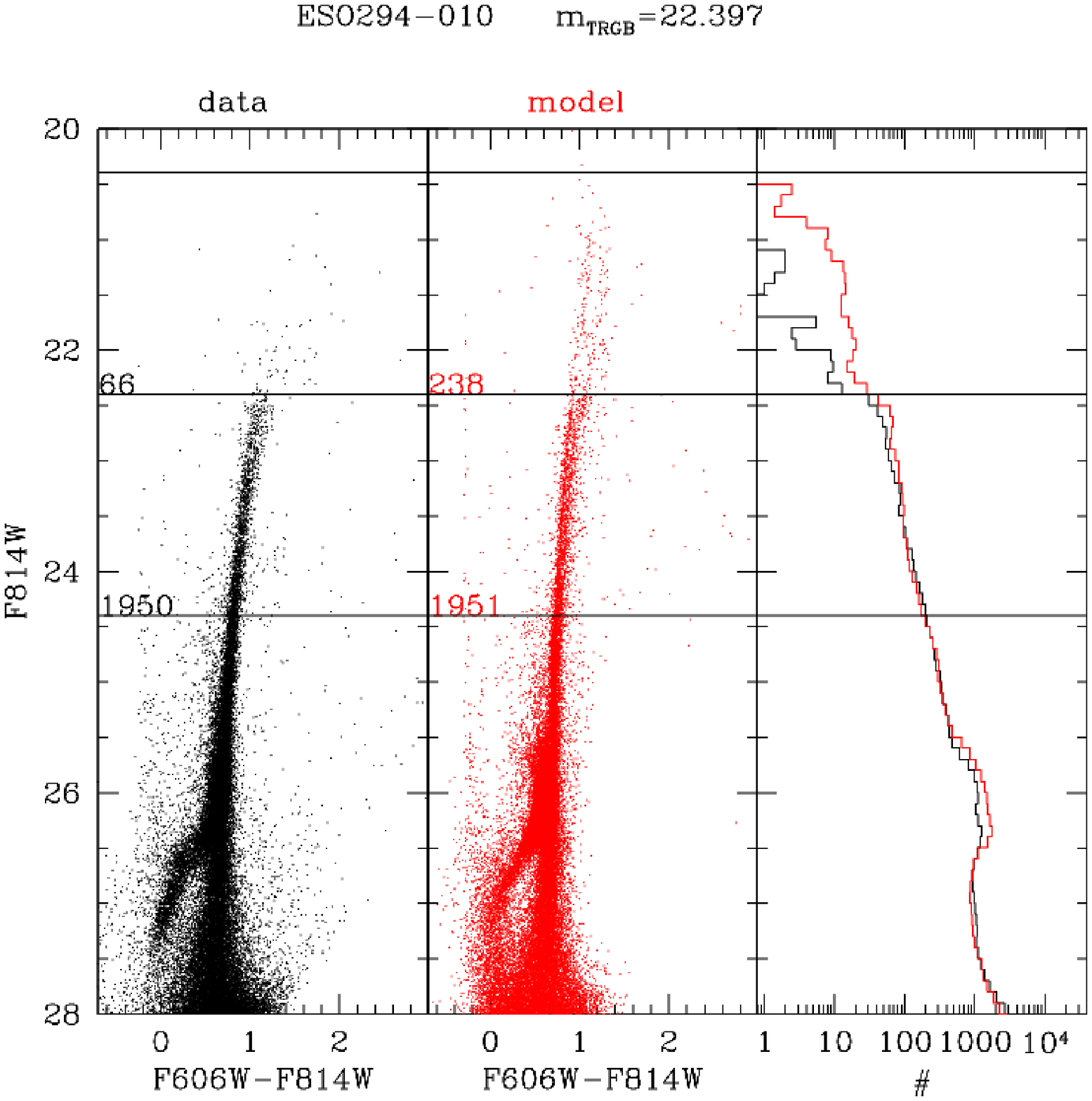}
\includegraphics[width=0.33\textwidth]{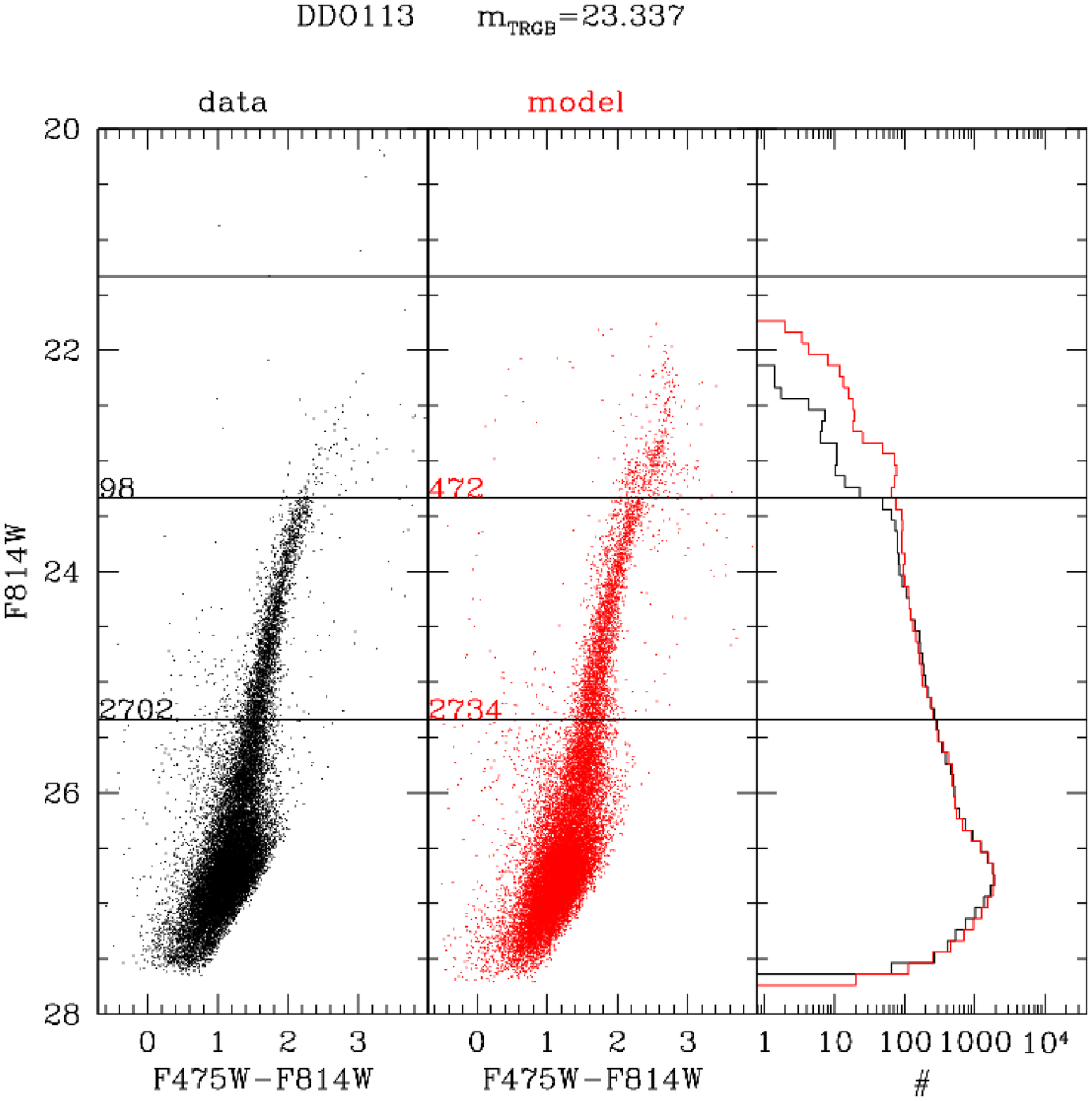}
\includegraphics[width=0.33\textwidth]{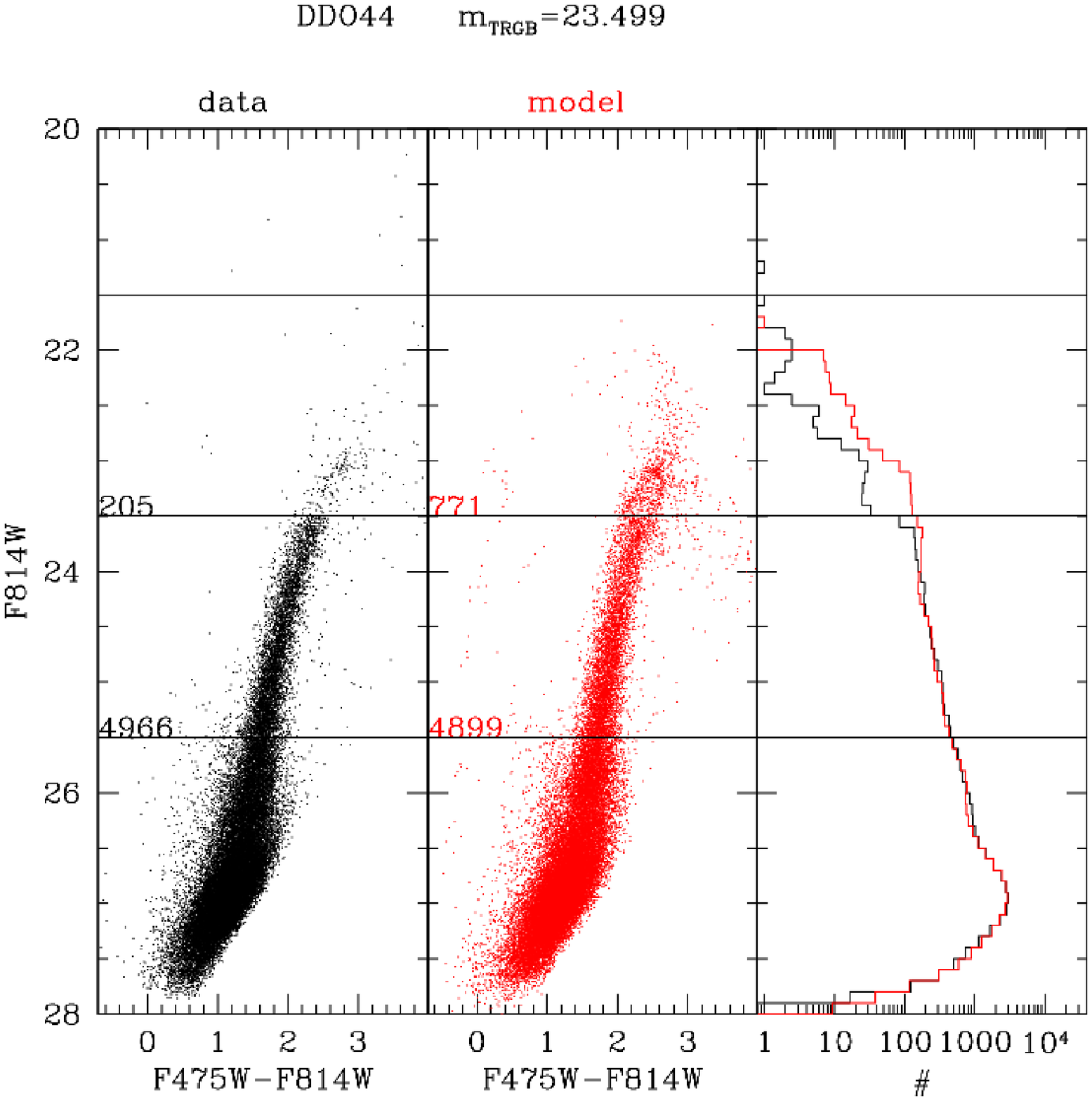}
\\
\includegraphics[width=0.33\textwidth]{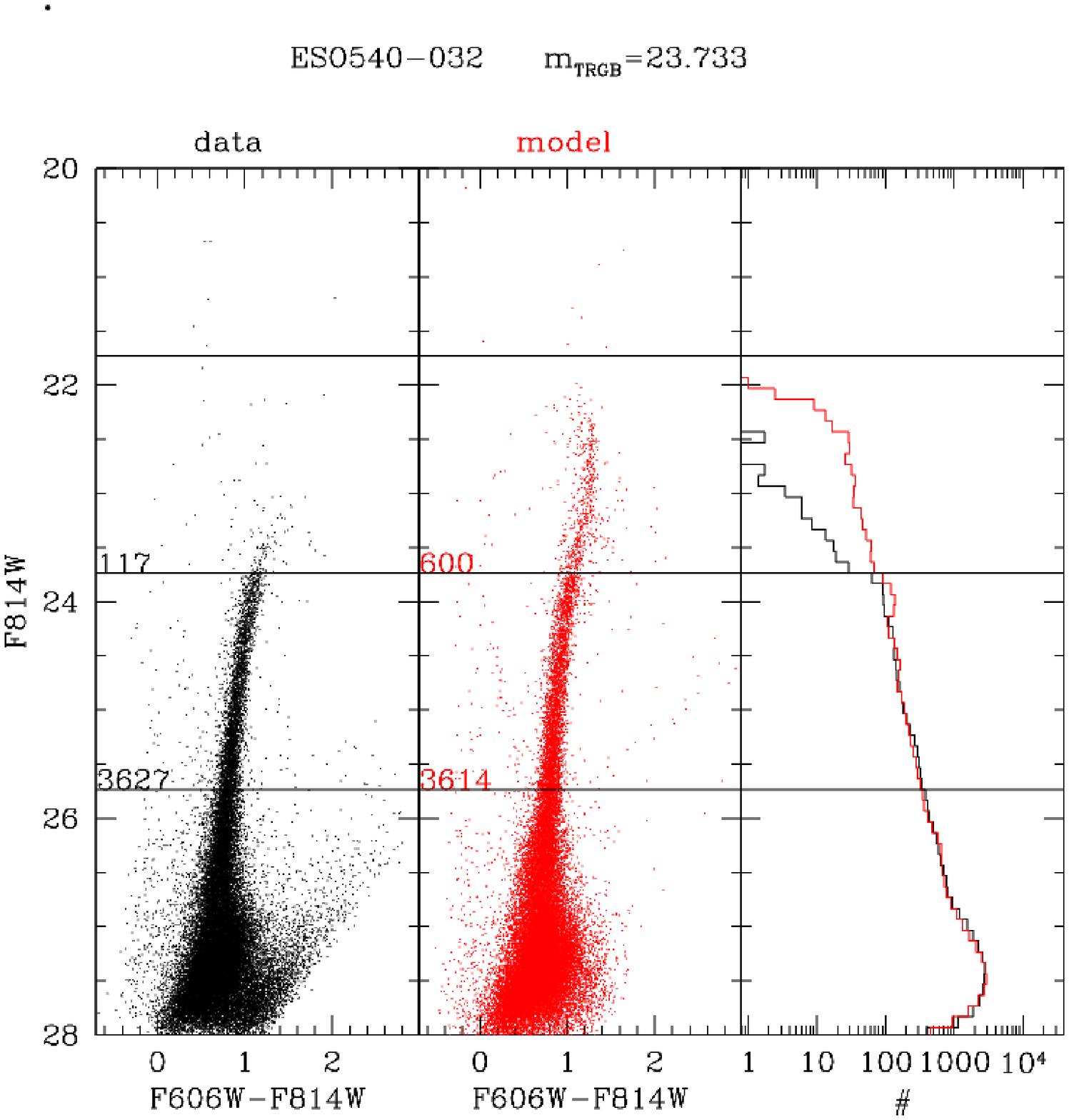}
\includegraphics[width=0.33\textwidth]{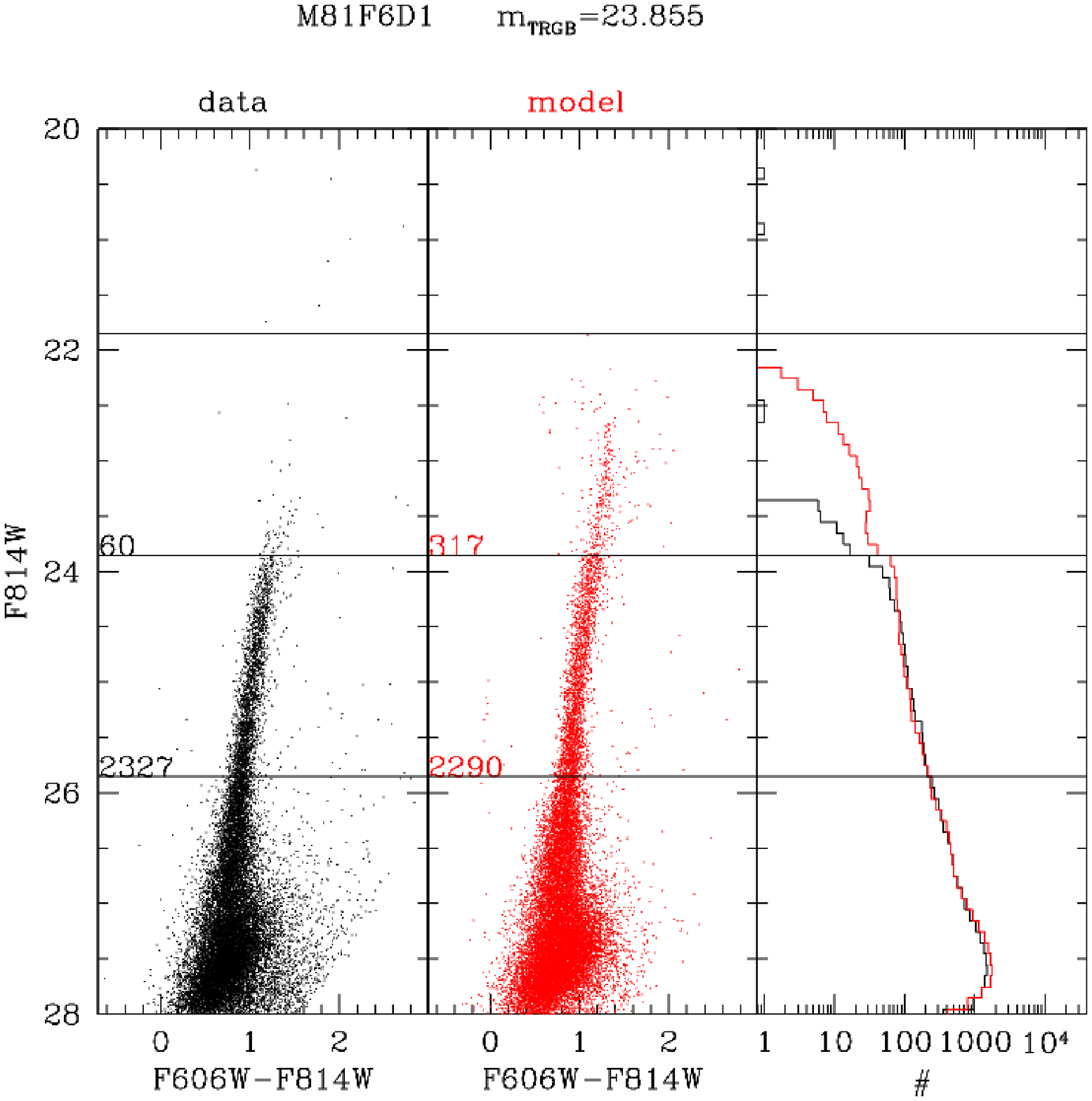}
\includegraphics[width=0.33\textwidth]{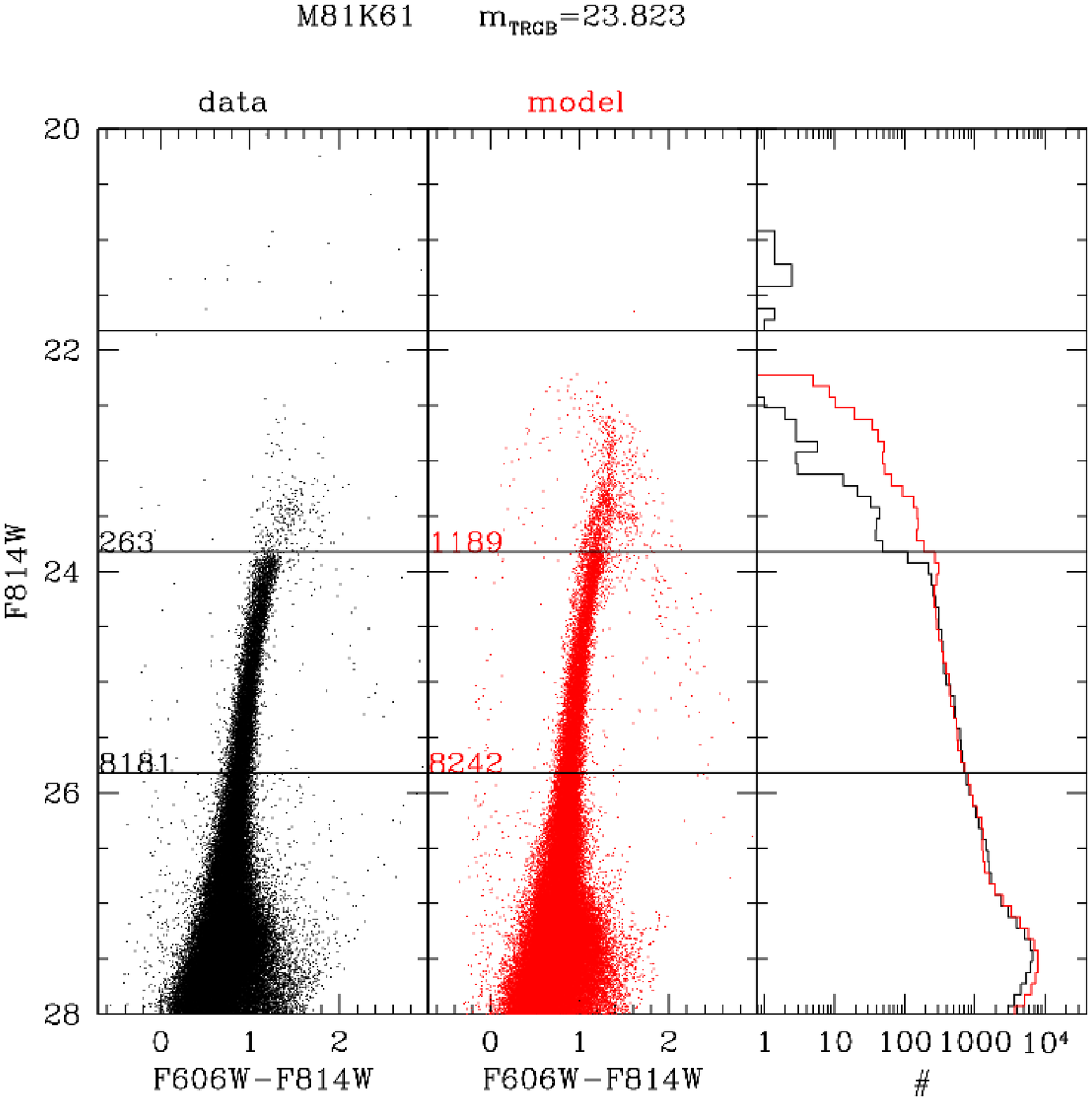}
\\
\includegraphics[width=0.33\textwidth]{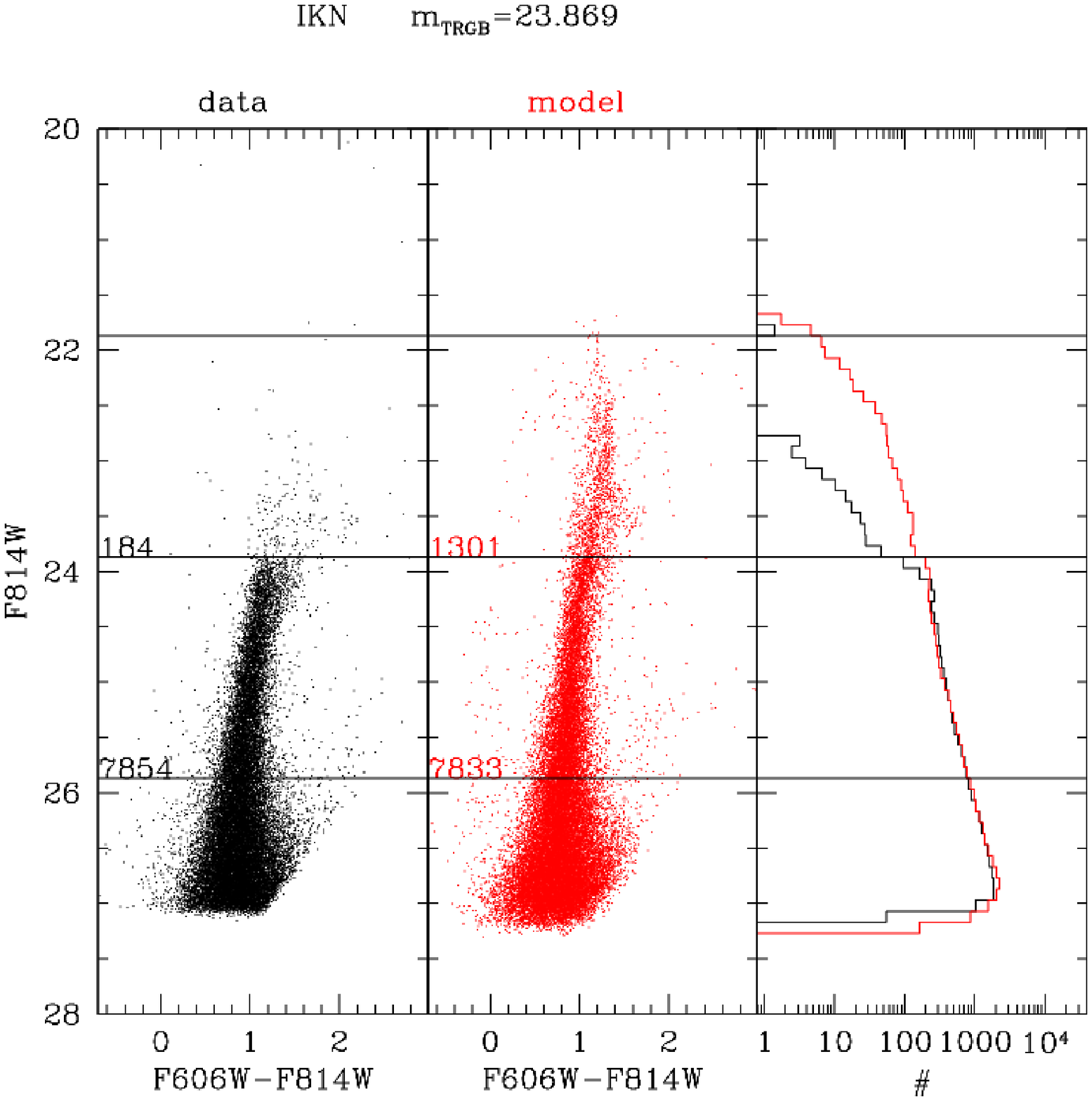}
\includegraphics[width=0.33\textwidth]{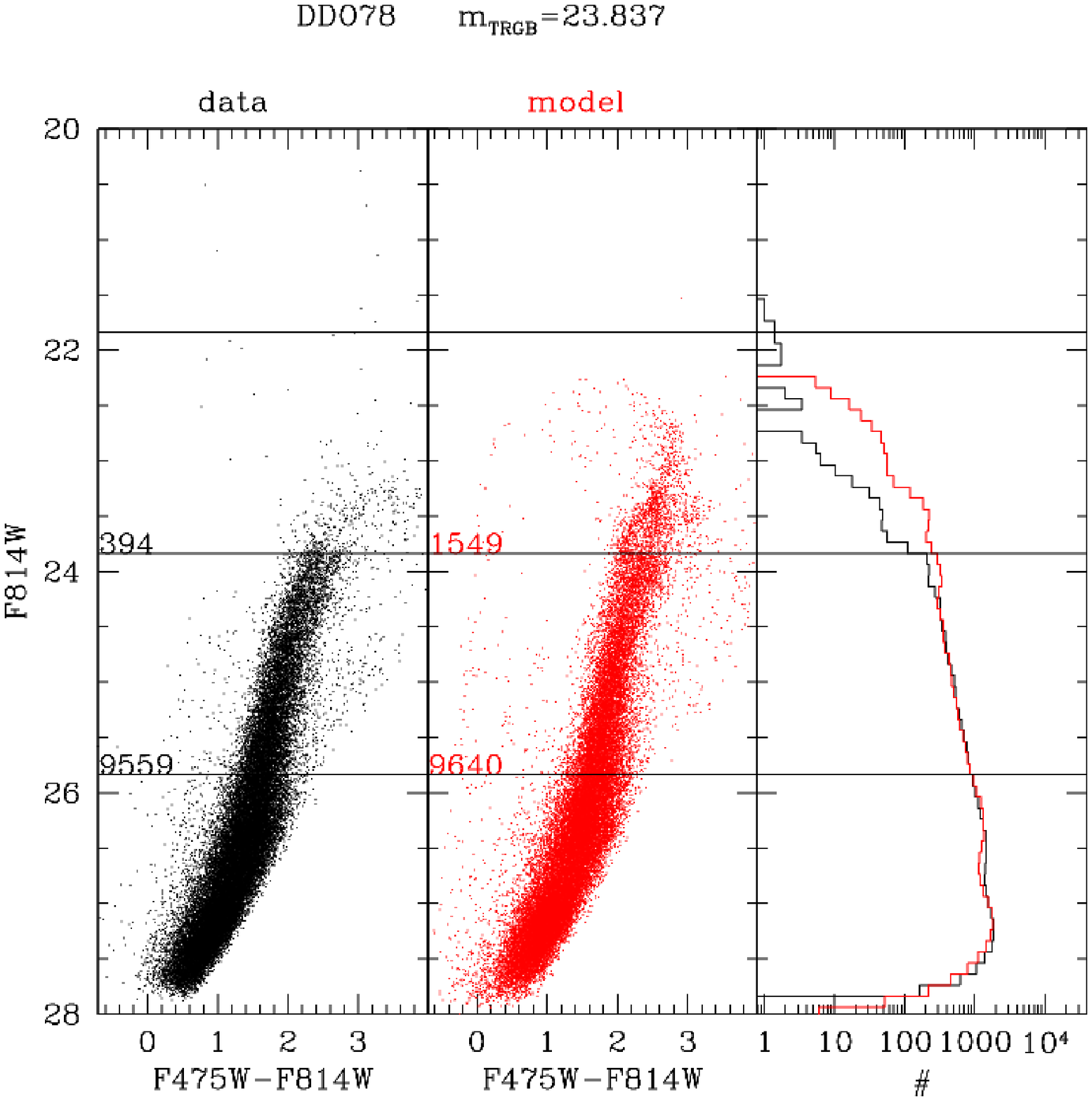}
\includegraphics[width=0.33\textwidth]{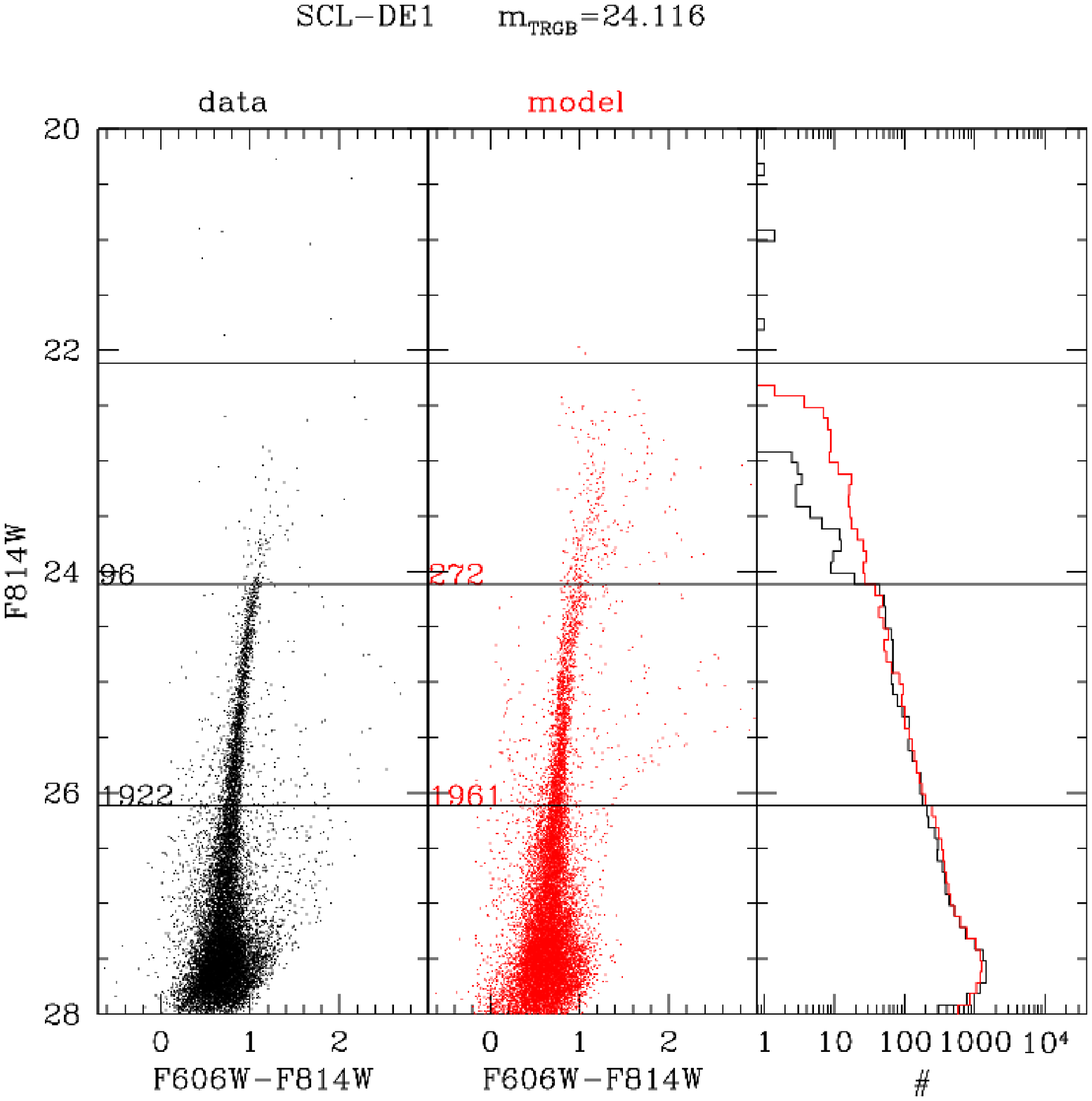}
\caption{The data and models used in this work, for all galaxies and
galaxy regions in our sample.  {\bf Left panels:} The observed CMD
from ANGST/ANGRRR. The horizontal lines mark the TRGB and upper and
lower magnitude limits of the stars considered to be in the upper RGB
and AGB. Their total numbers are marked inside the boxes.  {\bf Middle
panels:} The same as in the left panel, but for the model CMD from
TRILEGAL, with the default set of TP-AGB models \citep{Marigo_etal08}
and scaled to present the same number of upper RGB stars. {\bf Right
panels:} A comparison between the observed (black) and model (red or
grey) luminosity functions.}
\label{fig_ma08}
\end{figure*}
\begin{figure*}
\figurenum{\ref{fig_ma08} continued}
\includegraphics[width=0.33\textwidth]{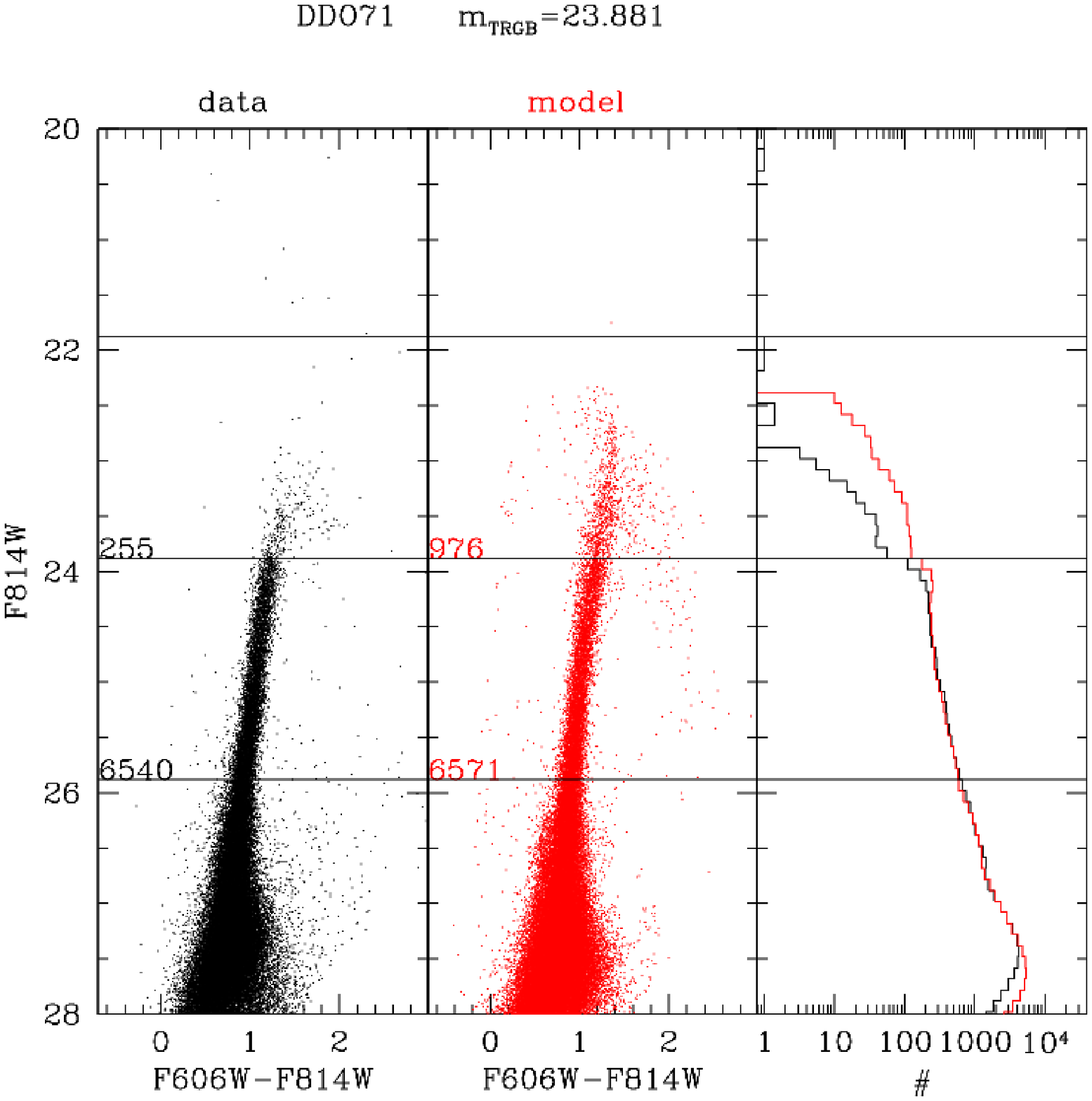}
\includegraphics[width=0.33\textwidth]{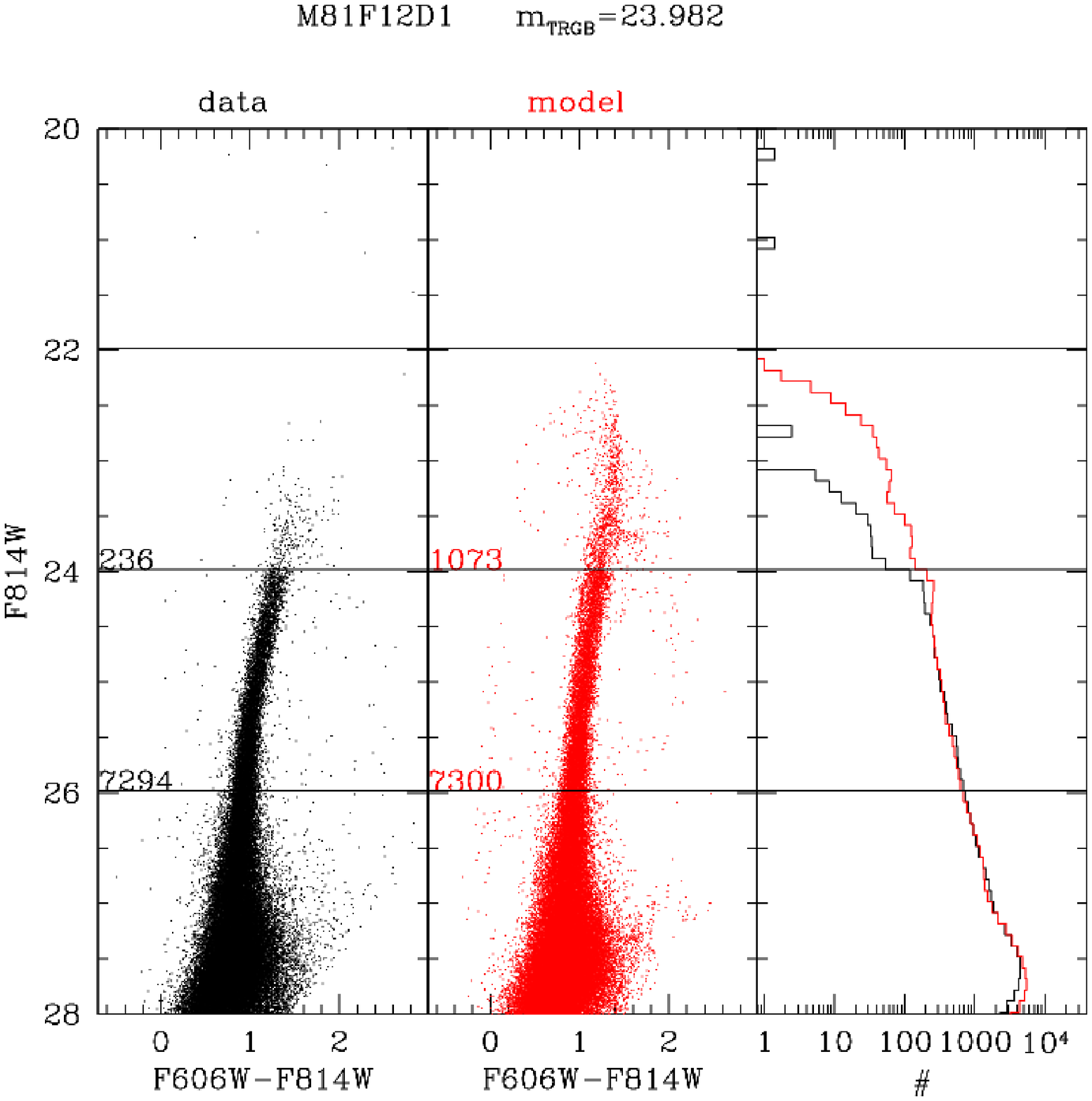}
\includegraphics[width=0.33\textwidth]{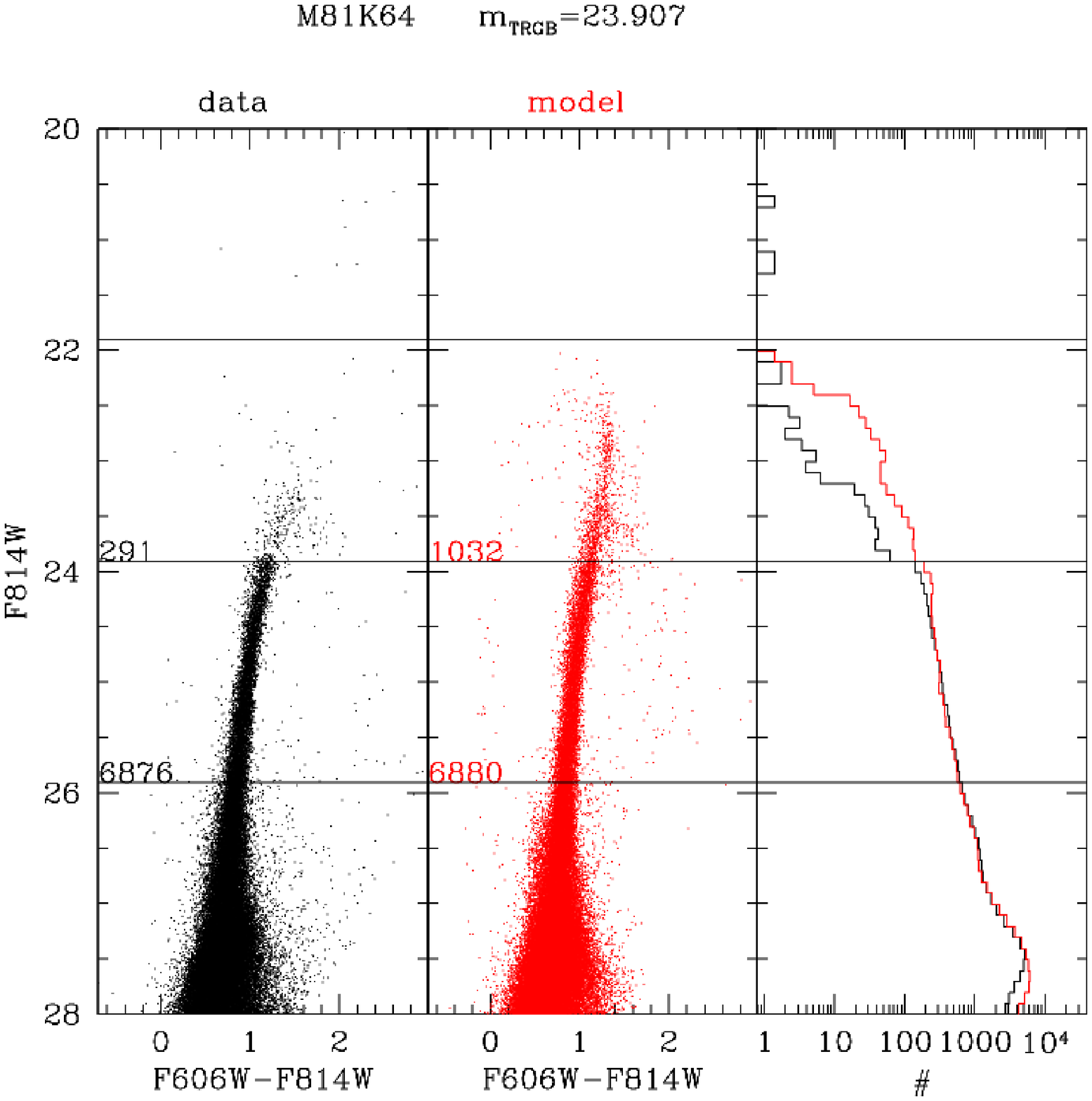}
\caption{}
\end{figure*}

We chose galaxies for our sample based on two primary criteria --
metallicity and mean stellar age.  The first restriction results from
our desire to analyze only the galaxies in which the ANGST/ANGRRR
optical data contains a nearly complete census of the AGB population.
This requirement naturally limits our work to galaxies with metal-poor
populations (say $\feh\la-1.2$).  At these metallicities, the upper
AGB develops at $\Teff>4000$~K and colors F475W$-$F814W$<2.4$, or
F606W$-$F814W$<1.2$, and still falls at the plateau of the BC-\Teff\
relations (Fig.~\ref{fig_bc}).  Bolometric corrections for
low-metallicity AGB stars are thus close to their minimum values,
ensuring that the brighter stars really do appear at the smaller
magnitudes. In contrast, for redder, high-metallicity AGB stars,
bolometric corrections become significant and increase steadily with
color.  Redder AGB stars thus become progressively less accessible in
the optical, and instead require near-infrared photometry to extend
the complete AGB counts towards the region of lower \Teff, which
corresponds to more metal-rich galaxies.

We further restrict the galaxy sample to regions that have a
simple-to-interpret SFH, with AGB stars limited to a well-defined
region of the age--metallicity plane. This request limits us to
galaxies dominated by old stars, for which there is no sign of recent
or intermediate-age star formation. Thanks to the steep mass--main
sequence lifetime relation, the absence of populations younger than
3~Gyr ensures that the evolved stars have initial masses confined to a
narrow interval, of roughly 0.8 to 1.4 \Msun. The AGB star counts in
these galaxies should then provide clear constraints on the evolution
of low-mass stars.

Based on these two conditions, we have selected a sample of galaxies
for which a visual inspection of the CMD revealed: (1) an almost
vertical RGB, with the TRGB at F475W$-$F814W$\la2$, or
F606W$-$F814W$\la1$ so as to indicate low metallicity and ${\rm
BC}_{\rm F814W}\la0.5$ (Fig.~\ref{fig_bc}), and low foreground
extinction, and (2) no evidence for recent and intermediate-age star
formation, as indicated by an absence of any younger main sequence and
helium burning sequence above the red clump/HB.  This latter
restriction eliminates galaxies with star formation in the most recent
$0.5$~Gyr, but may admit galaxies with some amount of star formation
at intermediate ages. A full analysis of the SFHs can be found in
Weisz et al.\ (in preparation).

Although our selection considered both ACS and WFPC2 data, it turns
out that the final sample contains ACS data only.

The left panels in Fig.~\ref{fig_ma08} illustrate the CMDs for all the
selected galaxies and galaxy regions. Their basic properties and
parameters are listed in Table~\ref{tab_sample}.


\begin{deluxetable*}{llllllllll}
\tablewidth{0pc}
\tabletypesize{\scriptsize}
\tablecaption{Basic parameters of our galaxy sample.}
\tablehead{
    \colhead{Target/name} &
    \colhead{Filters} &
    \colhead{$A_V$} &
    \colhead{$m_{\rm TRGB}$} &
    \colhead{$m_{\rm 50\% complete}$} &
    \colhead{$(m-M)_0$} &
    \colhead{$N_{\rm RGB}$} &
    \colhead{$N_{\rm AGB}$} &
    \colhead{$\frac{N_{\rm AGB}}{N_{\rm RGB}}$}
}
\startdata
ESO294-010 & F606W,F814W & 0.018 & 22.397$\pm$0.008 & 28.25 & 26.434  & 1950 & 66  & 0.034$\pm$0.004\\
DDO113     & F475W,F814W & 0.063 & 23.337$\pm$0.026 & 27.43 & 27.349  & 2702 & 98  & 0.036$\pm$0.004\\
DDO44      & F475W,F814W & 0.129 & 23.499$\pm$0.015 & 27.61 & 27.454  & 4966 & 205 & 0.041$\pm$0.003\\
ESO540-032 & F606W,F814W & 0.064 & 23.733$\pm$0.020 & 27.83 & 27.743  & 3627 & 117 & 0.032$\pm$0.003\\
M81F6D1    & F606W,F814W & 0.241 & 23.855$\pm$0.010 & 27.85 & 27.737  & 2326 & 60  & 0.026$\pm$0.003\\
M81K61     & F606W,F814W & 0.226 & 23.823$\pm$0.042 & 28.05 & 27.716  & 8181 & 263 & 0.032$\pm$0.002\\
IKN        & F606W,F814W & 0.181 & 23.869$\pm$0.019 & 26.97 & 27.786  & 7854 & 184 & 0.023$\pm$0.002\\
DDO78      & F475W,F814W & 0.066 & 23.837$\pm$0.017 & 27.55 & 27.818  & 9559 & 394 & 0.041$\pm$0.002\\ 
SCL-DE1    & F606W,F814W & 0.046 & 24.116$\pm$0.023 & 28.41 & 28.110  & 1922 & 96  & 0.050$\pm$0.005\\
DDO71      & F606W,F814W & 0.303 & 23.881$\pm$0.019 & 28.05 & 27.740  & 6540 & 255 & 0.039$\pm$0.002\\
M81F12D1   & F606W,F814W & 0.442 & 23.982$\pm$0.013 & 28.03 & 27.749  & 7294 & 236 & 0.032$\pm$0.002\\
M81K64     & F606W,F814W & 0.165 & 23.907$\pm$0.009 & 28.42 & 27.852  & 6876 & 291 & 0.042$\pm$0.003\\
total      &             &       &                  &       &         &65771 &2265 & 0.0344$\pm$0.0007
\enddata
\tablecomments{$A_V$ and $(m-M)_0$ come from the best MATCH solution.}
\label{tab_sample}
\end{deluxetable*}

\subsection{Selecting AGB and RGB stars}

To identify the location of RGB stars, we adopted the magnitude of the
TRGB ($m_{\rm TRGB}$) from \citet{Dalcanton_etal09}, who identified
the TRGB in uncrowded low extinction regions of each galaxy using the
edge-detection method of \citet{Mendez_etal02}. The TRGB magnitude can
also be converted into a distance modulus, using the foreground
extinction derived by \citet{Schlegel_etal1998}, and the absolute
magnitude of the TRGB in the isochrones of \citet{Marigo_etal08} at
the observed color of the RGB stars used to derive the TRGB.  The
TRGB-based distance modulus does not necessarily agree with the one
reported by MATCH, since the latter can be biased towards larger
distances in an attempt to better fit the magnitude of the
well-populated red clump, which is known to be faint in the
\citet{Marigo_etal08} isochrones currently used in MATCH.  In what
follows, we use the \dmo\ and \av\ derived from MATCH for generating
artificial CMDs. However, for isolating AGB stars, we use the
empirically determined value of $m_{\rm TRGB}$, as given in
Table~\ref{tab_sample}.

An interval of two magnitudes above the TRGB defines the ``AGB
sample''. In all cases, it includes the bulk of (if not all) stars
brighter than the TRGB. Their number, $N_{\rm AGB}$, is typically
between 60 and 400 per galaxy.

The RGB sample, instead, is defined over two magnitudes below the
TRGB. This sample has, typically, a completeness above 95~\%, and
about 30 times more stars than in the AGB sample.

The number ratios between AGB and RGB stars, $N_{\rm AGB}/N_{\rm
RGB}$, are presented in Table~\ref{tab_sample}, together with the
$1\sigma$ random errors. 

The AGB sample as above defined is contaminated by a few RGB stars at
its faintest magnitude bins, because of the scattering of stars to
brighter magnitudes by photometric errors and binaries. For the
galaxies with the smallest $N_{\rm AGB}/N_{\rm RGB}$, the effect is
such that this ratio can be increased by up to 20~\% with respect to
its true value. As we will see below, this is a marginal effect
considering the large discrepancies -- of a few times -- between
present models and the observations. Moreover, the scattering of RGB
stars to brighter magnitudes is fully taken into account in our
simulations (Sect.~\ref{sec_models}), so that no inconsistency results
from counting a few RGB stars in the AGB sample.  On the other hand,
the RGB sample contains both the initial sub-luminous section of the
TP-AGB phase and the excursions to low-luminosities driven by thermal
pulses, as well as stars leaving the early-AGB phase. Based on
evolutionary models, we estimate that this contamination, in the worst
cases, is just a few per cent. Moreover, the numbers of early-AGB
stars are expected to be quite insensitive from the uncertainties in
the mass-loss prescriptions, which instead plague the TP-AGB stars
above the TRGB.

\subsection{Uncertainties in the SFH and AGB/RGB ratio}

The measured $N_{\rm AGB}/N_{\rm RGB}$ ratios have values 
comprised between 0.023 and 0.050.  Random errors are, in all cases,
smaller than 20~\%. The mean value derived by adding all stars in the
sample is $0.0344\pm0.0007$. Clearly, the random noise in the
relative number of AGB stars is less of a concern in our dataset. The
main point of concern, instead is in the correctness of the SFHs for
the host galaxies, which is crucial for the interpretation of the
$N_{\rm AGB}/N_{\rm RGB}$ ratios.

As we show below, the current isochrones overpopulate the AGB
sequences, potentially biasing our SFHs to low star formation rates
and intermediate ages when these stars are included in the fit of the
CMD.  We therefore re-derive the SFHs excluding AGB stars, taking care
to keep all other aspects of the fitting identical.  Specifically, we
exclude CMD regions above the TRGB, and those with either
F606W$-$F814W$>3.5$ or F475W$-$F814W$>4.5$ (depending on the filters
used in the observations) from our fitting.  The resulting SFHs
correspond to the ``no-AGB'' cases in Table~\ref{tab_ratio}. They
differ little from the default ``with-AGB'' cases, indicating that the
structure of the red clump has far more influence on the inferred SFH
at intermediate ages with the weighting scheme currently employed by
MATCH.

We also have applied the ``Z-inc'' option to derive SFH. It consists
of limiting the number of free parameters in our fits by forcing the
fit to only attempt solutions where the metallicity remains constant
or increases with time.  The history returning the best fit to the CMD
is then selected. This process is important in some cases where the
photometry depth or the number of stars are insufficient to reliably
separate the effects of age from those of metallicity.

The initial columns in Table~\ref{tab_ratio} summarize the results in
terms of SFH, presenting the fraction of the SFH in the age intervals
from 0 to 1~Gyr, and from 1 to 3~Gyr, for all cases of SFH recovery
that were tested. It turns out, as expected, that all galaxies but
SCL-DE1 present a very small amount of their SFH at ages $<1$~Gyr, and
roughly 10~\% at $1-3$~Gyr. The SFHs for the different cases (with and
without AGB stars, default or Z-inc) do agree well considering the
typical error bars in this kind of measurement, again with the
exception of SCL-DE1 for which the Z-inc cases provide significantly
older SFHs.

\section{Modeling the data: method and results}
\label{sec_models}
\label{sec_results}

\subsection{Code and method}

To model the ANGST/ANGRRR data, we use a recent version of the
TRILEGAL code \citep{Girardi_etal05}, which generates multi-band mock
catalogs of resolved stellar populations for a given distribution of
distances and extinctions, following some specified star formation
history and age-metallicity relation.  The original code has been
expanded in several aspects to deal with the complex features of
TP-AGB stars \citep[][and later work]{GirardiMarigo07b}, including:
different bolometric corrections and \Teff-color relations for O-rich
and C-rich stars, luminosity ($L$) and \Teff\ variations driven by
thermal pulses, and obscuration by circumstellar dust. In order to
take these effects onto account, the code keeps track of a series of
stellar parameters, like the surface chemical composition, mass loss,
and the period and mode of the long-period variability. The dust
composition has been assumed to be the 60~\% silicate plus 40~\% AlOx
for O-rich stars, and 85~\% amorphous carbon plus 15~\% SiC for C-rich
stars from \citet{Groenewegen06}. 

The stellar evolutionary tracks are the same ones contained in
\citet{Marigo_etal08} isochrones and used by MATCH to derive 
the SFH of our galaxy sample. The transformations from $L$ and
\Teff\ to the HST photometry are described in \citet{Girardi_etal08}; 
however, we have updated it to use the latest transformations for
C-type stars from \citet{Aringer_etal09}, and the total ACS/WFC
throughput curves and zeropoints appropriate for post-July 2006
observations \citep{Mack_etal07,
Bohlin07}\footnote{http://www.stsci.edu/hst/acs/analysis/zeropoints}.

Alternative sets of TP-AGB tracks have been specifically calculated
for the present work, as described below in Sect.~\ref{sec_newtracks}.
TRILEGAL allows these tracks to be replaced quickly and with minimal
human effort.

We first start modelling the data for the complete sample of galaxies
in Table~\ref{tab_sample}, using the same stellar models which were
already used to derive the SFH, but using TRILEGAL instead of
MATCH. TRILEGAL simulates the photometry starting from the SFH file,
and shifts the data to the right distance modulus and extinction. The
simulations include stars up to 2 mags fainter than the faintest
observed one. 

In order to properly account for the real completeness and photometric
errors in our simulations, we proceed as follows. For each simulated
star, an artificial star of similar color and magnitude is randomly
extracted from the existing catalog of artificial stars from
HSTPHOT/DOLPHOT. If that artificial star has been detected by the
photometry software, the differences between the input colors and
magnitudes, and the output ones, are applied to the simulated star,
otherwise the same object is thrown away.

\subsection{Results using MG07 TP-AGB tracks} 
\label{sec_resmg07}

Results from this exercise are shown in Fig.~\ref{fig_ma08}. Notice
that the simulations are forced to have a number of bright RGB stars
(within 2 mag of the TRGB) consistent with the observed one within
$1\sigma$. One can notice that these simulations resemble very much
the observations, except for the TP-AGB regime, for which there is a
clear excess of simulated stars, by factors that can be as large as 6
or 7. The numerical results are tabulated in Table~\ref{tab_ratio}
below. It is also evident that the simulated AGB stars reach much
brighter magnitudes than the observed ones. The excess of AGB stars is
probably related to what has been already detected by
\citet{Gullieuszik_etal08}, \citet{Held_etal10}, and
\citet{Melbourne_etal10}.

\input{rgbagb_table2}

We have checked that these results depend little on the use that MATCH
makes of AGB stars. Indeed, as shown in Table~\ref{tab_ratio}, the SFH
derived when one hides the AGB stars in MATCH are very similar to
those in which they are included, and do not produce great changes in
the predicted AGB/RGB ratios. Moreover, the comparison with models in
which the metallicity is forced to not decrease with the galaxy age
(the Z-inc option in MATCH) also produces similar results. We can only
conclude that the problem resides in the AGB stellar models, and not
in the observed samples or process of SFH recovery.

The observed galaxies always present, overall, a small fraction of
intermediate age stars (Table~\ref{tab_ratio}), or some
moderately-high metallicities, so that one might think that the
results do not actually correspond to old metal-poor populations as it
was intended to be. We know however that in dwarf galaxies the
youngest (and more metal-rich) star formation is, as a rule,
concentrated in the galaxy centers \citep[e.g.][and references
therein; also Gilbert et al. in preparation]{Weisz_etal08,
Stinson_etal09}.  Therefore, for a subsample of our dwarf galaxies, we
have selected the external rings for which the SFH is expected to be
older than for the overall galaxy. The results are amended at the end
of Table~\ref{tab_ratio}, for the no-AGB no-Z-inc option
only. Although these outer regions are indeed slightly older than the
entire galaxies, they show essentially identical $N_{\rm AGB}/N_{\rm
RGB}$ ratios as the inner regions, suggesting that old populations
dominate the CMDs at all radii. Therefore, the basic result for the
excess of predicted AGB stars, by a factor of about 4, still remains
in these data.

\subsection{What is the problem with current TP-AGB models?}
\label{sec_problem}

The above-mentioned results were obtained with the use of MG07 TP-AGB
evolutionary tracks, which constitute a quite non-standard grid of
such models. They have considered, for the first time, features like
the changes in molecular opacities and mass-loss rates as the surface
composition of TP-AGB stars passes from O- to C-rich, and the increase
in the mass-loss rates as the long-period pulsation switches from the
first overtone to the fundamental mode. Added to these improved
prescriptions, there were attempts to calibrate the poorly known
parameters of the models via the fitting of observational data. To be
explicitly considered in the fitting were (1) the lifetimes of AGB
stars as a function of stellar mass as derived from star counts in
Magellanic Cloud clusters, and (2) the C-star luminosity functions in
both Clouds. After this calibration, other quantities were checked, as
for instance the integrated colors of Magellanic Cloud clusters, and
the initial--final mass relation derived from white dwarfs in the
solar neighborhood. All these comparisons revealed an overall
improvement in fits to data, even if there remained some problems,
like for instance the integrated $V-K$ and $J-K$ colors being
apparently too red as compared to the cluster data in the age interval
from $10^8$ to $4\times10^8$~yr \citep[see the appendix
in][]{Marigo_etal08}.

One can separate the prescriptions used to build MG07 models into two
broad groups: those which depend on the composition of stellar
atmospheres -- and especially on the C/O ratio -- and those which do
not. The first set of prescriptions is probably of secondary relevance
for the present work, since we are dealing with stars in metal-poor
and old stellar systems, most of which do not suffer convective
dredge-up on the TP-AGB and hence are expected to be predominantly
O-rich (of spectral types K and M). For the stars we are considering
here, the TP-AGB lifetime and termination luminosity are crucially
determined by the process of mass loss.

For O-rich stars, MG07 adopted a mass-loss prescription heavily based
on \citet{BowenWillson91} dynamical model atmospheres for
fundamental-mode pulsators, and with a dependence on metallicity as
derived from \citet{Willson00}. The initial period of TP-AGB mass-loss
as first-overtone pulsators was based on a set of relations derived
from very few exploratory models by \citet{Bowen88}. The procedure
however was not straightforward, involving a large number of
assumptions and fitting relations, caused by the different assumptions
and ranges of parameters (e.g., the different underlying
radius--luminosity--mass relation) between the original models from
Bowen and Willson, and the synthetic TP-AGB models. A detailed review
of the literature, together with the results of Fig.~\ref{fig_ma08},
suggests a complete revision of this scheme, as we now describe.

\subsection{New TP-AGB tracks} 
\label{sec_newtracks}

Our main problem is to reduce the lifetimes of TP-AGB tracks which,
due to their low mass and metallicity, present relatively low
luminosities and hot temperatures. Under such conditions, the
classical dust-driven winds typically found in more luminous and cool
AGB stars might be less efficient. In any case, we need to introduce
some description for the ``pre-dusty'' winds (with rates $\dot M_{\rm
pre-dust}$), i.e. before radiation pressure on dust grains becomes the
main driving agent of mass loss \citep{ElitzurIvezic01}.  Among the
few such formulas which have been proposed in the literature, we
decided to test the semi-empirical one by \citet{SchroderCuntz05}
\begin{equation}
\dot M_{\rm pre-dust} = 
\eta \frac{L\, R}{M}\displaystyle \left(\frac{T_{\rm
eff}}{4000\,{\rm K}}\right)^{3.5} 
\left(1+\frac{g_{\odot}}{4300 \, g}\right)\,\,[M_{\odot}\,{\rm yr}^{-1}]
\end{equation} 
where $R$, $M$, and $L$ are the stellar radius, mass, and luminosity
expressed in solar units; $g$ and $g_{\odot}$ indicate the stellar and
solar surface gravity, respectively.  This formula is a kind of
\citet{Reimers75} law and was originally derived to describe the 
mass loss suffered by red giants under the assumption that the stellar
wind originates from magneto-acoustic waves operating below the
stellar chromosphere.  The fitting parameter $\eta$ is set to
$0.8\times 10^{-13}$, a value calibrated by \citet{SchroderCuntz05} to
reproduce the morphology of horizontal branches in globular clusters.

The \citet{SchroderCuntz05} relation is employed from the beginning of
the TP-AGB phase until a critical stage is met, i.e., the attainment
of the minimum mass-loss rate, $\dot M_{\rm dust}^{\rm min}$, required
for the development of a dust-driven wind. At each time step $\dot
M_{\rm dust}^{\rm min}$ is evaluated numerically following the
analysis by \citet{GailSedlmayr87}, from the condition
\begin{equation}
\label{eq_mdotmin}
\alpha=\frac{L\, \kappa}{4\pi c G M} \, = 1\, ,
\end{equation}
which expresses the balance between the outward force caused by
radiation pressure on dust and the inward gravitational pull of the
star.  Here $c$ is the light speed, $G$ is the gravitational constant,
and $\kappa$ is the flux-averaged mass extinction coefficient of the
gas-dust mixture. Following \citet{FerrarottiGail06} it is reasonable
to assume that in the dust-driven outflow
\begin{equation}
\kappa= \kappa_{\rm gas} + \sum_{i} f_i \kappa_{{\rm dust},i}
\end{equation} 
where the opacity contributions from the gas, $\kappa_{\rm gas}$, and
dust species, $\kappa_{{\rm dust},i}$, are expressed as Rosseland
means, while $f_i$ represents the condensation degree of the $i^{\rm
th}$ dust species under consideration.

In our calculations we consider different dust types depending on the
star chemical type: silicates (pyroxene, olivine and quartz) and iron
in the case of M-stars (C/O~$<~1$), iron in the case of S-stars
(C/O~$\sim~1$), and carbon, silicon carbide, and iron in the case of
C-stars (C/O~$>~1$).  The dust opacities are evaluated through the
relations provided by \citet{GailSedlmayr99},
\citet{FerrarottiGail01}, and \citet{FerrarottiGail02}, and the
corresponding condensation degrees are estimated with the analytic
fits proposed by \citet{Ferrarotti03}.

These latter depend on the current $\dot M$, so that the minimum
mass-loss rate of a dust-driven wind, $\dot M_{\rm dust}^{\rm min}$,
is found iteratively through Eq.~(\ref{eq_mdotmin}).

As soon as $\dot M_{\rm pre-dust} \ge \dot M_{\rm dust}^{\rm min}$,
the star enters the dust-driven wind regime, which is treated 
according to two formalisms:
\begin{itemize}
\item case {\bf A}: based on \citet{BowenWillson91}, but relaxing 
the metallicity dependence suggested by \citet{Willson00};
\item case {\bf B}: based on \citet{Bedijn88}, but with a somewhat
different calibration of the parameters.
\end{itemize}

Case {\bf A} for dust-driven winds is the same as in MG07, but for the
correction factor describing the explicit metallicity dependence,
which is left out in the new TP-AGB models.  Nonetheless, an intrinsic
metallicity effect still remains in the computed mass-loss rates, via
the stellar surface parameters $R$ and $T_{\rm eff}$. In fact at lower
metallicities the atmospheres of AGB stars tend to be hotter and more
compact.  In practice, case {\bf A} is meant to explore the hypothesis
that the dust-driven winds may depend only mildly on the initial metal
content.

Case {\bf B} closely resembles the approach developed by
\citet{Bedijn88}, to which the reader is referred for all details.
Briefly, assuming that the wind mechanism is the combined effect of
two processes, i.e., radial pulsation and radiation pressure on the
dust grains in the outermost atmospheric layers, \citet{Bedijn88}
derived a formalism for the mass-loss rate as a function of basic
stellar parameters, $M$, $R$, $T_{\rm eff}$, and the photospheric
density $\rho_{\rm ph}$.  Similarly to \citet[][see his Fig. 1 and
appendix B]{Bedijn88} the free parameters have been calibrated on a
sample of Galactic long-period variables with known mass-loss rate,
pulsation period, stellar mass, radius, and effective temperature.
More details about the fit procedure will be given elsewhere (Marigo
et al., in preparation).

It should be noted that, in our calculations, $R$, $T_{\rm eff}$, and
$\rho_{\rm ph}$ are derived from numerical integrations of complete
envelope models extending from the photosphere down to the degenerate
C-O core \citep[see][for details]{Marigo_etal99}.  A key prerogative
of our TP-AGB code is that at each time step, low-temperature
opacities are computed, for the first time, on-the-fly with the
{\AE}SOPUS tool \citep{MarigoAringer09}, thus assuring a full
consistency with the surface chemical composition. In this way we
avoid the loss in accuracy that otherwise must be paid when
interpolating on pre-computed opacity tables (the standard approach in
stellar evolution models).  This is important since low-temperature
opacities are crucial in determining the position of a giant in the
Hertszprung--Russell diagram.  The opacities of carbon-rich stars
(C/O$~>1$), for instance, may be significantly higher, on average,
than in oxygen-rich stars (C/O$~<1$), so that the atmospheres of
carbon-rich stars are usually less dense (and more extended) than
those of oxygen-rich stars
\citep{Marigo02}.

\begin{figure}
\resizebox{1.\hsize}{!}{\includegraphics[]{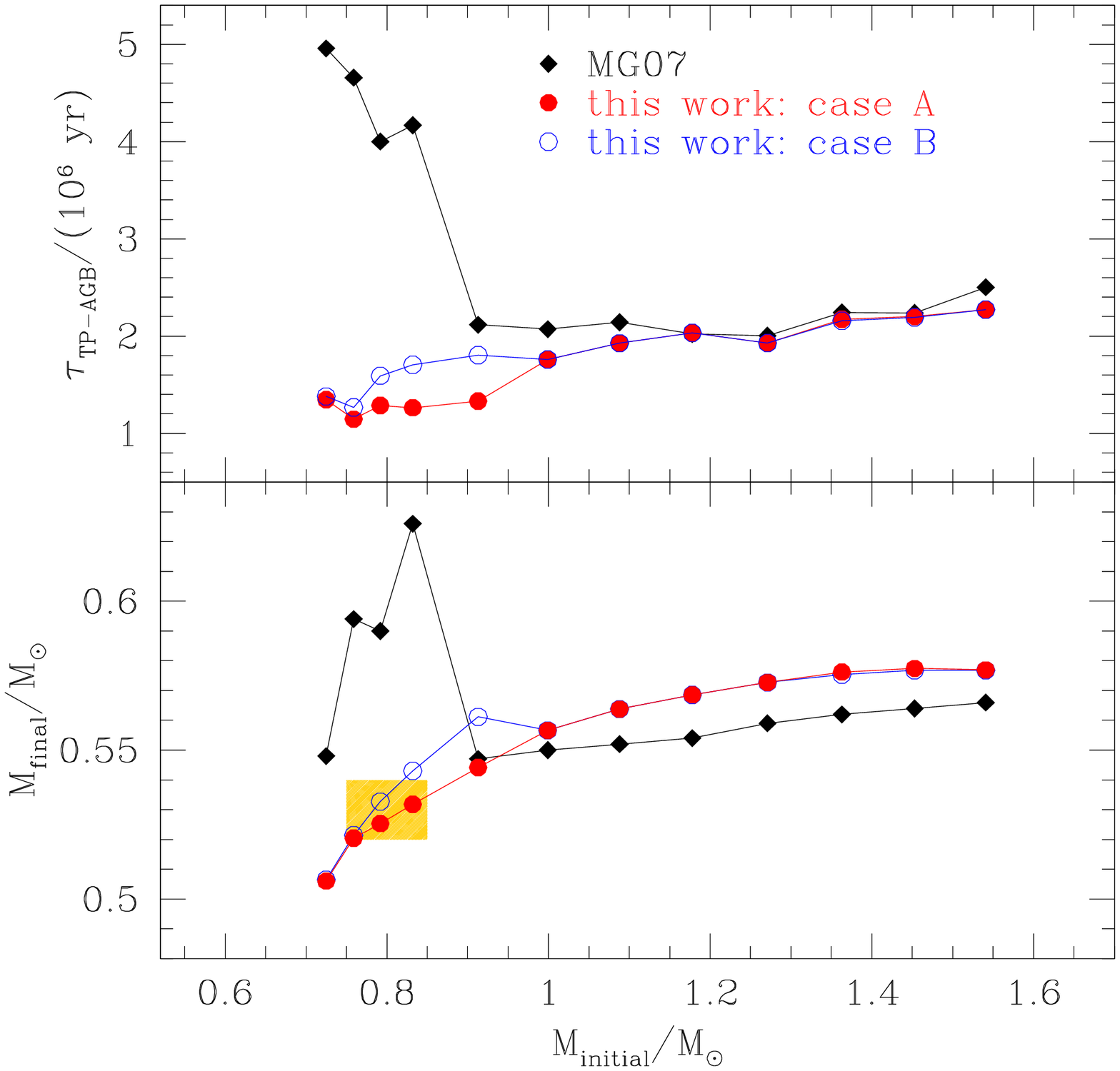}}
 \caption{Comparison of AGB models with initial metallicity $Z=0.001$
 and for three different mass-loss prescriptions, as described in the
 text.  Top panel: TP-AGB lifetimes as a function of the initial
 stellar mass. Bottom panel: Final core masses left after the ejection
 of the envelope as a function of the initial stellar mass.  The
 hatched rectangle shows the initial-final mass relation for the
 Galactic globular custer M4, according to the recent white dwarf
 measurements by \citet{Kalirai_etal09}.}
\label{fig_agb_taumf}
\end{figure}

Figure~\ref{fig_agb_taumf} compares the MG07 results with the new
ones, in terms of (i) duration of the TP-AGB phase and (ii)
initial--final mass relation.  It is evident the significant reduction
(by a typical factor of $3-4$) of the TP-AGB lifetimes for initial
masses $M_{\rm initial} \la 0.9\, M_{\odot}$, as a consequence of the
new prescriptions for the mass-loss rates. The shortening of the
TP-AGB lifetimes derives from two concurring factors, i.e.: the
non-negligible mass-loss rates already prior the onset of the
dust-driven wind, and the higher efficiency during the dusty regime
compared to MG07.

\begin{figure*}
\resizebox{0.45\hsize}{!}{\includegraphics[]{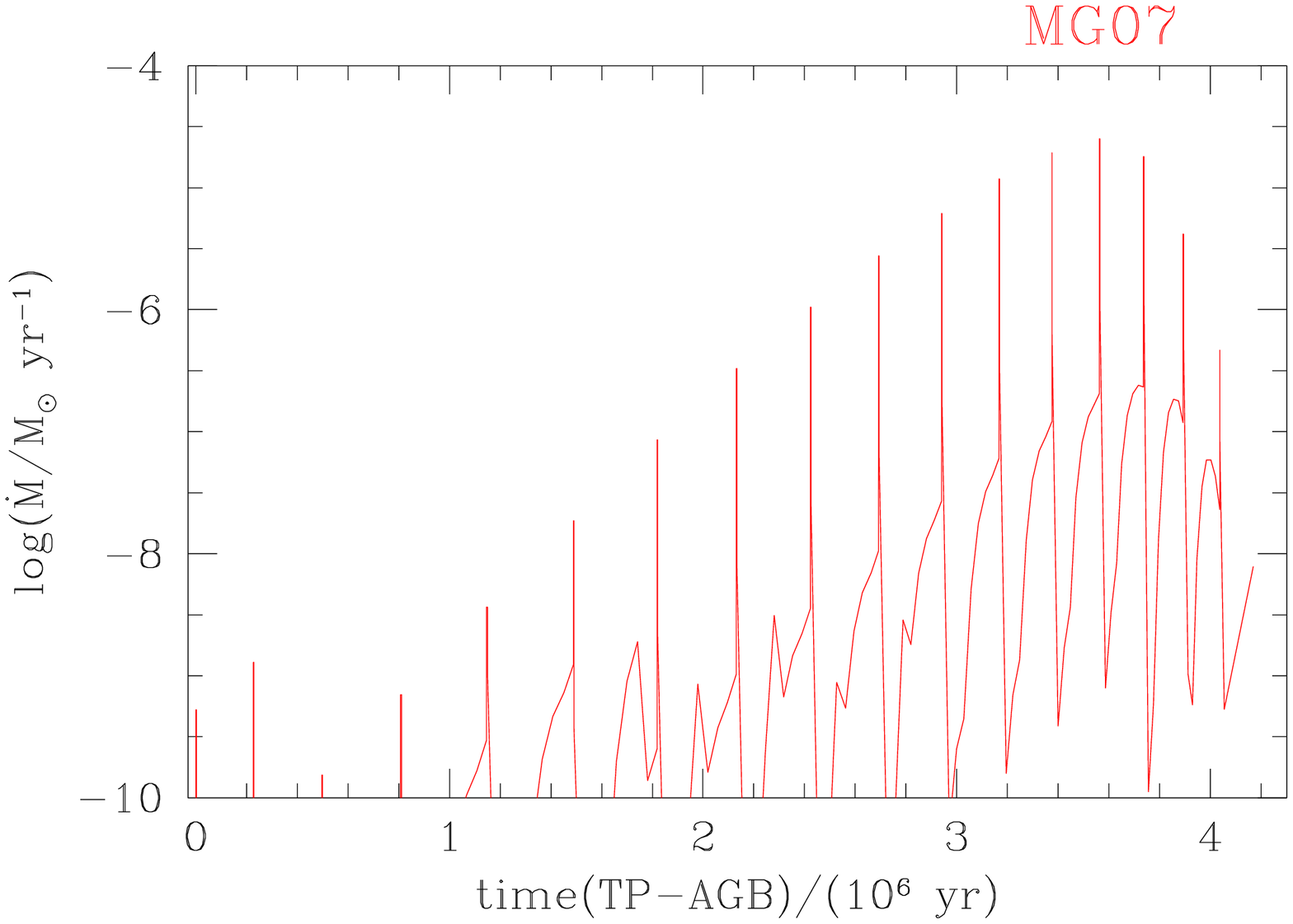}}
\resizebox{0.45\hsize}{!}{\includegraphics[]{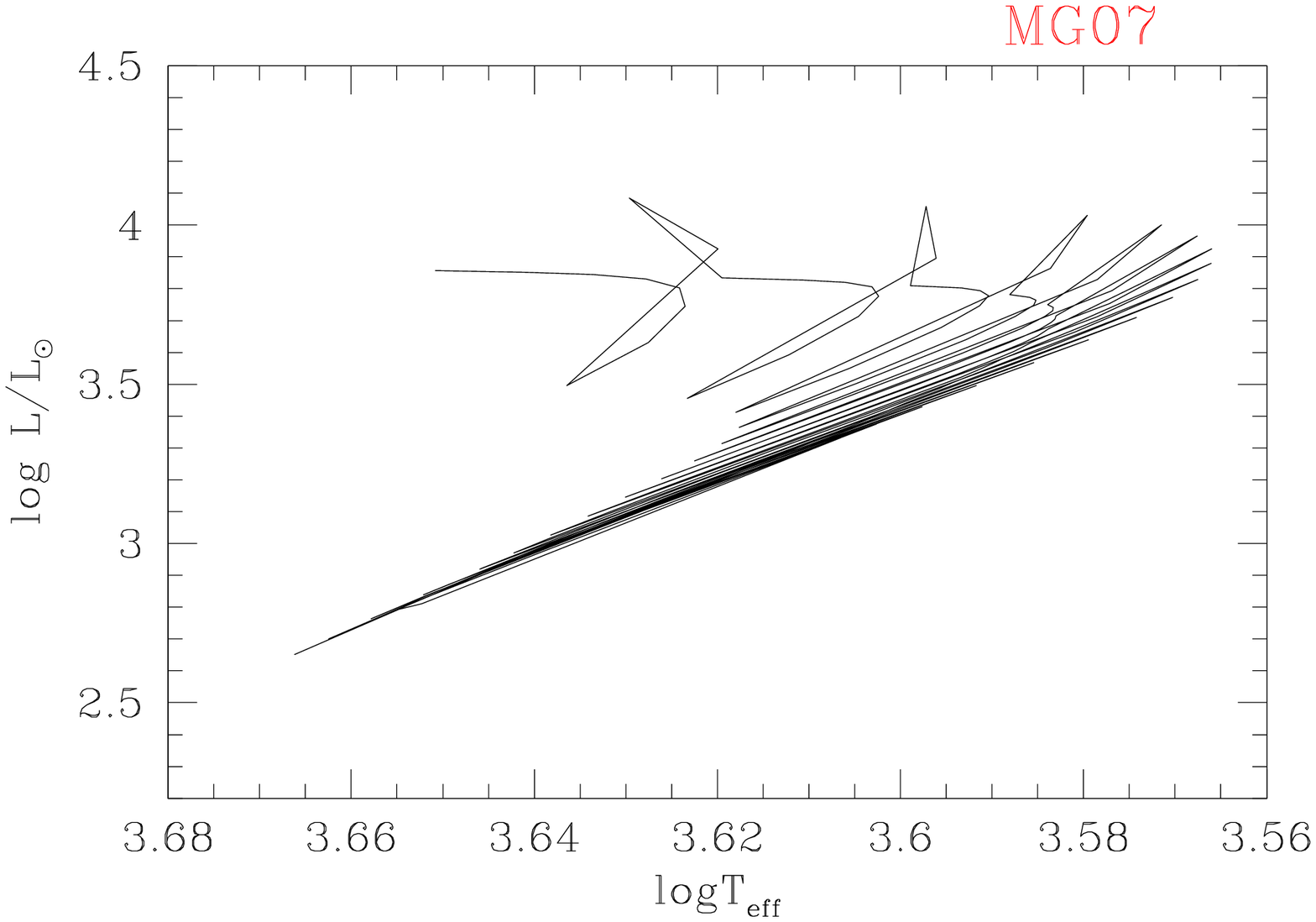}}
\\
\resizebox{0.45\hsize}{!}{\includegraphics[]{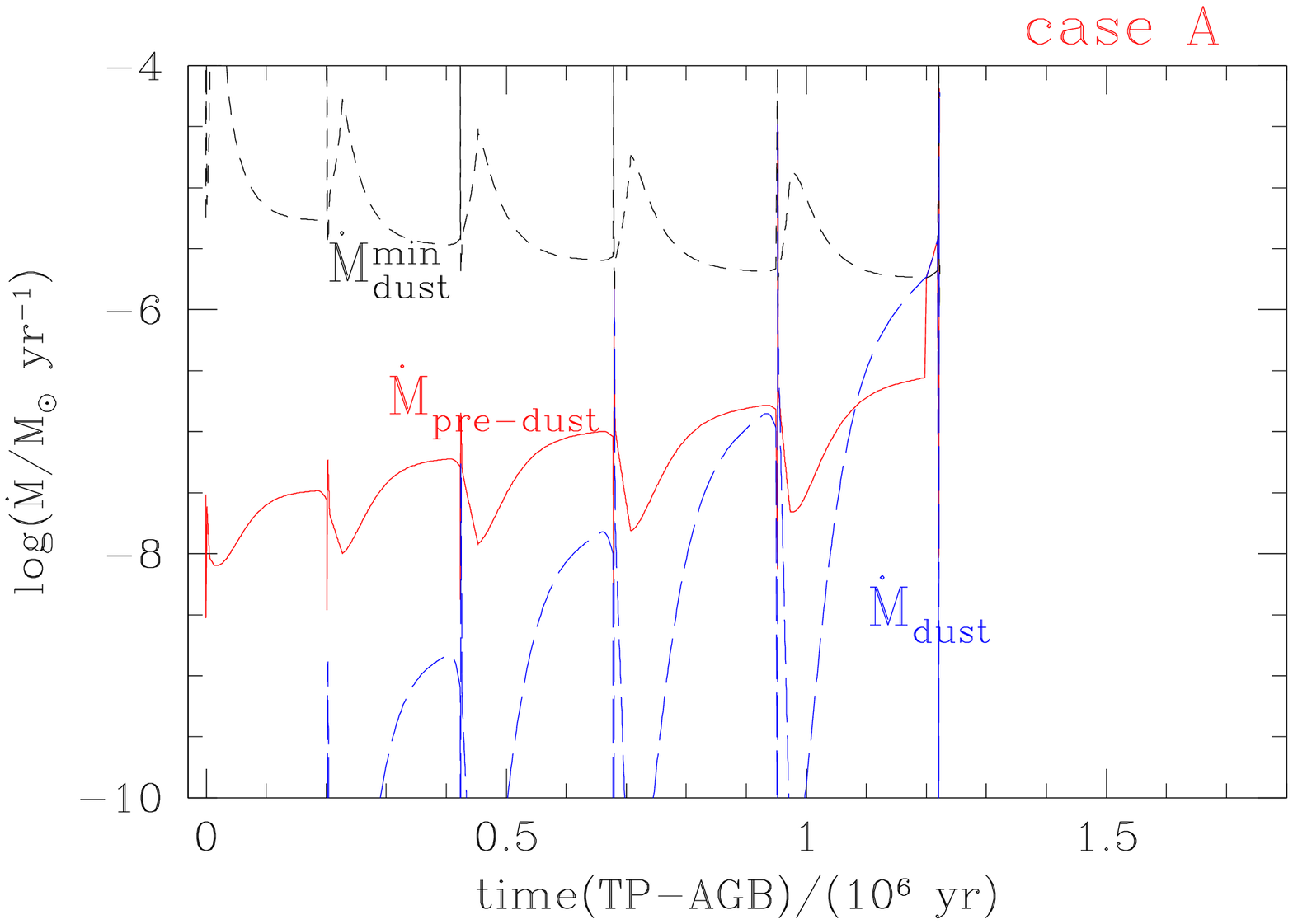}}
\resizebox{0.45\hsize}{!}{\includegraphics[]{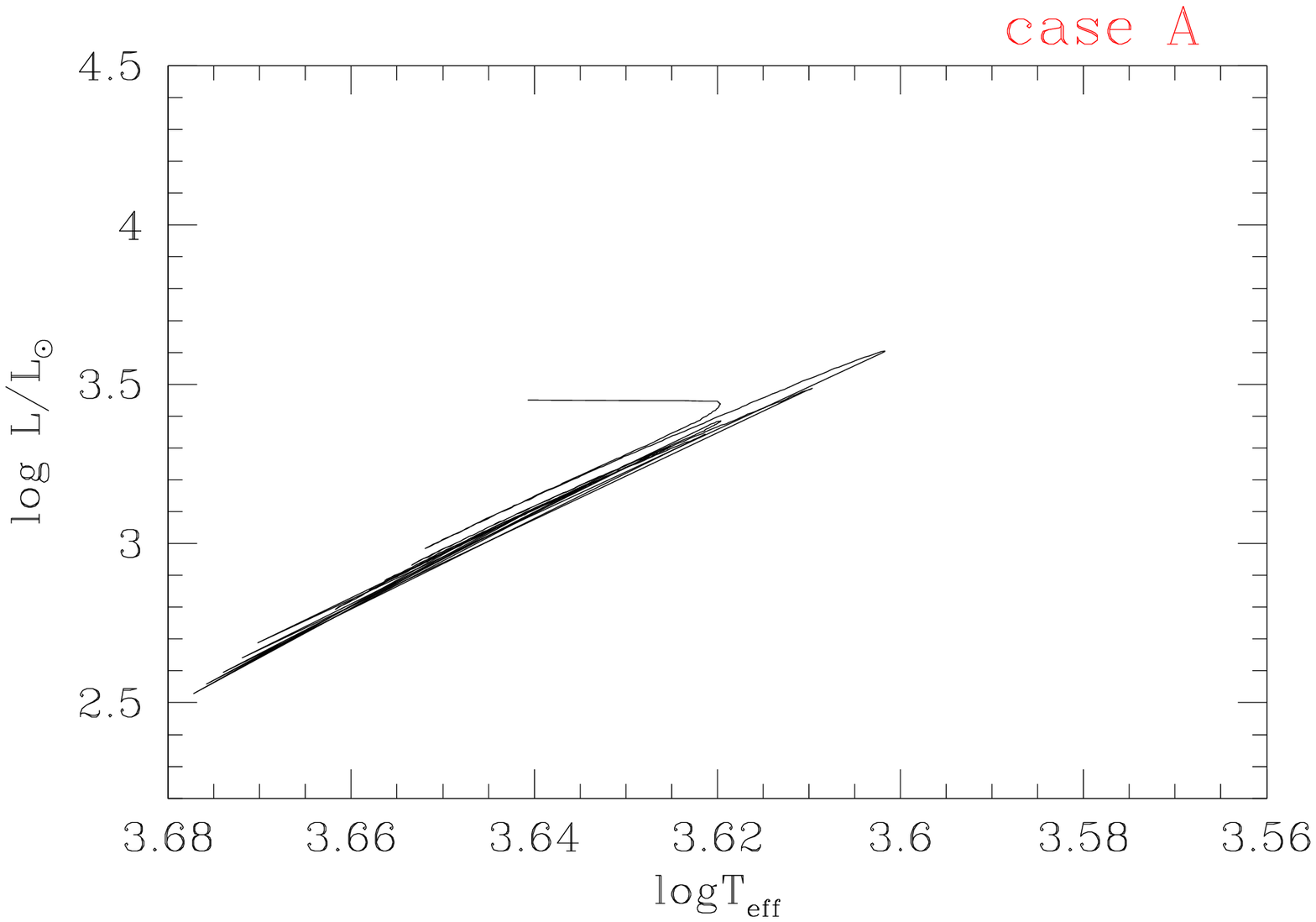}}
\\
\resizebox{0.45\hsize}{!}{\includegraphics[]{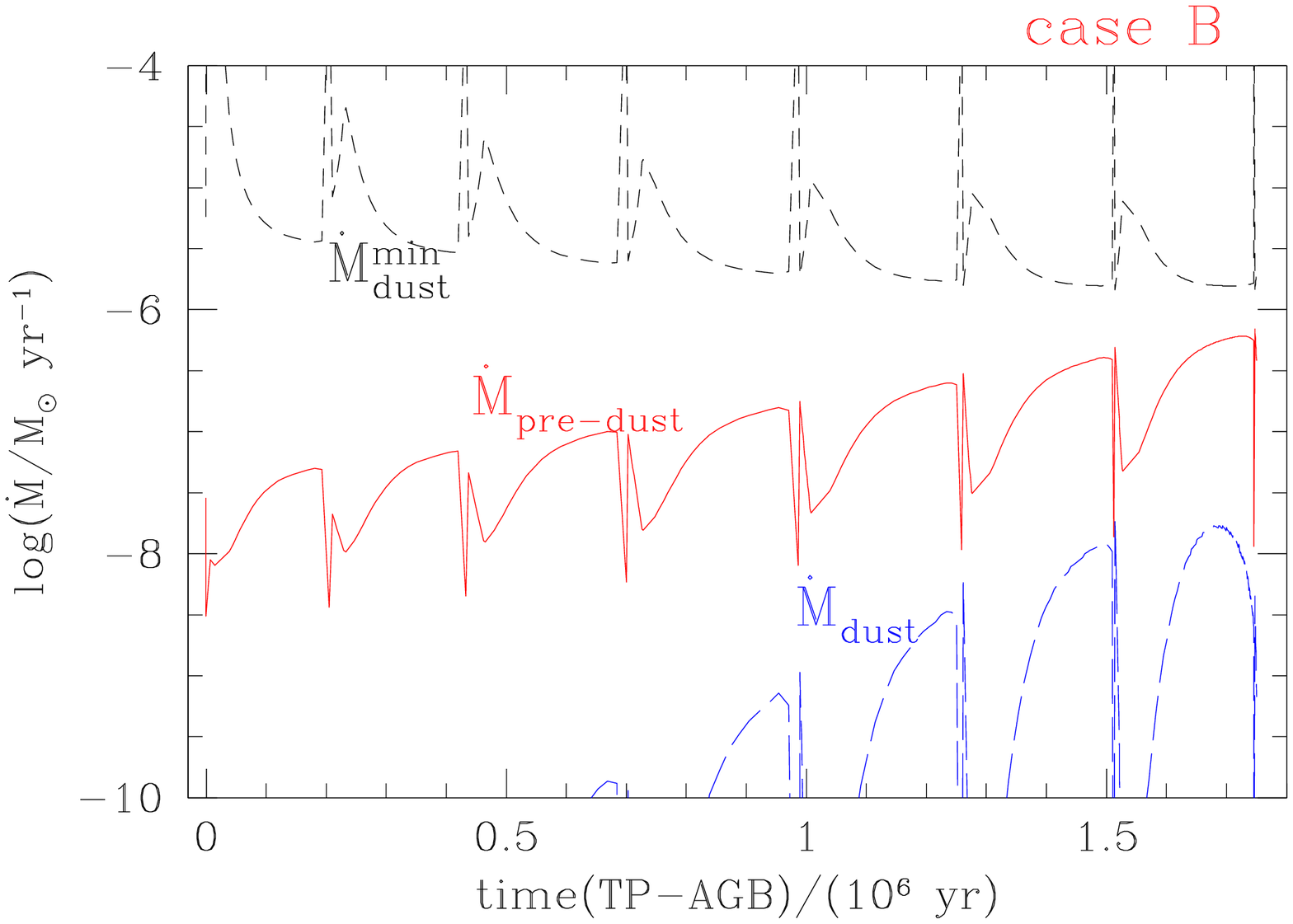}}
\resizebox{0.45\hsize}{!}{\includegraphics[]{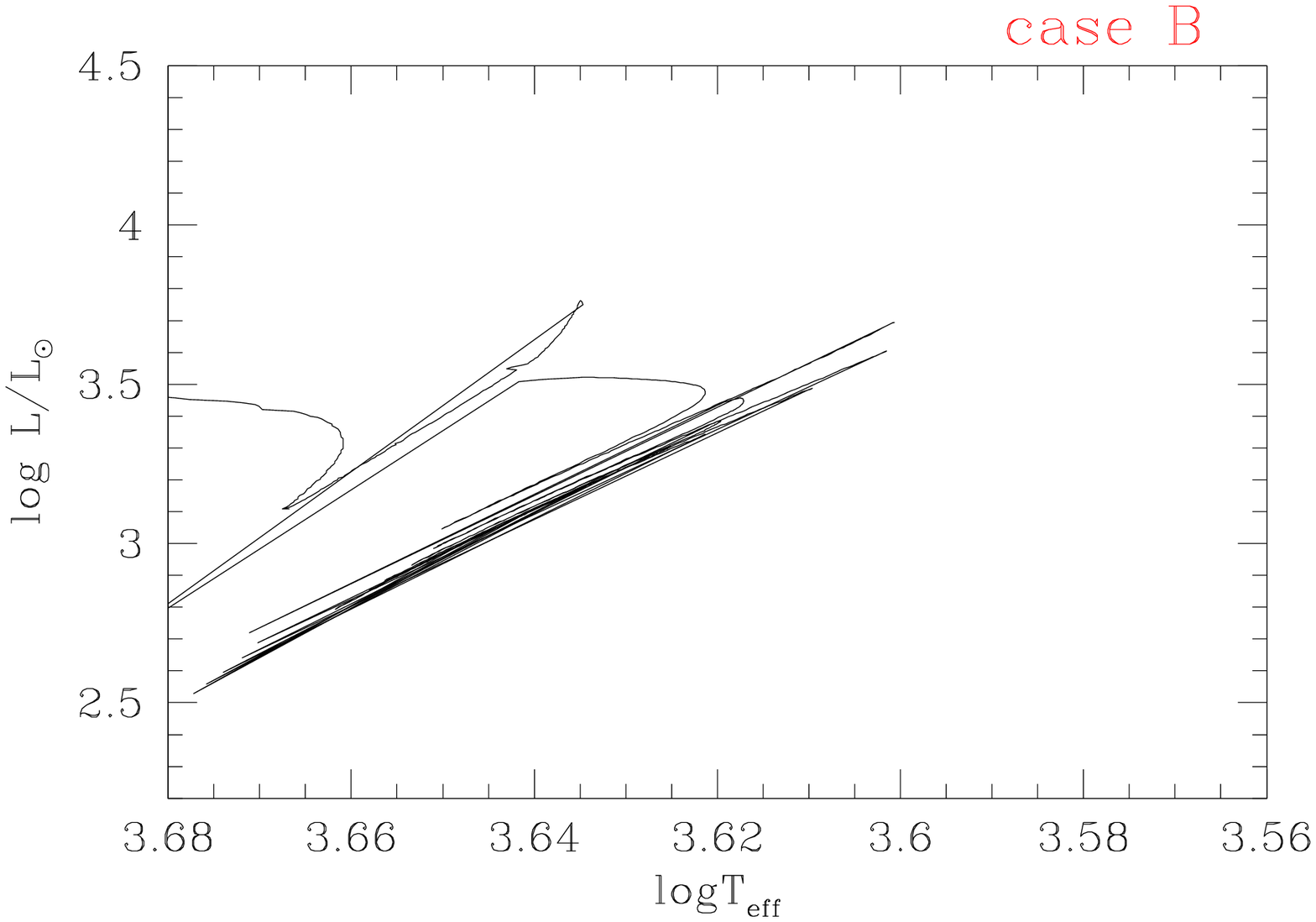}}
\caption{Evolution during the TP-AGB phase of a model with 
$M_{\rm init} =0.832\, M_{\odot}$ (corresponding to a mass at the
first thermal pulse $M_{\rm 1TP} =0.80\, M_{\odot}$) and initial
metallicity $Z=0.001$, of both in the of the mass-loss rate versus
time (left panels) and in the Hertzprung--Russell diagram (right
panels).  The top panels show the results for MG07 models, while the
middle and bottom panels diplay the new calculations, for both
formalisms of mass-loss.  The pulsation-dust-driven mass loss is
represented with a dashed blue line, and $\dot M_{\rm dust}^{\rm min}$
is plotted with a dotted black line.  As long as $\dot M_{\rm dust}$
keeps lower than $\dot M_{\rm dust}^{\rm min}$, the current mass-loss
rate is given by $\dot M_{\rm pre-dust}$ following the
\citet{SchroderCuntz05} formalism. Refer to the text for more
explanation. }
\label{fig_agbmdot}
\end{figure*}

This fact is illustrated in Fig~\ref{fig_agbmdot}, showing the
predicted evolution of the mass-loss rate for a model with initial
mass $M_{\rm init} =0.832\, M_{\odot}$ and metallicity $Z=0.001$,
according to the three different choices of mass-loss formalisms
discussed here. The evolutionary time is counted since the first
thermal pulse up to the ejection of almost the entire envelope
(calculations are stopped as soon as the envelope mass falls below
$0.01\, M_{\odot}$).  One critical feature common to cases {\bf A} and
{\bf B} is that, despite the low metallicity, the mass-loss rate $\dot
M_{\rm pre-dust}$ predicted by the \citet{SchroderCuntz05} relation is
already efficient, quickly increasing from $\approx 10^{-8}$ to
$\approx 5\times 10^{-7}\, M_{\odot}\,{\rm yr}^{-1}$ in a few thermal
pulses.  These rates are high enough to determine the ejection of the
entire stellar mantle even before the development of the dust-driven
wind (case {\bf B}; bottom panel), or to favor its earlier onset
(case {\bf A}; middle panel).  In general the use of mass-loss
prescription {\bf A} leads to somewhat shorter TP-AGB lifetimes than
case {\bf B}.

Consequently, compared to MG07, the models with the new mass loss
descriptions predict a sizable reduction of the final masses, that are
now mostly comprised in the interval $0.51 \la (M_{\rm final}/M_{\odot})
\la 0.55$ for $0.70 \la (M_{\rm initial}/M_{\odot}) \la 0.90$ (bottom
panel of Fig.~\ref{fig_agb_taumf}).  The most important result is that
with no attempt to tune the mass loss, the new TP-AGB models are able
to naturally recover the empirical initial-final mass relation of
population~II stars (bottom panel of Fig.~\ref{fig_agb_taumf}) as
derived by \citet{Kalirai_etal09} from the first direct mass
measurements of individual white dwarfs in the Galactic globular
cluster M\,4.  We have also verified that all metallicity sets in the
range $0.0001 \le Z \le 0.001$ intersect the empirical data.

At larger initial masses, $0.9 \la M_{\rm initial}/M_{\odot} \la 1.5$,
instead, there is a general agreement between the three sets of
models, which are all predicted to make the transition to the C-rich
domain (C/O $>1$), given the higher efficiency of the third dredge-up
at the low metallicities here considered. In this case the short
TP-AGB lifetimes are controlled by the higher mass-loss rates for
C-stars (following the reduction of $T_{\rm eff}$ as soon as C/O
exceeds unity), while the flattening of the initial-final mass
relation is mainly shaped by the deep dredge-up events. Interestingly,
the three mass-loss prescriptions adopted here converge to similar
results.

A general finding is that low-mass low-metallicity models ($M_{\rm
initial}\la 1.0\, M_{\odot}$) would remain O-rich for all their TP-AGB
evolution, experiencing non-negligible mass loss already before
reaching the conditions to activate a dust-driven wind (not excluding
that some passive dust could actually be present), so that the
mass-loss mechanism on the TP-AGB should be ascribed to a different
(unknown) driver, e.g., magneto-acoustic waves within the chromosphere
\citep[as discussed in][]{SchroderCuntz05}. In contrast, models with
larger stellar masses, at roughly $M \ge 1.0\, M_{\odot}$, would
quickly become C-rich as a consequence of the third dredge-up; under
these conditions the long-period pulsation and dust condensation would
become efficient enough to trigger a radiative wind, with consequent
strong enhancement of the mass loss and quick termination of the AGB
phase.  This prediction is in agreement with the results of detailed
models of dust formation \citep{FerrarottiGail06}.

Finally, we remark that we apply the above-described mass-loss
formulas solely during the TP-AGB. We recognize that it would be
important to explore the effect they have also for the RGB and
early-AGB phases, but this is difficult in practice. It would require
the calculation of more extended grids of evolutionary tracks,
properly evaluating the mass loss during the evolution on the
early-AGB, and including additional sets of horizontal branch models
of smaller masses than presently available. This is very time
consuming. Moreover, the evolutionary effects expected would be much
smaller than those described above.

\subsection{Results with the new TP-AGB tracks}

\begin{figure*}
\includegraphics[width=0.33\textwidth]{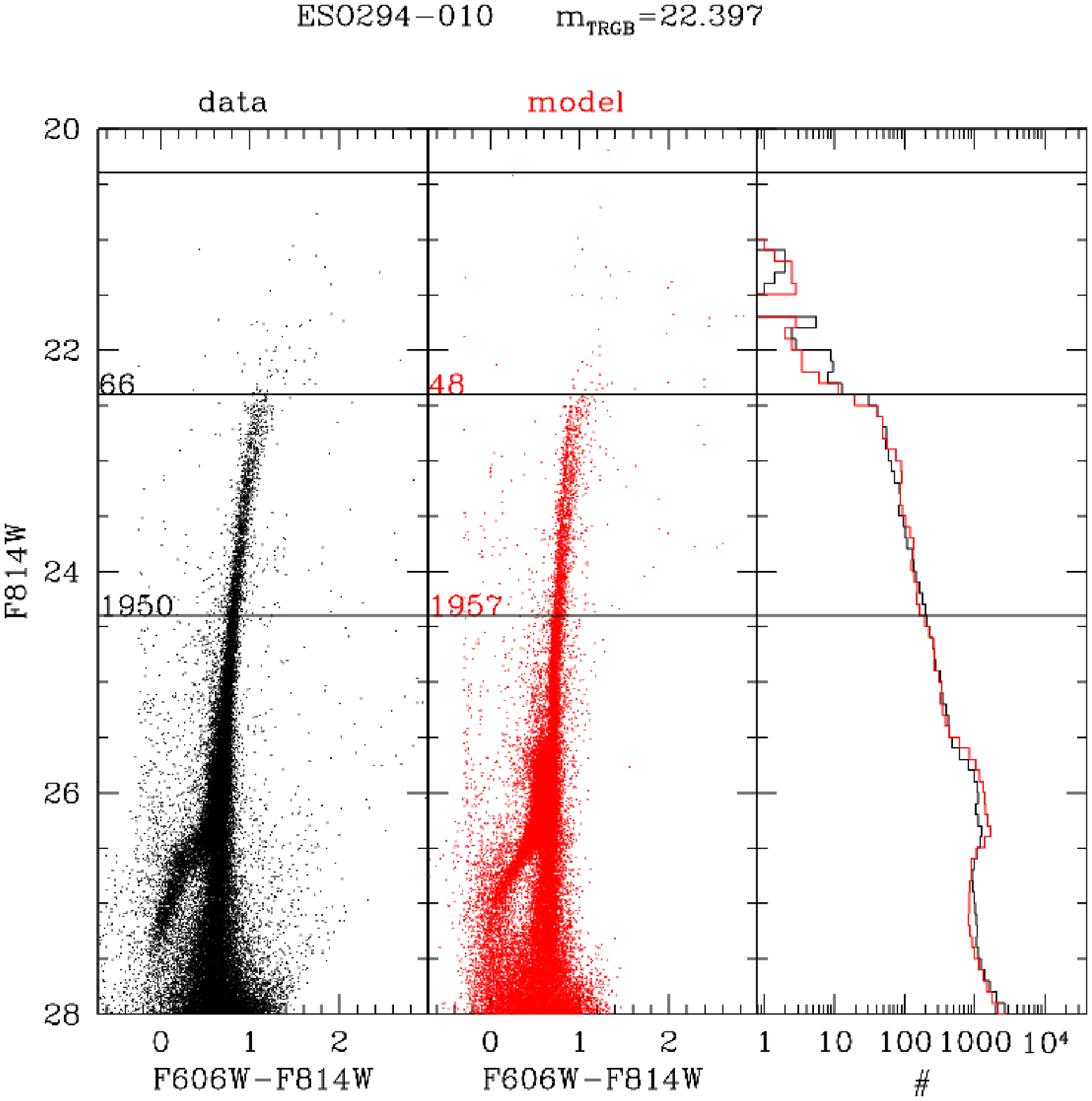}
\includegraphics[width=0.33\textwidth]{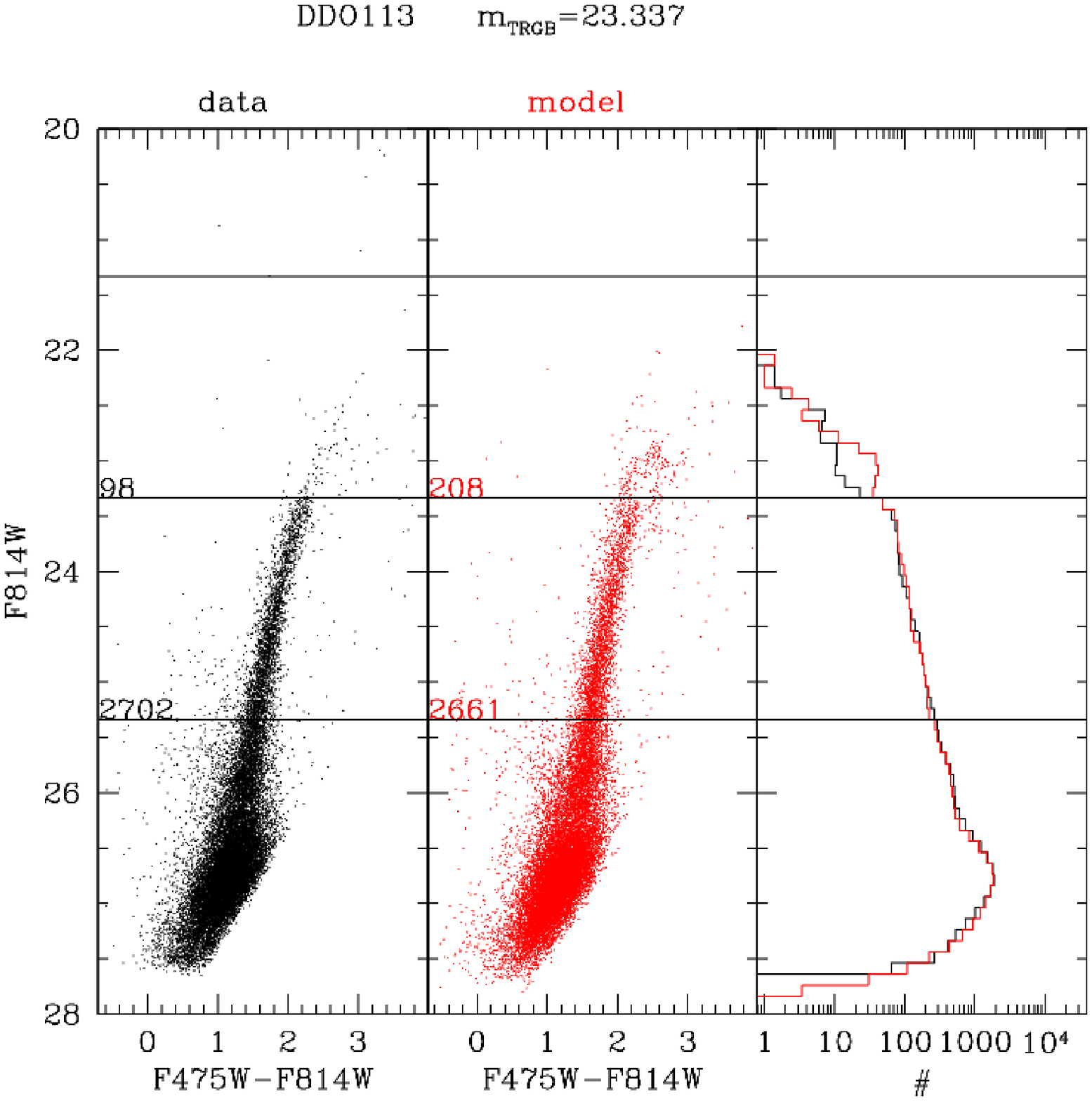}
\includegraphics[width=0.33\textwidth]{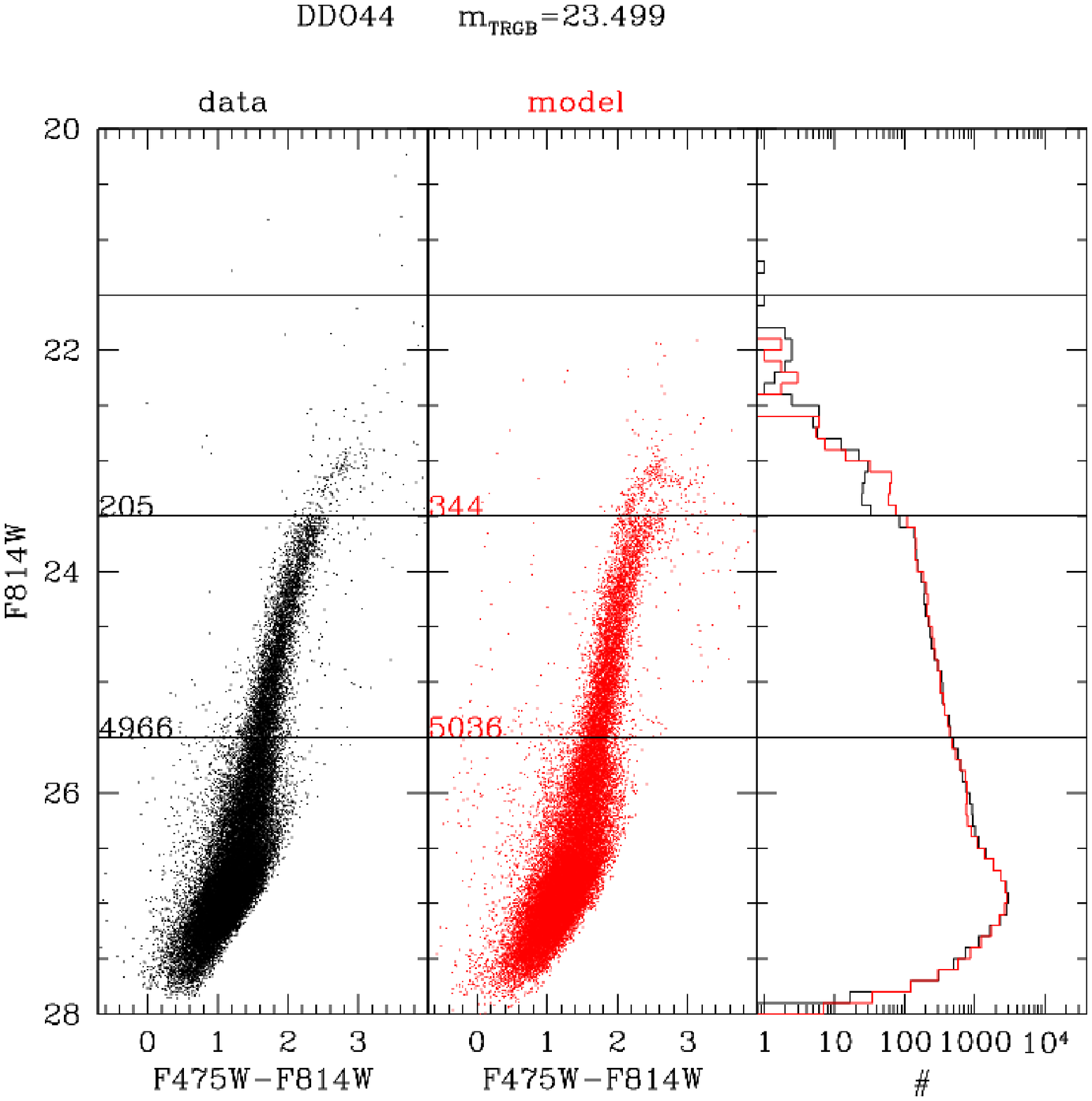}
\\
\includegraphics[width=0.33\textwidth]{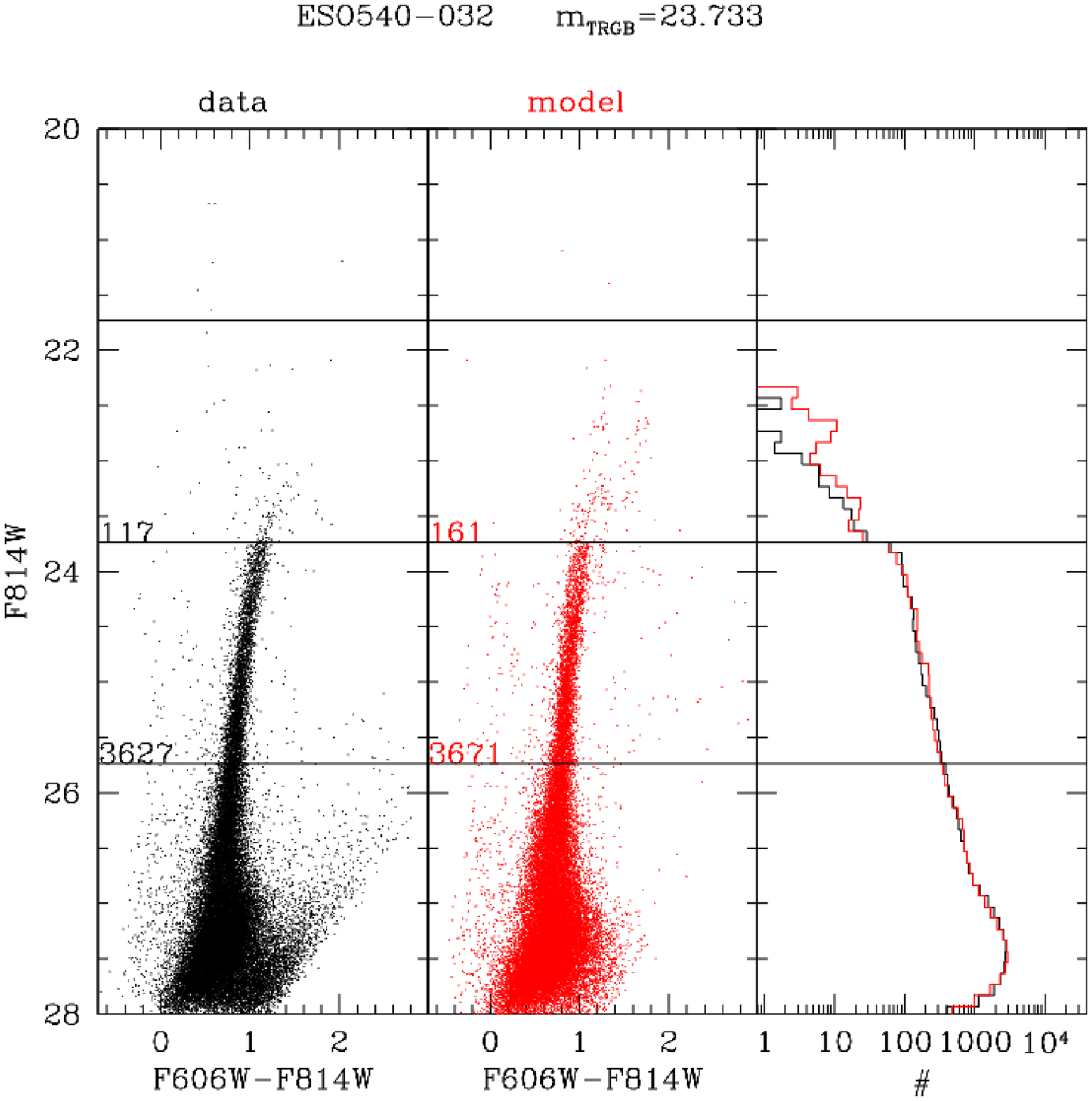}
\includegraphics[width=0.33\textwidth]{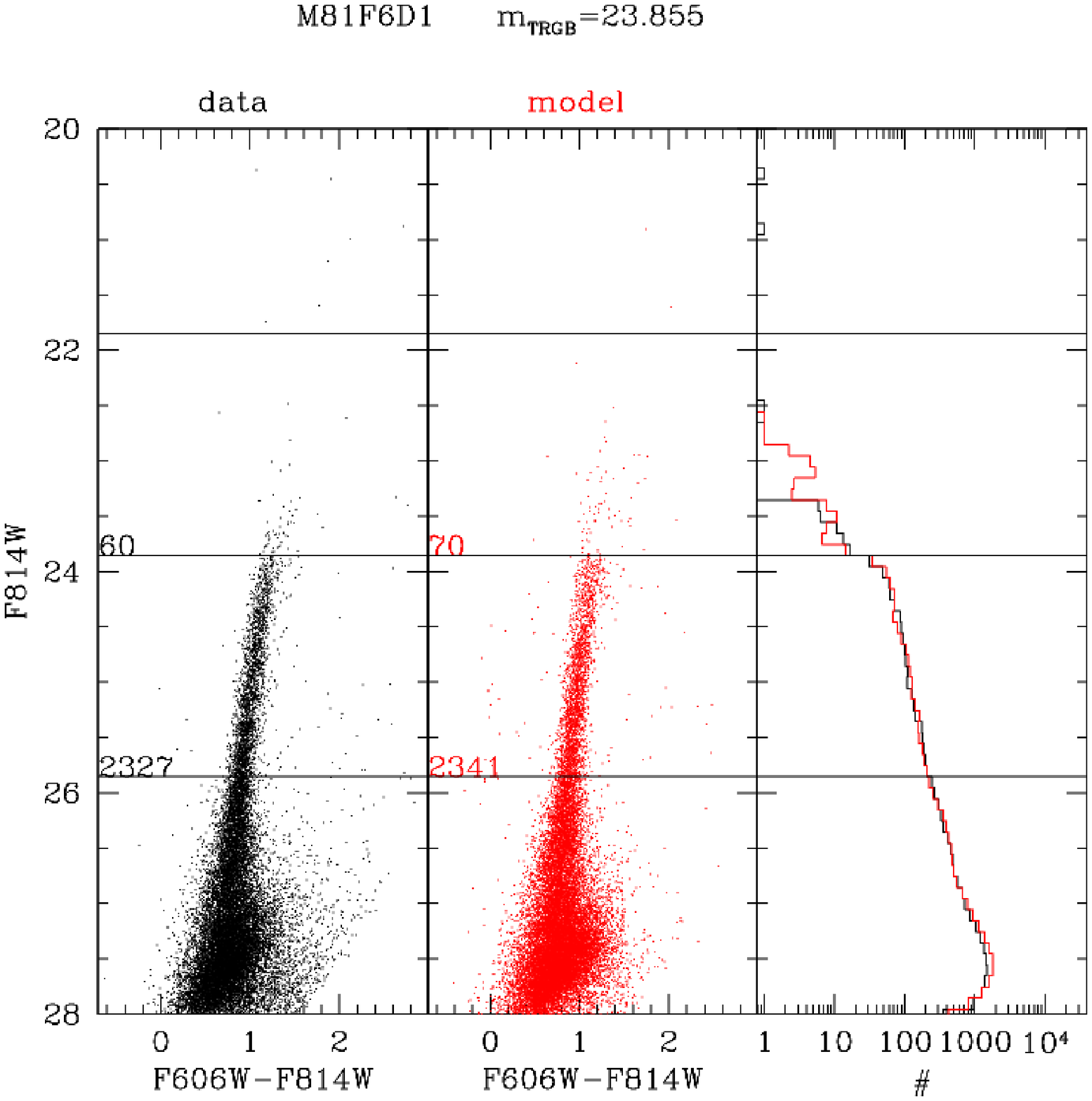}
\includegraphics[width=0.33\textwidth]{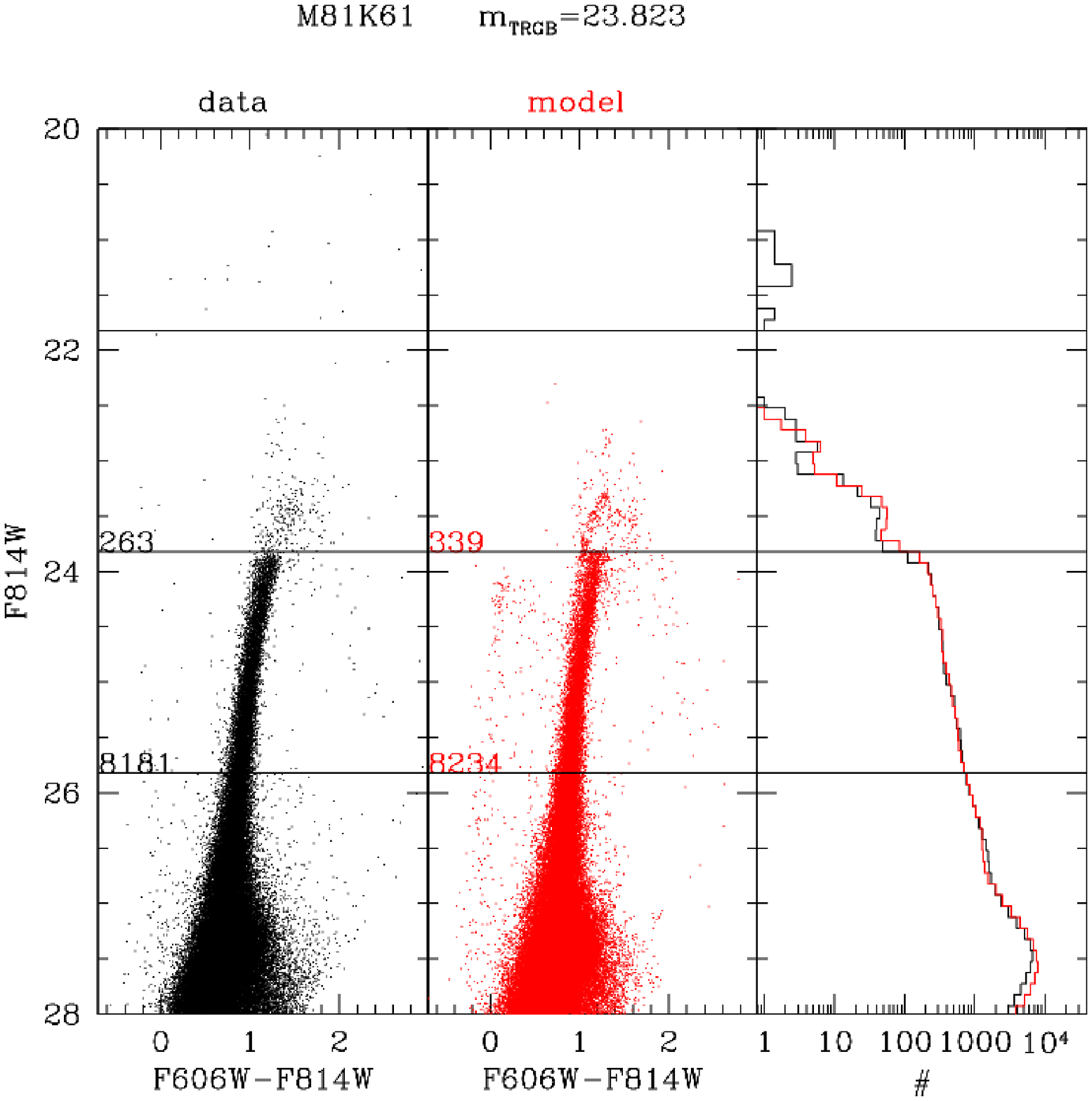}
\\
\includegraphics[width=0.33\textwidth]{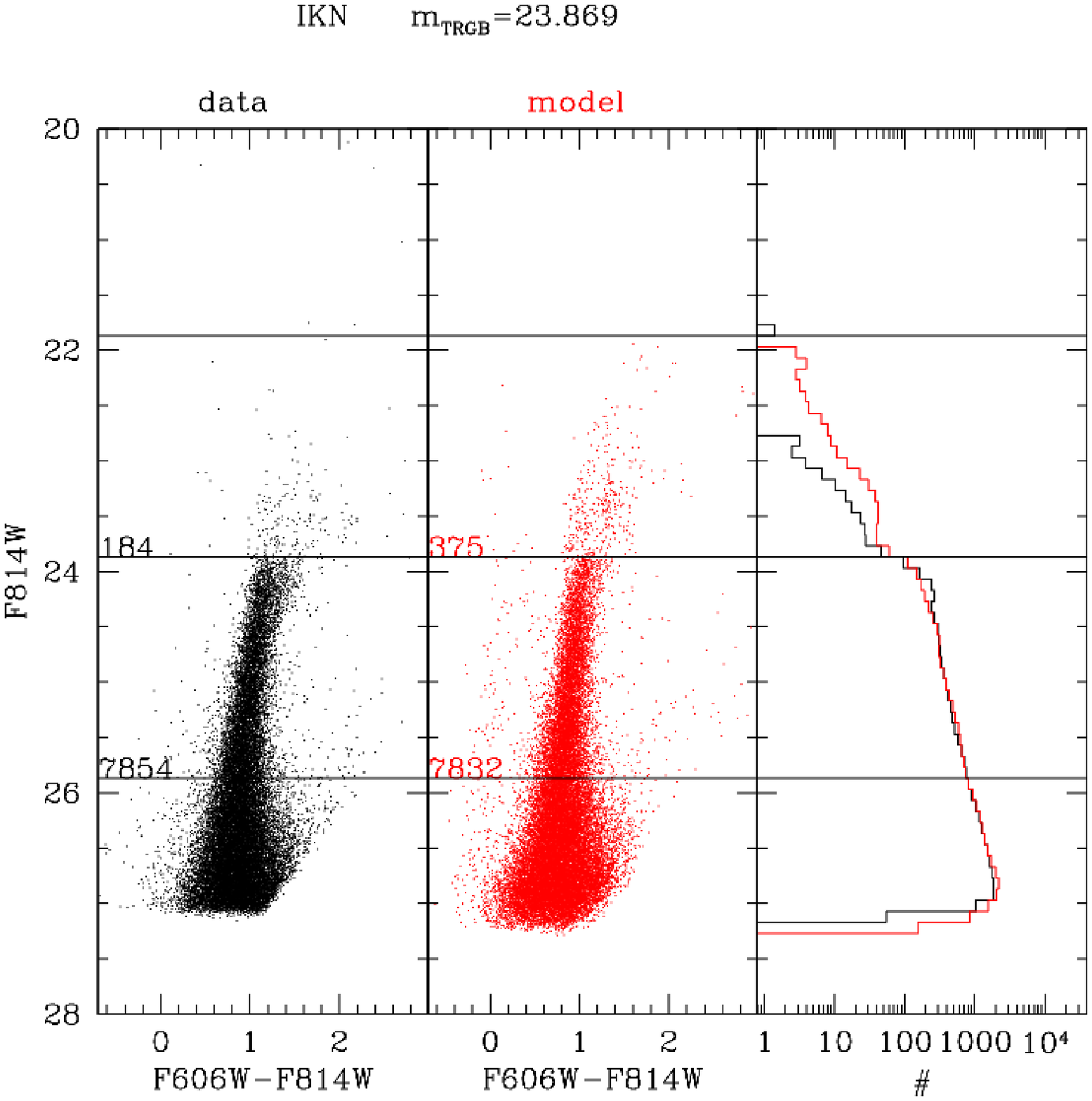}
\includegraphics[width=0.33\textwidth]{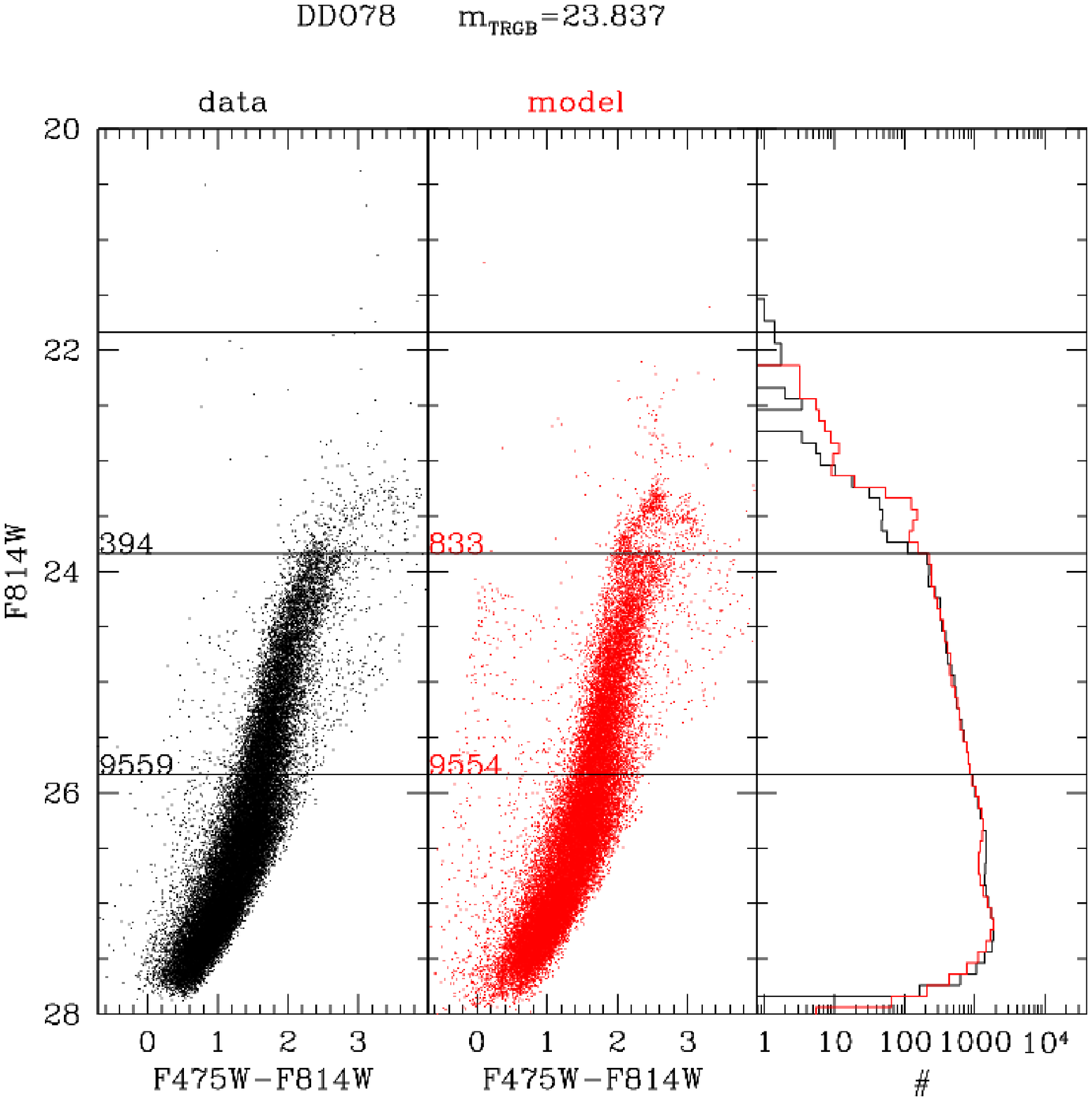}
\includegraphics[width=0.33\textwidth]{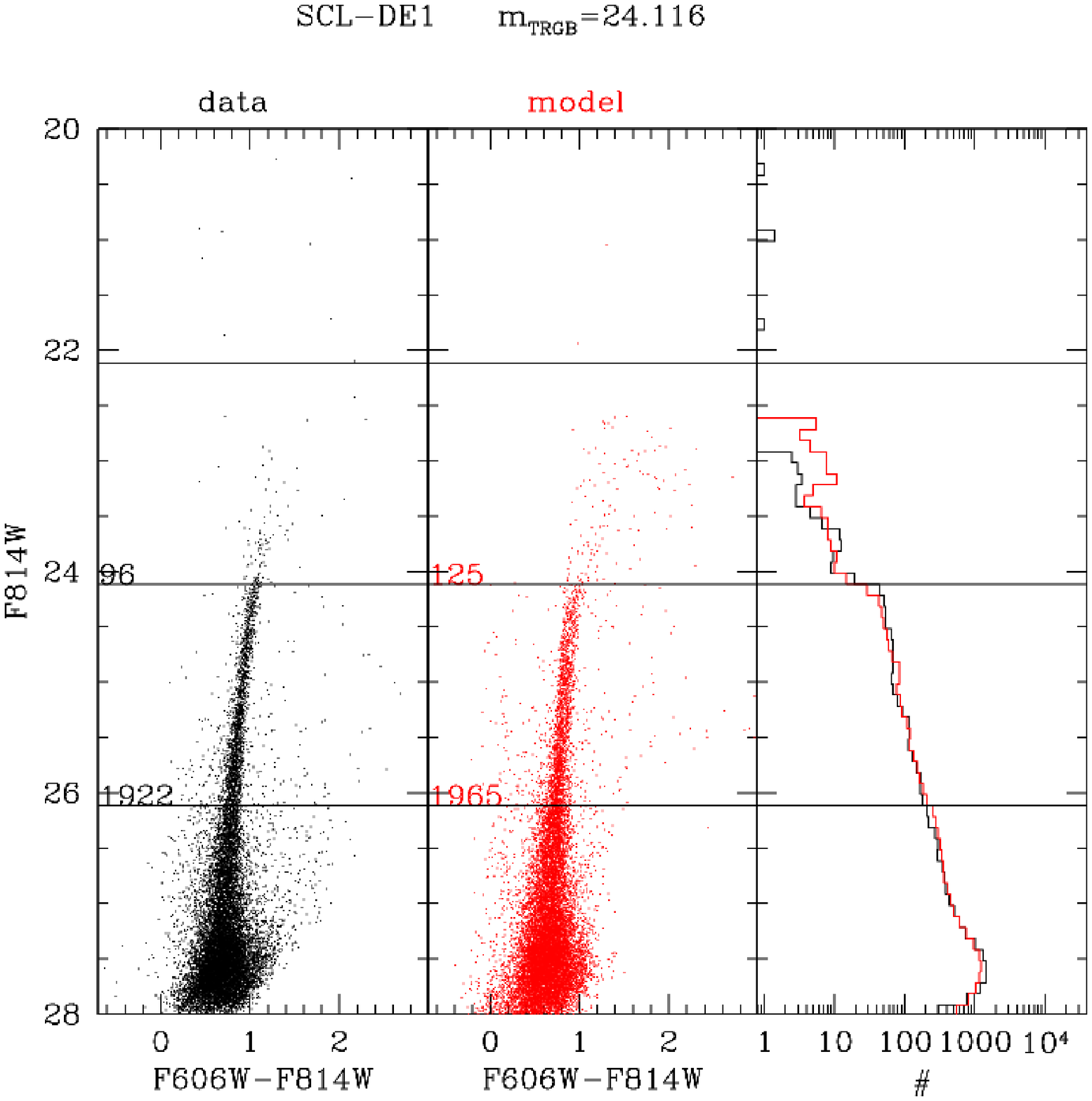}
\caption{The same as Fig.~\ref{fig_ma08}, but now using our best
TP-AGB models, for case A mass-loss. }
\label{fig_ma08_mod2}
\end{figure*}
\begin{figure*}
\figurenum{\ref{fig_ma08_mod2} continued}
\includegraphics[width=0.33\textwidth]{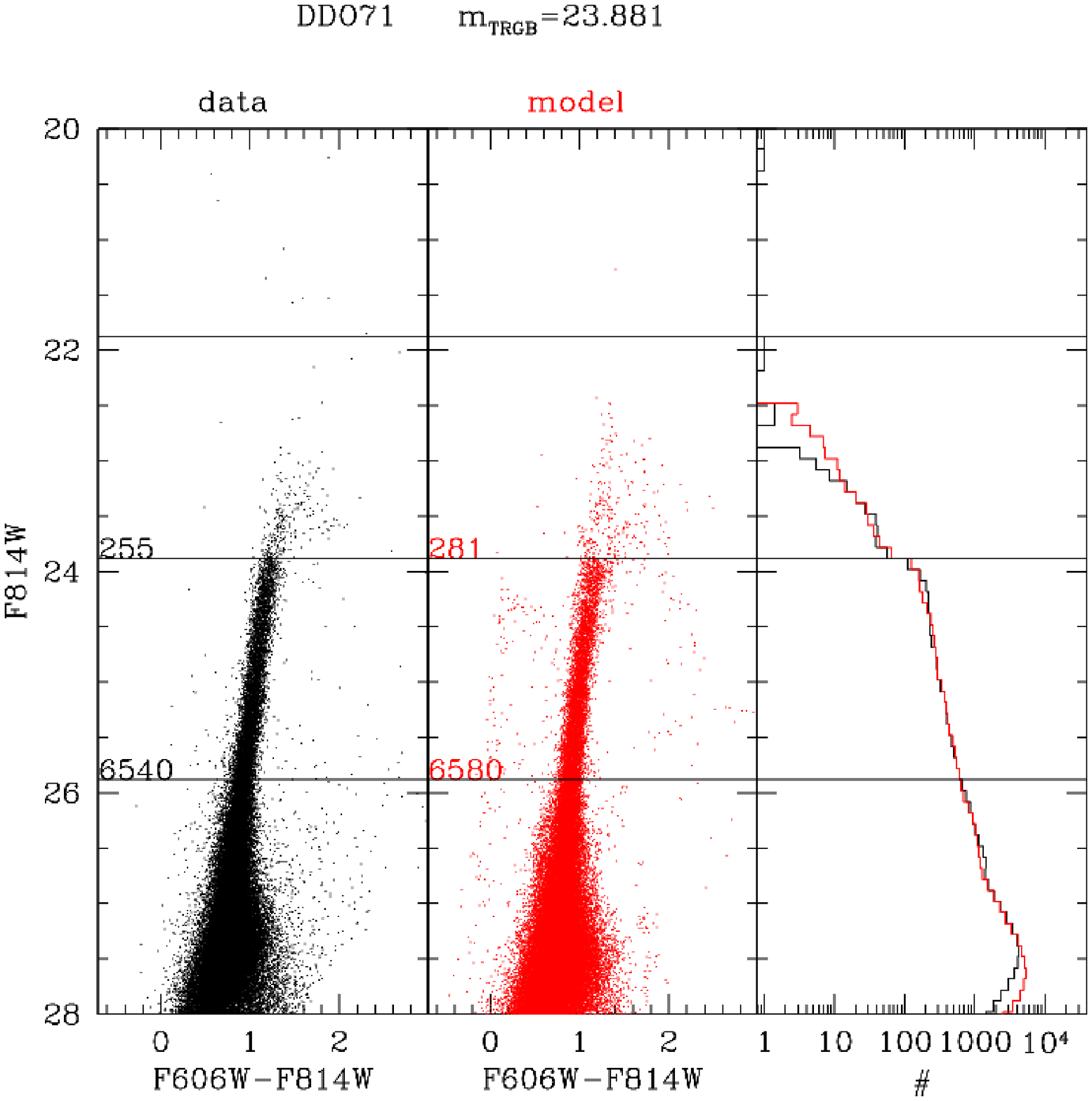}
\includegraphics[width=0.33\textwidth]{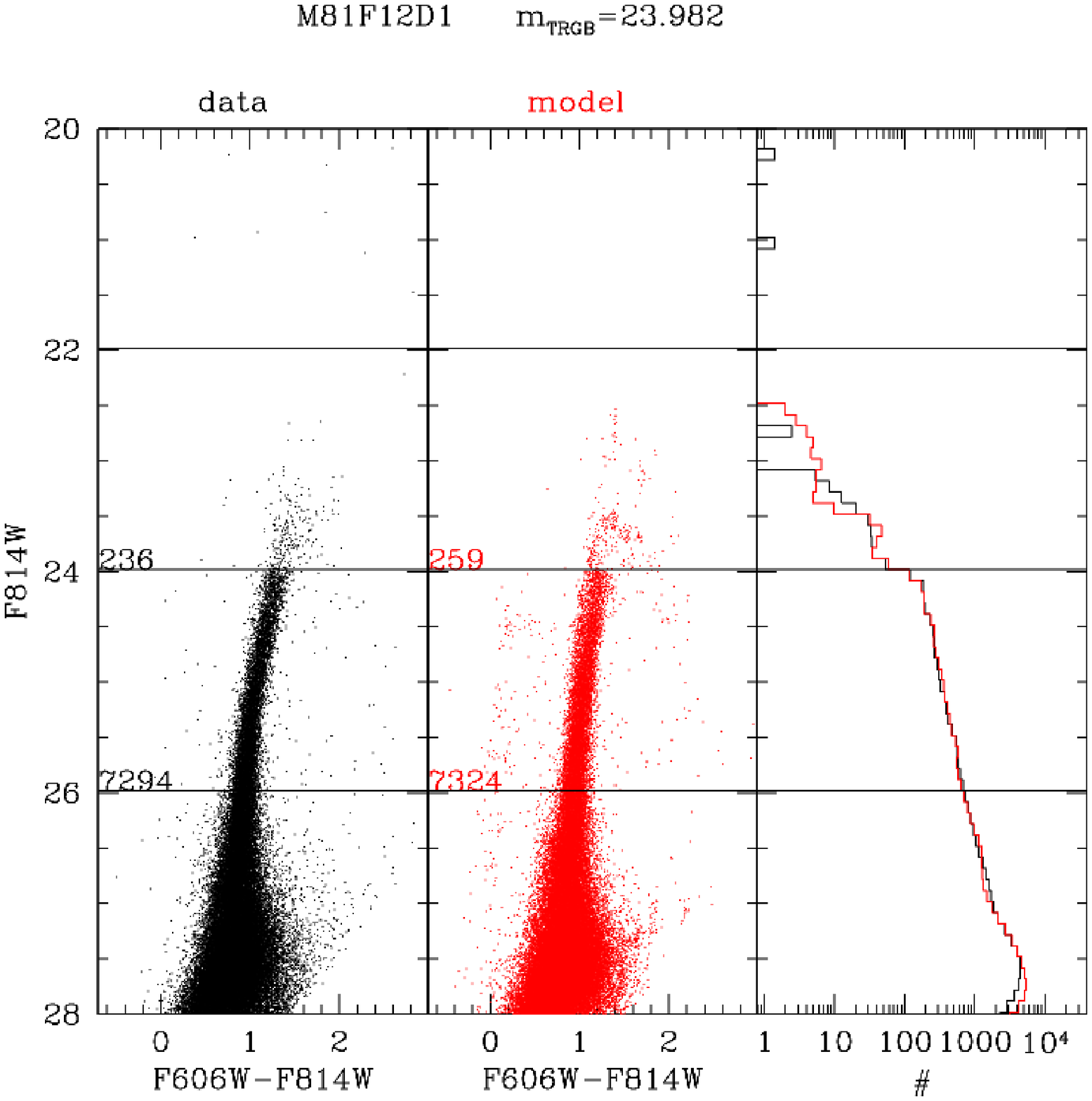}
\includegraphics[width=0.33\textwidth]{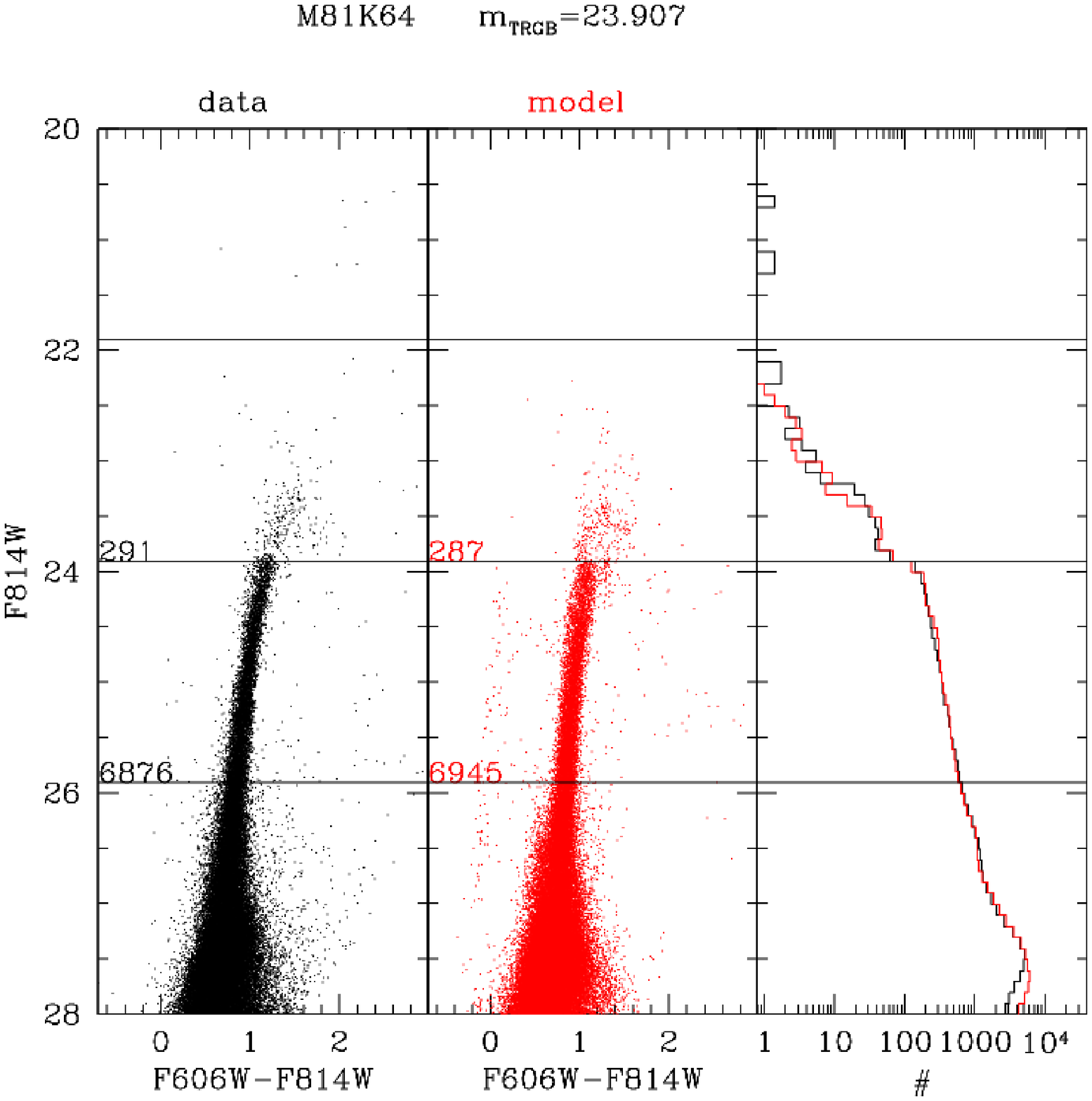}
\caption{}
\end{figure*}

\begin{figure*}
\includegraphics[width=0.33\textwidth]{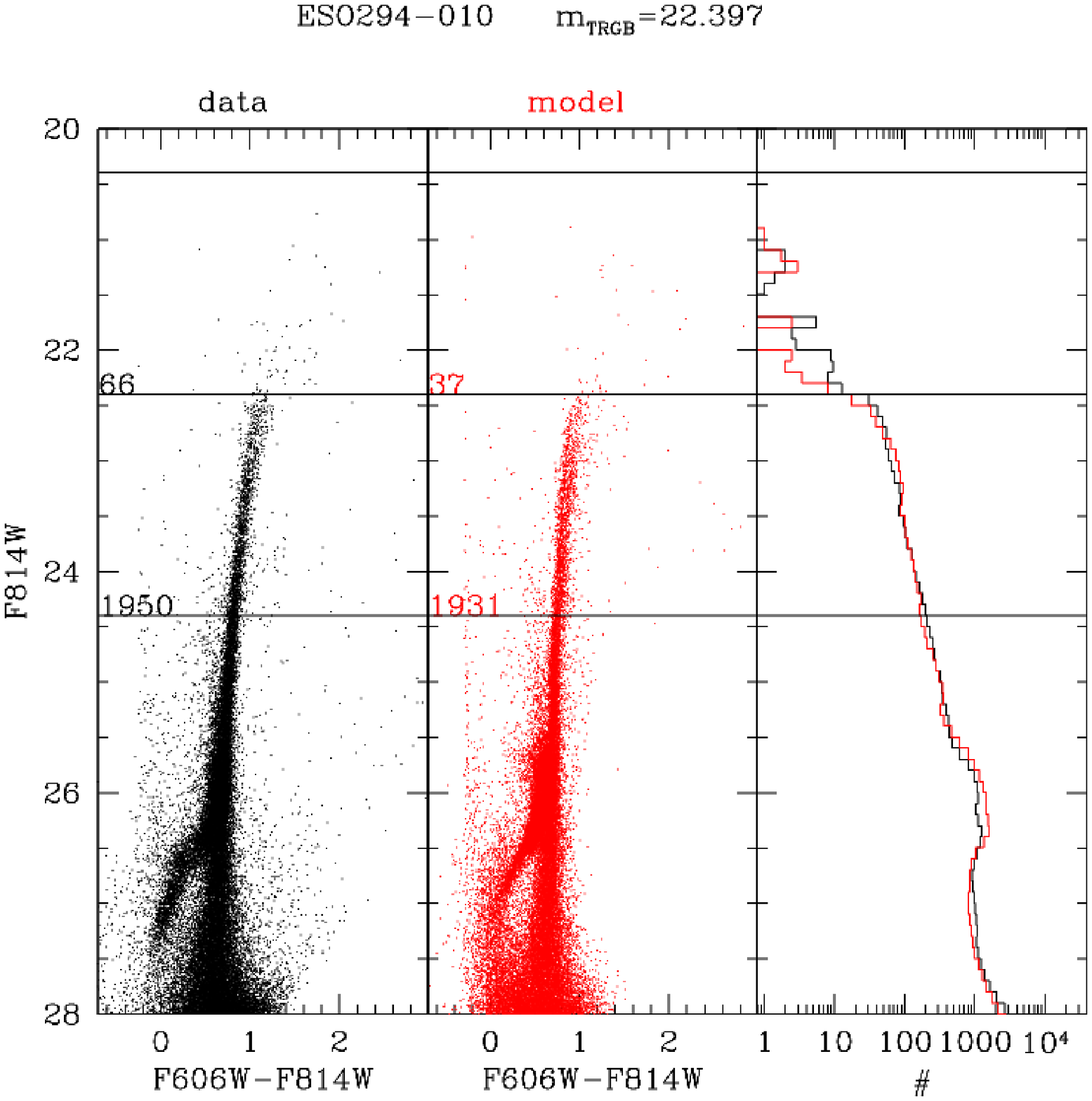}
\includegraphics[width=0.33\textwidth]{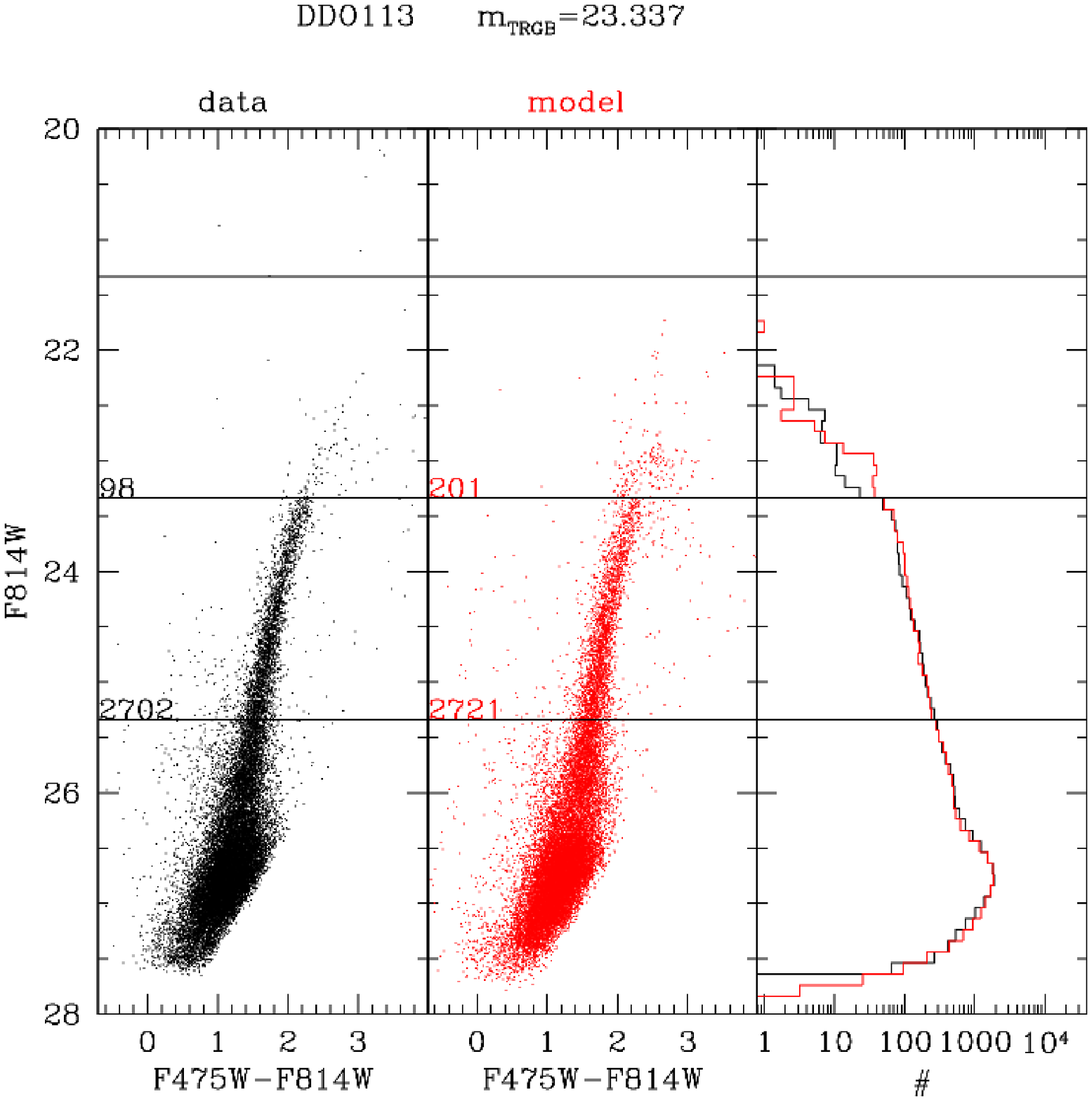}
\includegraphics[width=0.33\textwidth]{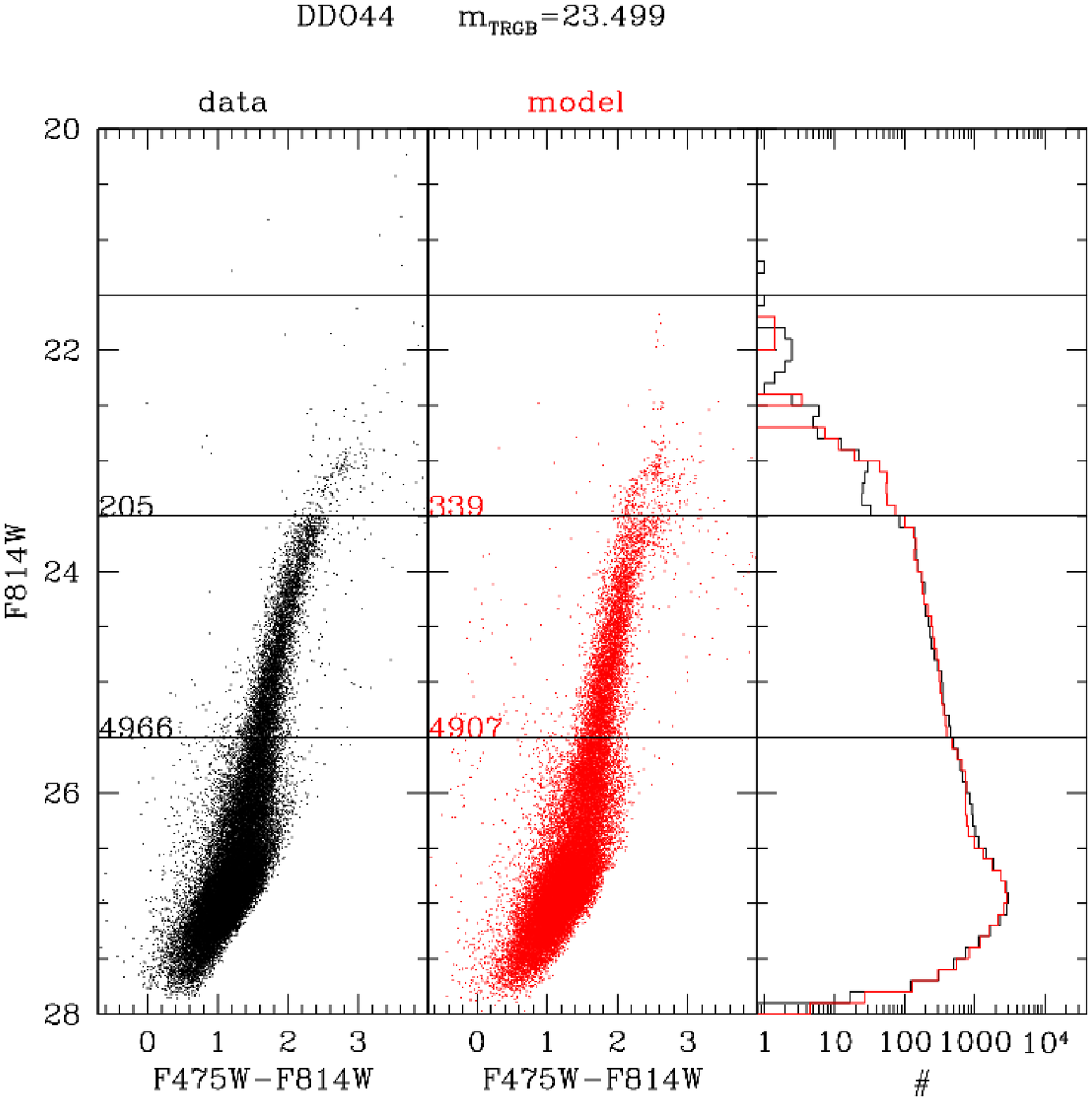}
\\
\includegraphics[width=0.33\textwidth]{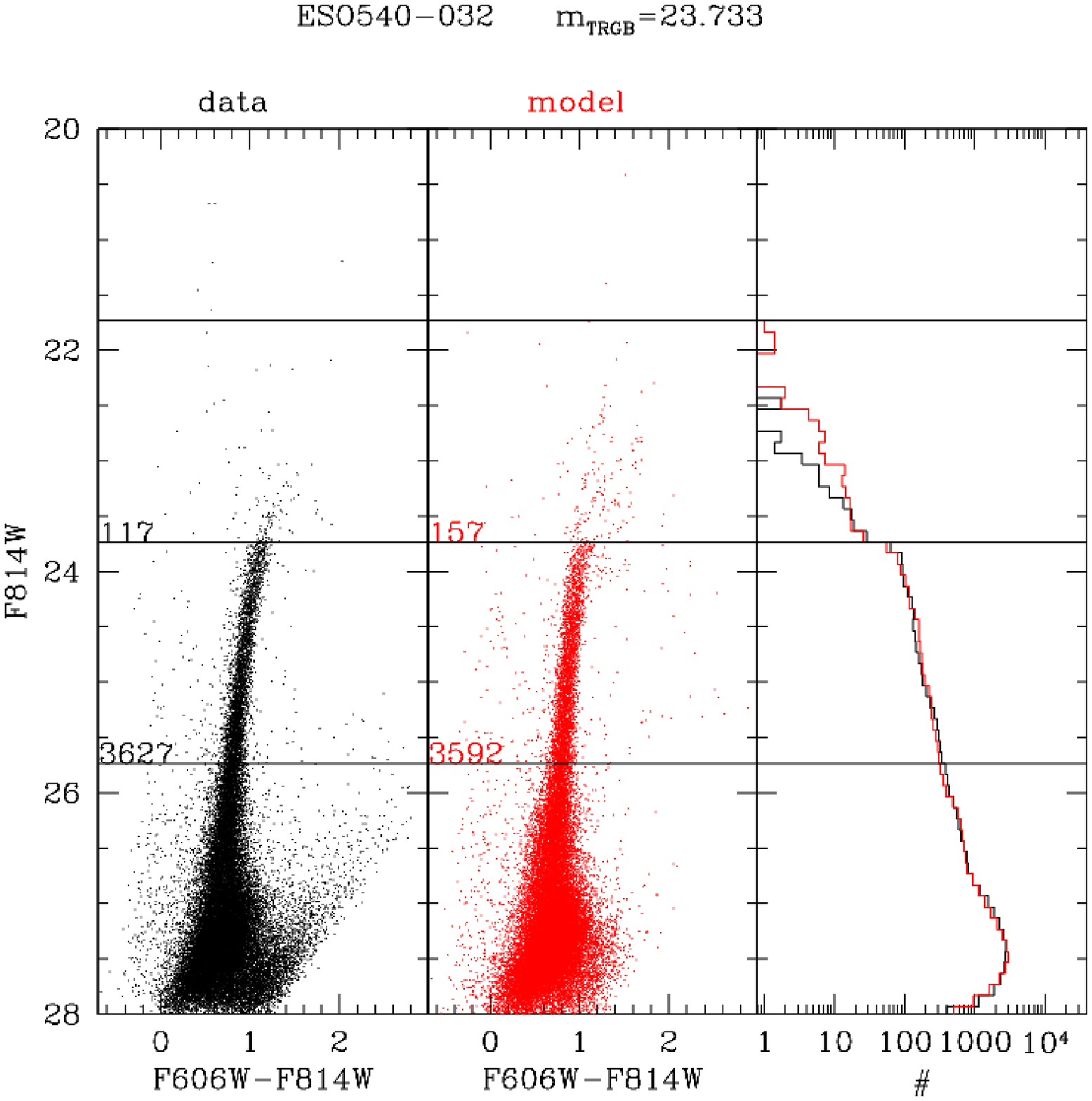}
\includegraphics[width=0.33\textwidth]{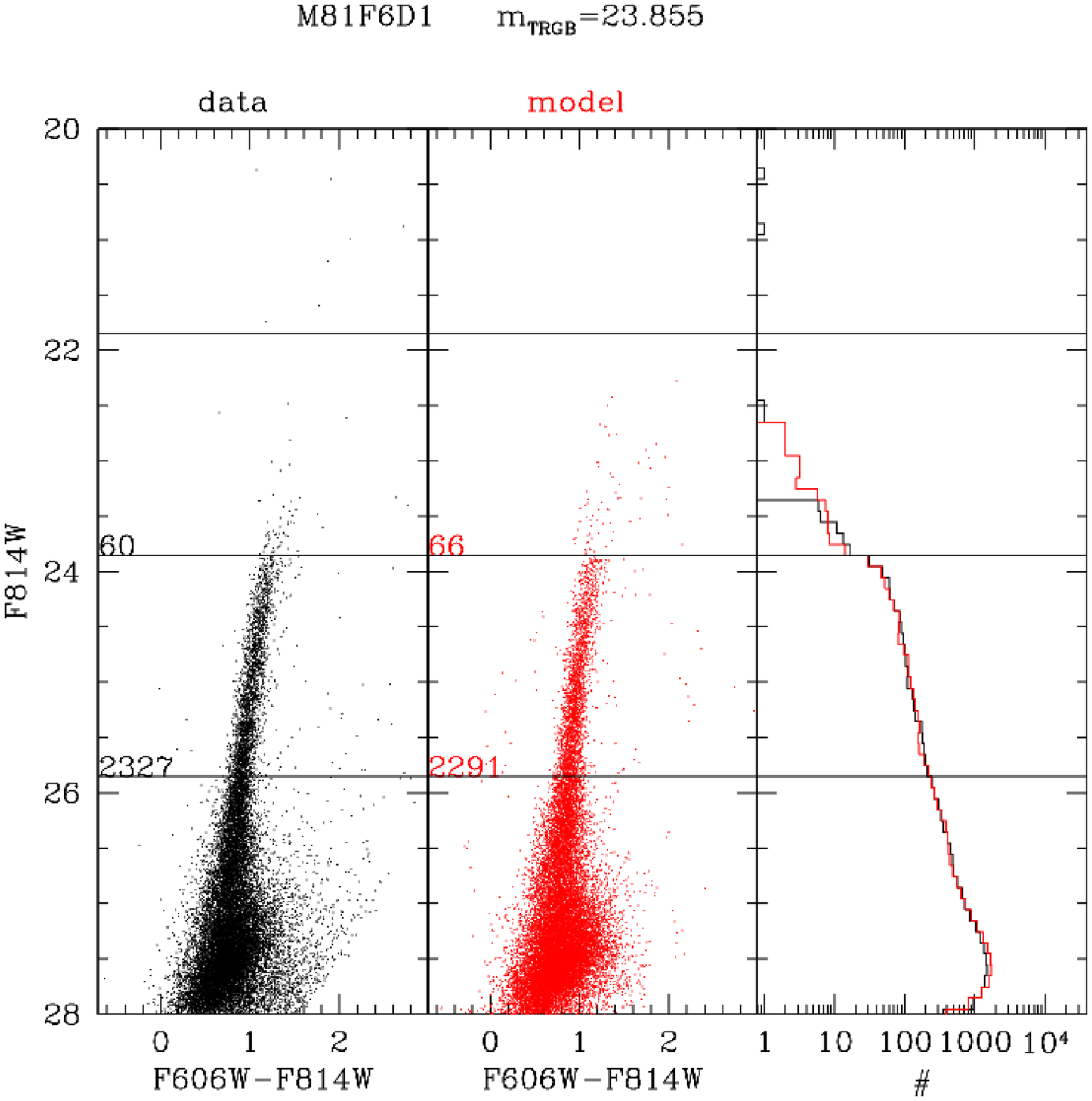}
\includegraphics[width=0.33\textwidth]{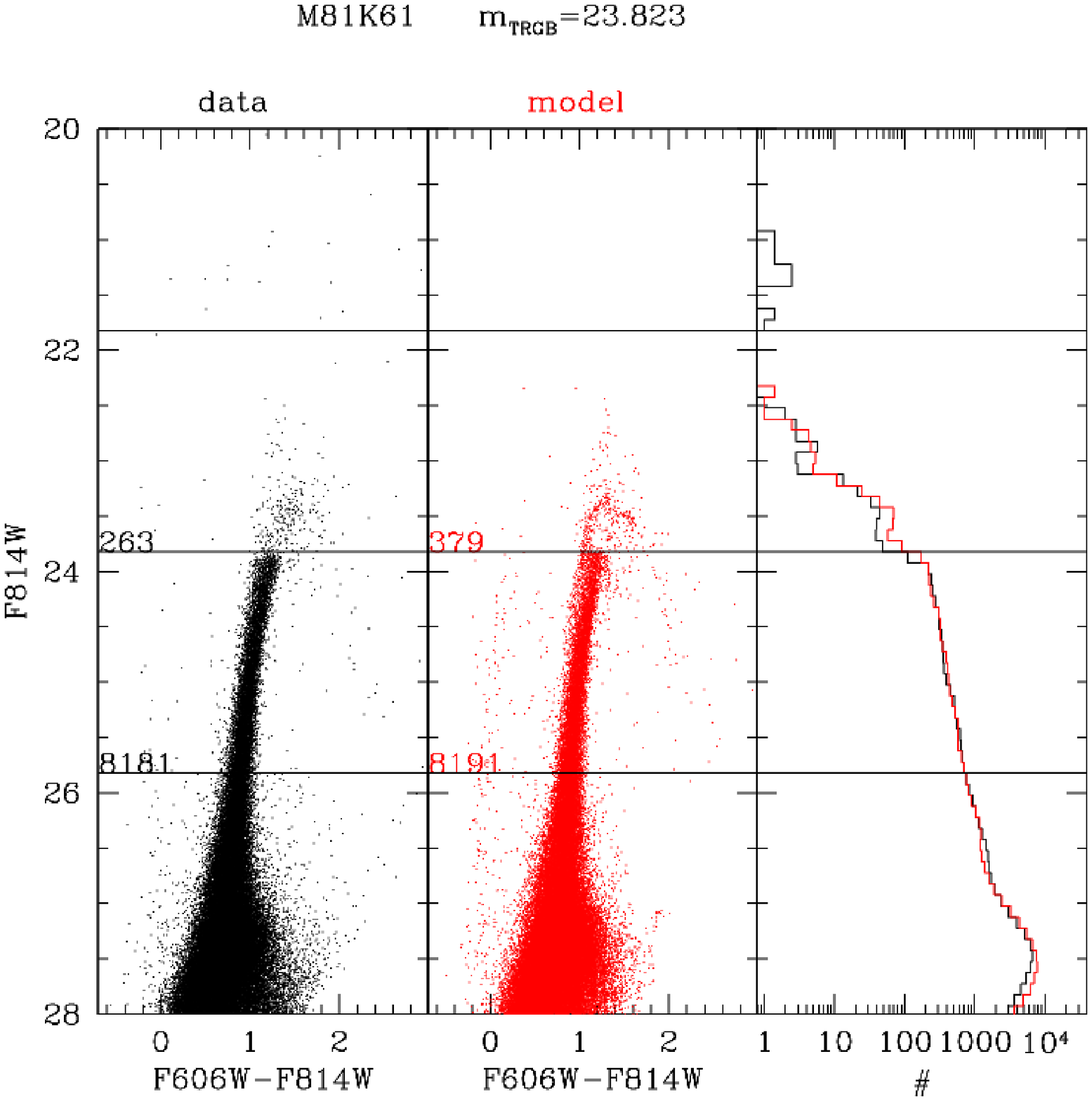}
\\
\includegraphics[width=0.33\textwidth]{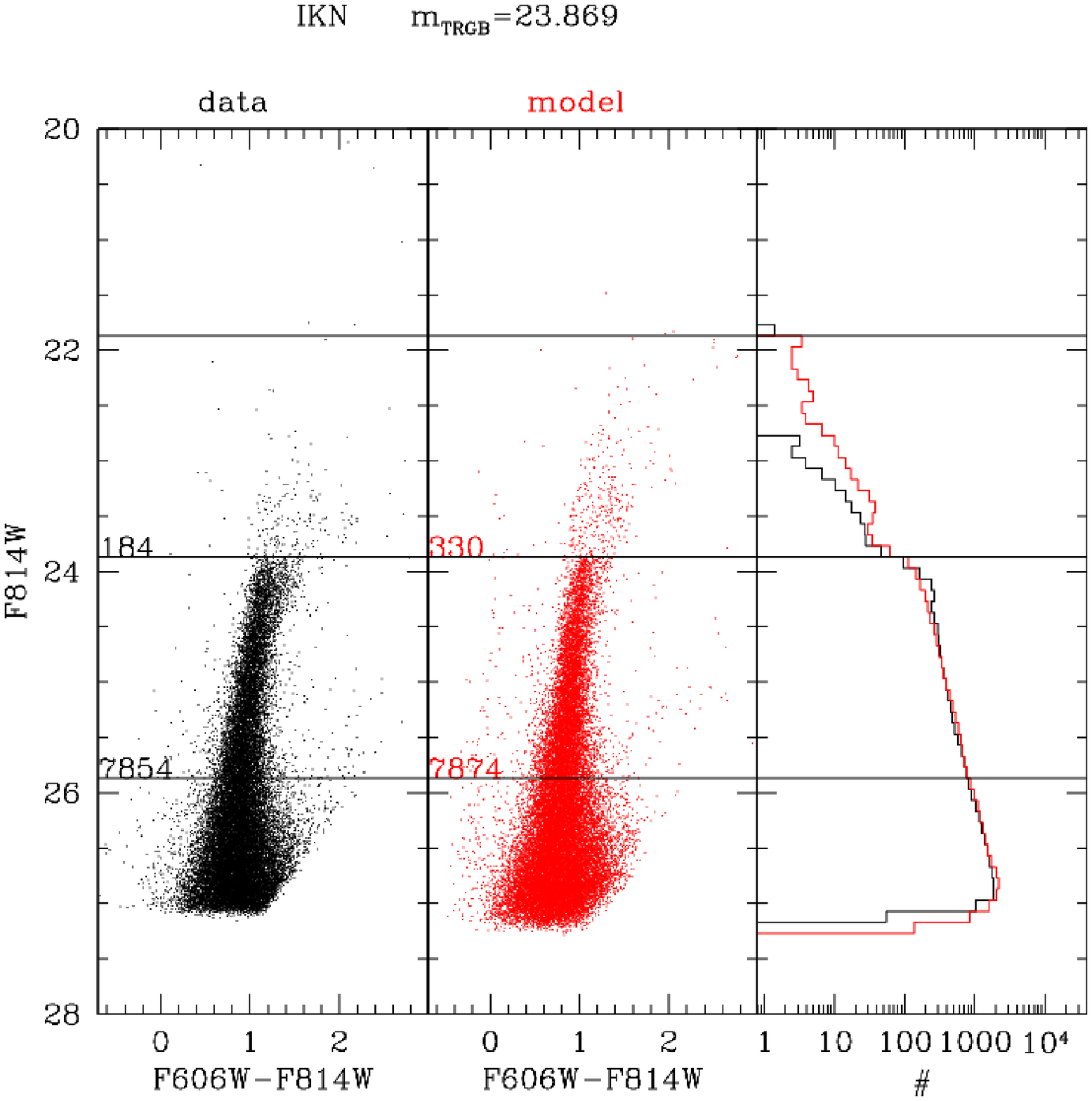}
\includegraphics[width=0.33\textwidth]{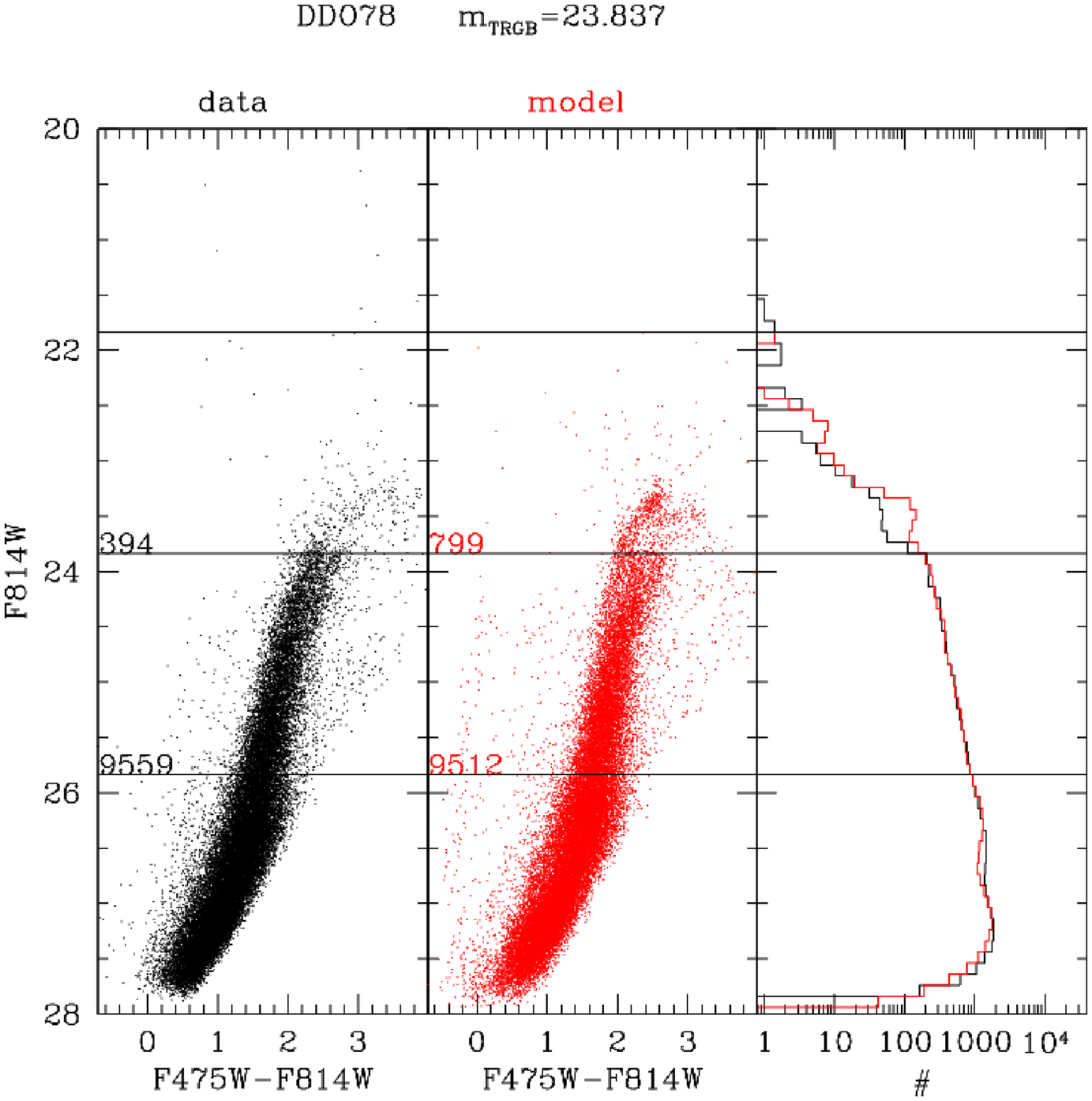}
\includegraphics[width=0.33\textwidth]{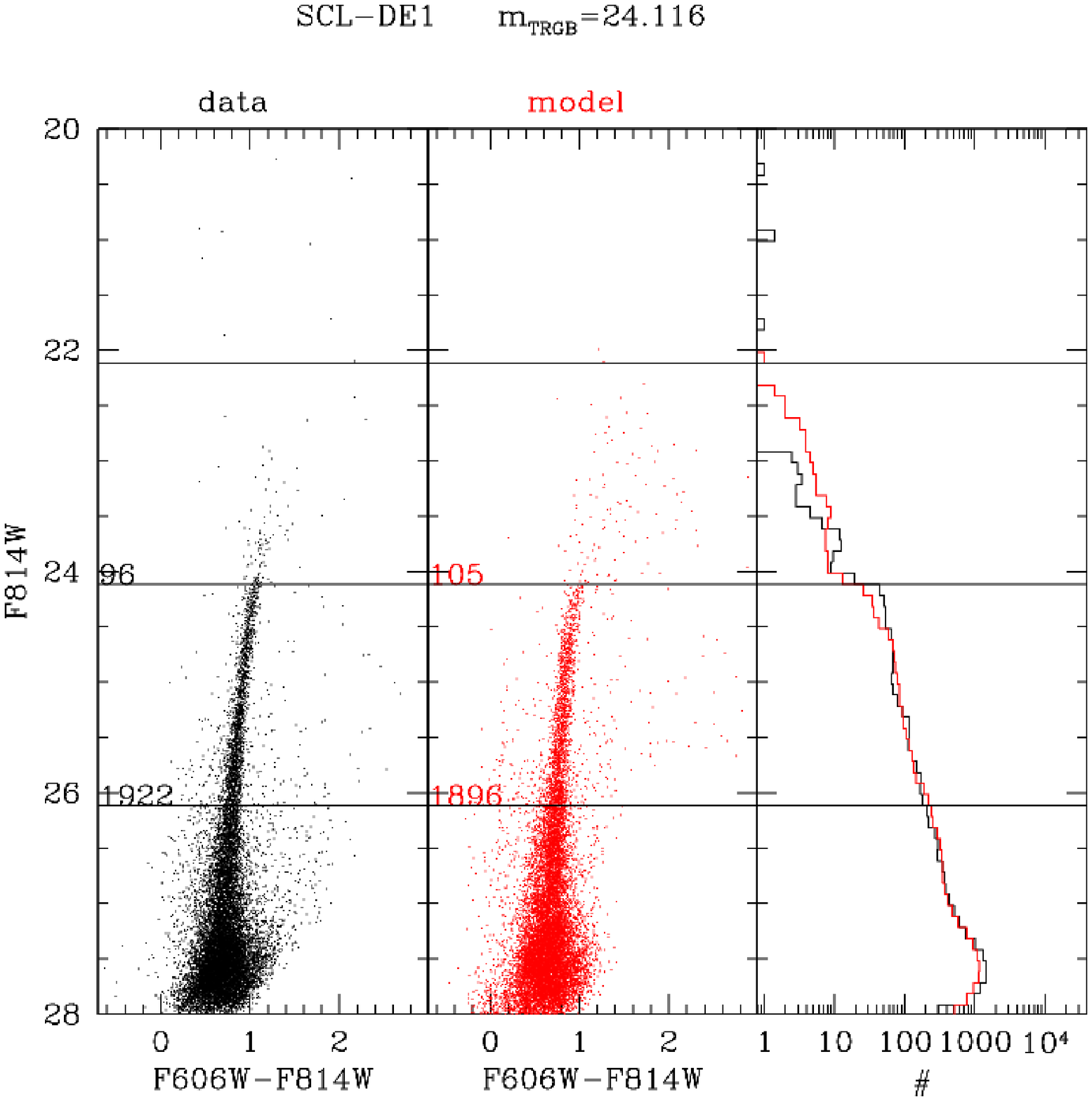}
\caption{The same as Fig.~\ref{fig_ma08}, but now using the TP-AGB models
for case B mass-loss. }
\label{fig_ma08_mod3}
\end{figure*}
\begin{figure*}
\figurenum{\ref{fig_ma08_mod3} continued}
\includegraphics[width=0.33\textwidth]{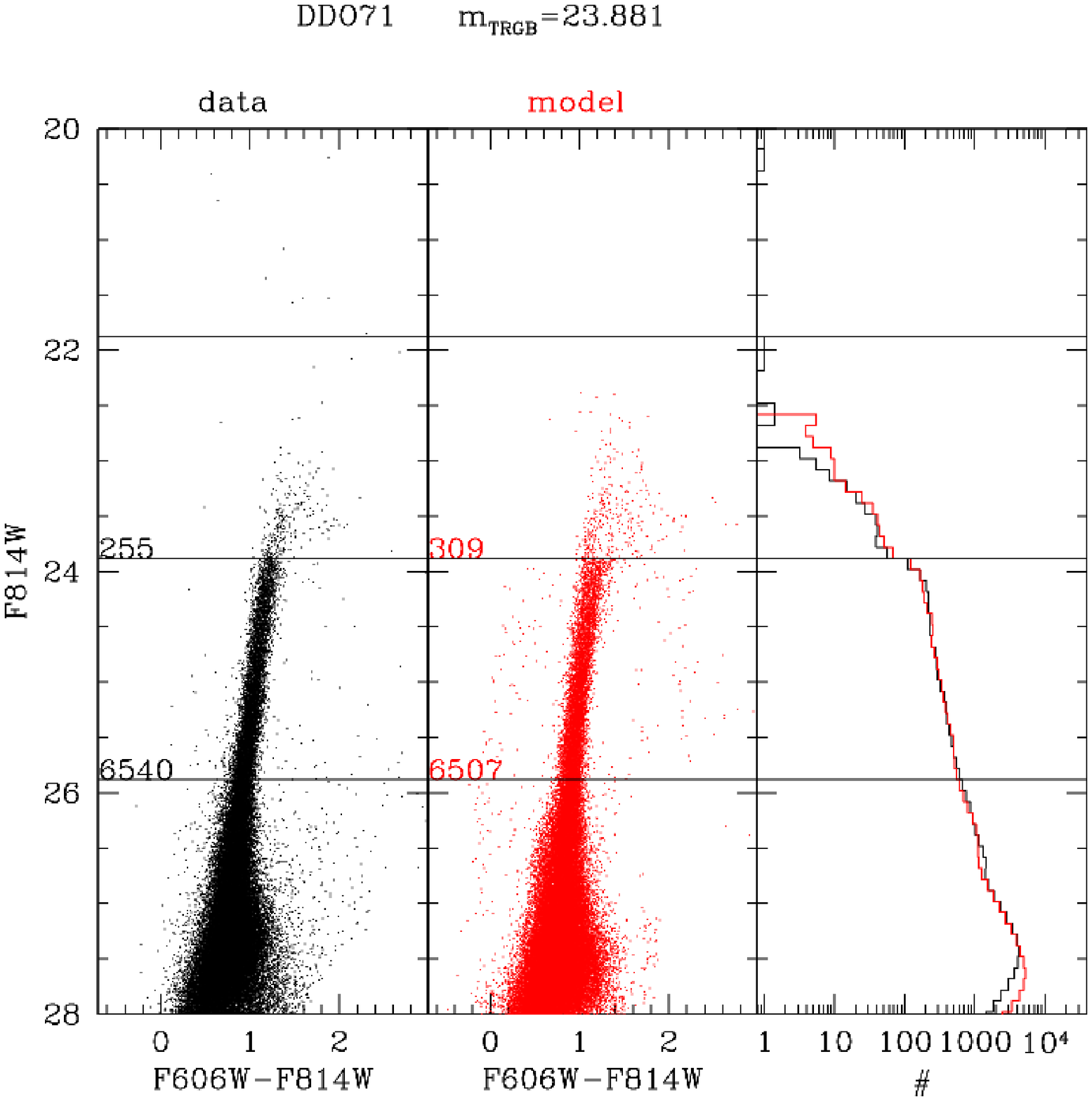}
\includegraphics[width=0.33\textwidth]{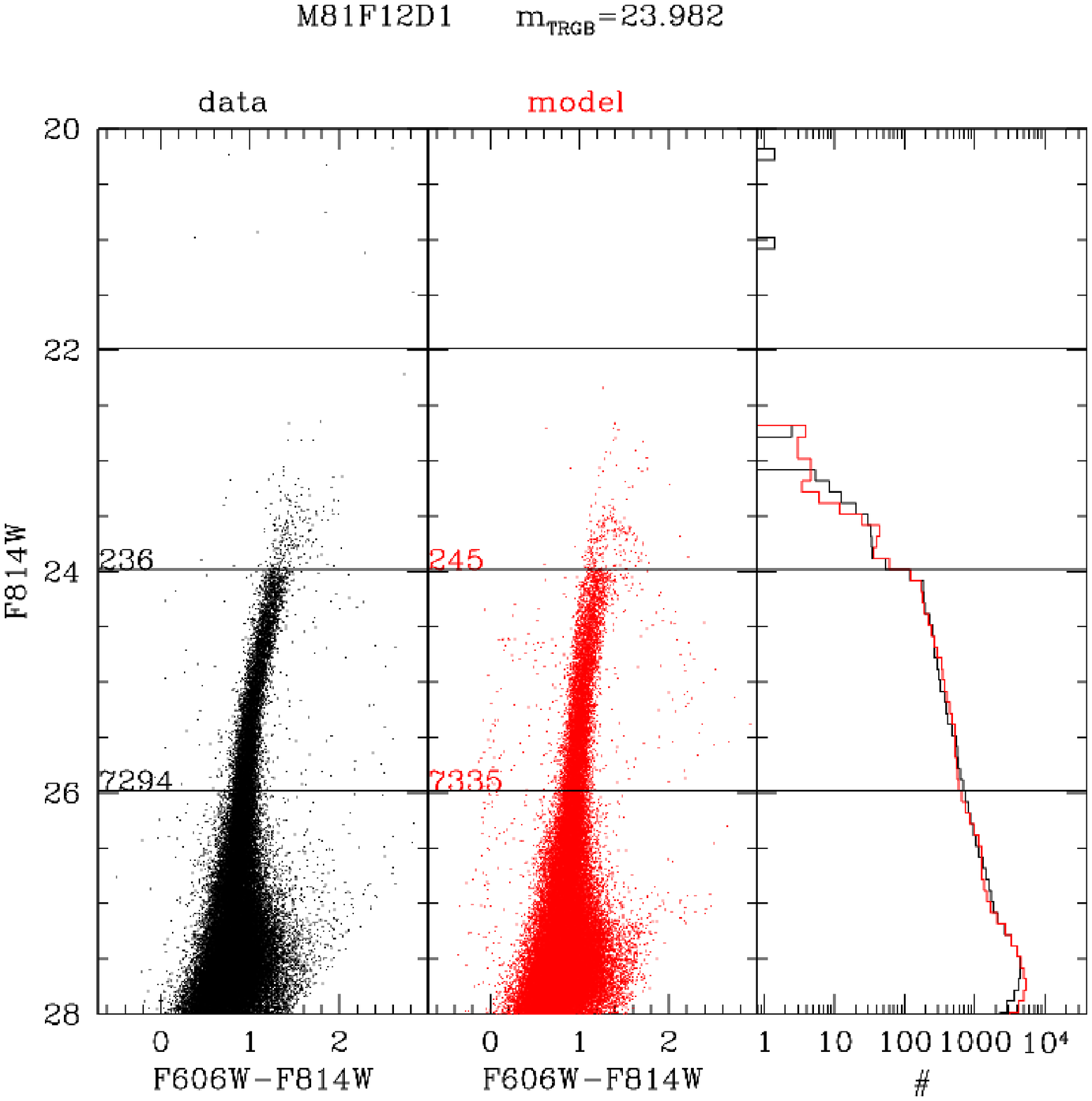}
\includegraphics[width=0.33\textwidth]{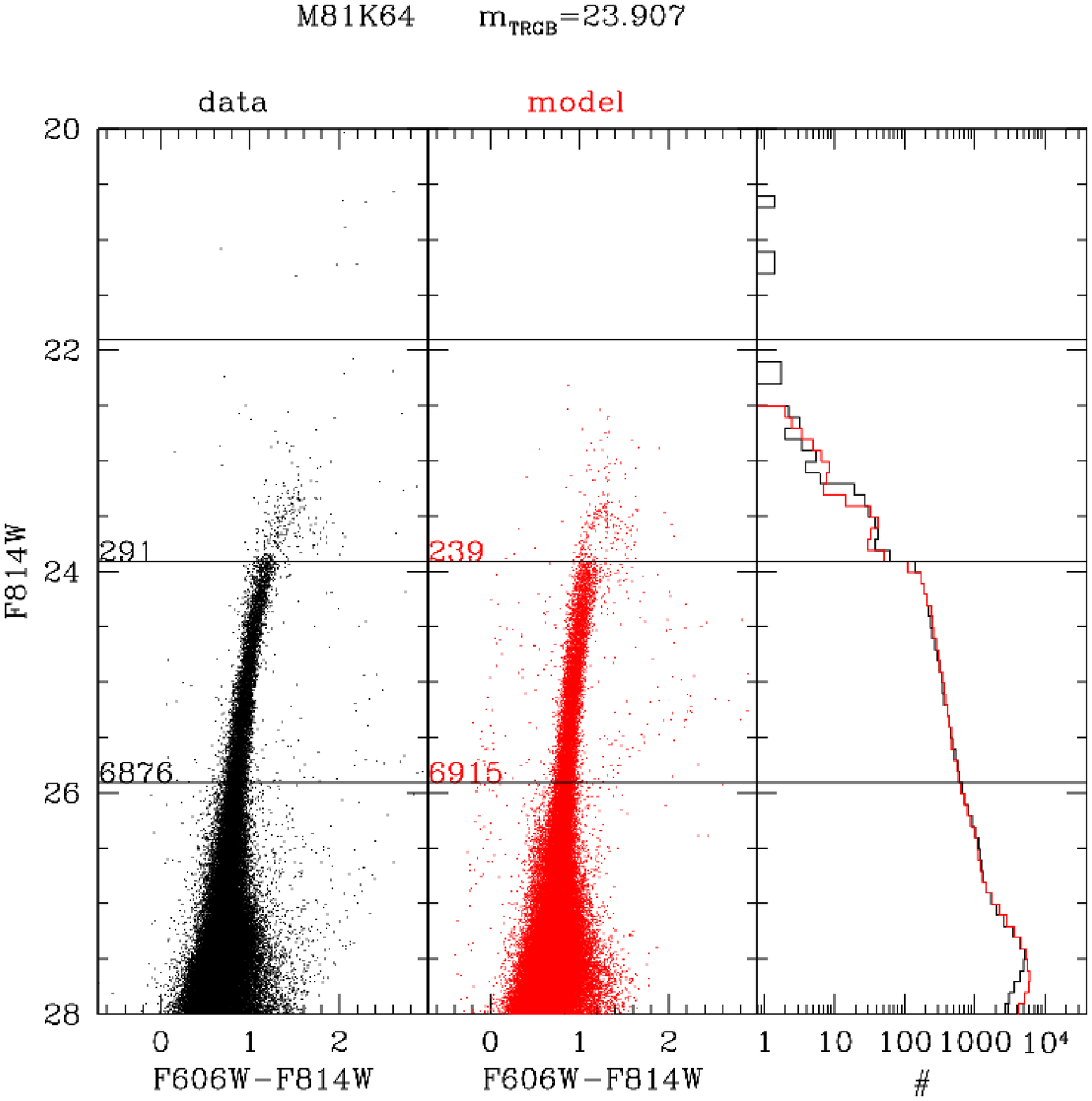}
\caption{}
\end{figure*}

Figures~\ref{fig_ma08_mod2} and \ref{fig_ma08_mod3} show the final
result of using the new TP-AGB tracks in the simulation of ANGST
galaxies, for cases A and B, respectively. It is evident that the
reduction of TP-AGB lifetimes, and consequently of the TP-AGB
termination luminosity, leads to a much better description of the data
in both cases. For some galaxies such as ESO294-010,
M81F6D1 and M81K64, the model-data agreement can be qualified as
excellent, with simulated AGB numbers within the 67~\% confidence
level of Poisson fluctuations in the data, and a good description of
the luminosity function for AGB stars.

In a few other cases, however, the agreement is still not completely
satisfactory, and the models tend to overestimate the AGB numbers by
factors of up to 2, as for DDO113, DDO44, IKN, and DDO78. Notice that
in all these cases the observed AGB/RGB ratio is very small and close
to its minimum value, which suggests that these galaxies are really
dominated by very old populations. Therefore, we consider that the
mismatch in the AGB/RGB ratios from the models might be attributable
to the errors in the SFHs and metallicities of these galaxies, added
of course to the residual mismatches in the stellar models of higher
mass, and in the prescription for dust obscuration.

Finally, for the galaxies ESO0540-032, M81K61, SCL-DE1, DDO71, and
M81F12D1, there is in general just a modest excess (smaller than
$\sim50$~\%) in the predicted numbers of AGB stars. It is remarkable
that this excess is generally more evident for the bright section of
the AGB sequence, whereas close to the TRGB the numbers of predicted
and observed AGB stars are in quite good agreement. We interpret the
mismatch for bright AGB stars as being likely attributable to the
younger population in these galaxies -- with the causes being either
in a slight overestimation of the lifetimes of the more massive TP-AGB
models or in a modest excess in the SFH derived from ANGST data for
ages $\la3$~Gyr.

From those galaxies in which there is a good model--data agreement, we
can conclude the following: They indicate that the TP-AGB lifetimes
above the TRGB are of the order of 1.2 to 1.8~Myr
(cf. Fig.~\ref{fig_agb_taumf}) for the low-mass low-metallicity AGB
stars ($M\la1~\Msun$, $\feh\la-1.2$) which are typically being
sampled. The need for such a reduction in TP-AGB lifetimes, with
respect to those in MG07 models, has already been indicated by
\citet{Gullieuszik_etal08} in their study of the old metal-poor Leo~II
dSph, and to a lesser extent also by \citet{Held_etal10} and
\citet{Melbourne_etal10} while studying metal-poor galaxies with
somewhat younger SFHs. On the other hand, it is quite reassuring that
the final core masses of our best-fitting AGB models are between 0.51
and 0.55~\Msun\ (again for low-mass stars, with say $M\la1~\Msun$),
which is in good agreement with the observed masses of white dwarfs in
globular clusters of the MW \citep[of $0.53\pm0.01$~\Msun,
cf.][]{Kalirai_etal09}. This latter indication also agrees with the
low initial--final mass relation derived from the number counts of hot
white dwarfs -- mostly from the thin disk -- in GALEX wide area
surveys \citep{Bianchi_etal10}. Thus, we find that the dramatic
reduction in the lifetimes and final masses of our TP-AGB models are
well in line with the indications from independent data.

Whereas we have tested two different prescriptions for the mass loss
rate at the dust-driven phase, both turn out to provide essentially
the same results, whith case~A reproducing our data for old metal-poor
galaxies just slightly better than case~B. However, the differences
between these two prescriptions are expected to increase for stars of
masses $\ga2$~\Msun, which reach significantly higher
luminosities. Therefore, they need to be tested in galaxies containing
young populations, which will be the subject of forthcoming papers.

\section{Closing remarks}
\label{sec_closing}

In this work, we have derived strong constraints on the lifetimes of
low-mass metal-poor AGB stars, via the study of their numbers and
luminosity functions in galaxy regions with a predominantly-old and
metal-poor SFH. The AGB lifetimes at magnitudes brighter than the TRGB
turn out to be between 1.2 and 1.8~Myr for stellar masses of
$M\la1~\Msun$, at metallicities $\feh\la-1.2$. Indeed, TP-AGB
evolutionary tracks with these lifetimes are able to reproduce quite
well the low AGB/RGB number ratios in a subsample of these galaxies,
as well as their luminosity functions.  In some other cases, the
agreement is still not perfect, but the new models perform
significantly better than the previous MG07 ones.  Although we have
tested just two possible alternative prescriptions for mass loss, they
both produce a quite similar reduction in the AGB lifetimes, as well
as similar final masses for the resulting white dwarfs.

Whereas the constraints we derive may apply to quite a limited range
of stellar masses and metallicities, they also represent an important
step towards a robust calibration of the lifetimes of TP-AGB stars at
all masses and metallicities. Indeed, with this single point in
mass-metallicity space we can discard some of the already-proposed
formulas for the mass loss on the AGB, for instance the
\citet{BowenWillson91} and \citet{Willson00} one as implemented in
MG07. Moreover, we can use this result as a reference point to start a
more thorough calibration of the AGB lifetimes at regimes of
increasing masses and metallicities, making use of additional galaxies
with the presence of younger SFH and more metal-rich giant
branches. Such an extension of the AGB calibration, however, cannot
proceed via the analysis of optical data only, as performed in this
paper. It requires {\em at least} the use of near-infrared photometry
in order to sample the most massive and evolved AGB stars, as well as
those of higher metallicities. Indeed, we are presently collecting
near-infrared data with the new HST/WFC3 camera (IR channel), for some
dozens of ANGST galaxies with a well-measured SFH. Forthcoming papers
will discuss the problem of improving AGB models in the light of these
new data.

The present TP-AGB tracks surely represent an improvement in our sets
of isochrones and tools to simulate stellar populations. For this
reason they are being included as an alternative to the MG07 models,
in the CMD\footnote{http://stev.oapd.inaf.it/cmd} and
TRILEGAL\footnote{http://stev.oapd.inaf.it/trilegal} web interfaces,
which provide isochrones and simulated photometry of resolved stellar
populations in a wide variety of filter systems. 

\acknowledgments

L.G. and P.M. acknowledge financial support from contract ASI-INAF
I/016/07/0 and INAF/PRIN07 CRA 1.06.10.03.  B.F.W., J.J.D., P.R. \&
D.R.W.  acknowledge financial support from HST GO-10915.
K.M.G. acknowledges support from HST GO-10945. P.R. acknowledges the
Achievement Rewards for College Scientists (ARCS)
Fellowship. J.J.D. and P.R. were partially supported by AR-10945,
GO-11718, and GO-11307.

This work is based on observations made with the NASA/ESA Hubble Space
Telescope, obtained from the data archive at the Space Telescope
Science Institute. Support for this work was provided by NASA through
grant number GO-10915 and GO-10945 from the Space Telescope Science
Institute, which is operated by AURA, Inc., under NASA contract NAS
5-26555.  This research has made use of the NASA/IPAC Infrared Science
Archive and the NASA/IPAC Extragalactic Database (NED), which are both
operated by the Jet Propulsion Laboratory, California Institute of
Technology, under contract with the National Aeronautics and Space
Administration.  This research has made extensive use of NASA's
Astrophysics Data System Bibliographic Services.



{\it Facilities:}  \facility{HST (ACS)}, \facility{HST (WFPC2)}.






\input{ms_corr.bbl}
\end{document}

%% file: rgbagb_table2.tex
\begin{deluxetable*}{l|lllll|llll
}
\tabletypesize{\tiny}
\tablecaption{The SFR and AGB/RGB ratio}
\label{tab_ratio}
\tablehead{
    \colhead{Target/name} & \multicolumn{5}{c}{Which SFH} &
    \multicolumn{4}{c}{${N_{\rm AGB}}/{N_{\rm RGB}}$} \\
\cline{2-6}
\cline{7-10}
     &
    \colhead{AGB?} &
    \colhead{Zinc?} &
    \colhead{$\frac{{\rm SFR}_{\rm <1Gyr}}{{\rm SFR}_{\rm tot}}$} &
    \colhead{$\frac{{\rm SFR}_{\rm 1-3Gyr}}{{\rm SFR}_{\rm tot}}$} &
    \colhead{$\langle\feh\rangle$} &
    \colhead{observed} &
    \colhead{Ma08} &
    \colhead{case {\bf A}} &
    \colhead{case {\bf B}} 
}
\startdata
ESO294-010     
        & Y & N    & 0.01 & 0.07 & -1.68 & 0.034$\pm$0.004 & 0.149& 0.020 & 0.020 \\ 
        & Y & Y    & 0.01 & 0.07 & -1.63 &                 & 0.131& 0.021 & 0.019 \\ 
        & N & N    & 0.02 & 0.08 & -1.75 &                 & 0.133& 0.025 & 0.019 \\ 
        & N & Y    & 0.02 & 0.09 & -1.74 &                 & 0.155& 0.021 & 0.022 \\ 
DDO113       
        & Y & N    & 0.02 & 0.23 & -1.55 & 0.036$\pm$0.004 & 0.147& 0.068 & 0.073 \\ 
        & Y & Y    & 0.01 & 0.17 & -1.57 &                 & 0.144& 0.070 & 0.062 \\ 
        & N & N    & 0.03 & 0.16 & -1.44 &                 & 0.162& 0.076 & 0.068 \\ 
        & N & Y    & 0.03 & 0.17 & -1.42 &                 & 0.155& 0.066 & 0.075 \\ 
DDO44      
        & Y & N    & 0.01 & 0.21 & -1.20 & 0.041$\pm$0.003 & 0.147& 0.070 & 0.069 \\ 
        & Y & Y    & 0.01 & 0.19 & -1.13 &                 & 0.131& 0.039 & 0.050 \\ 
        & N & N    & 0.02 & 0.11 & -1.13 &                 & 0.150& 0.064 & 0.067 \\ 
        & N & Y    & 0.02 & 0.21 & -1.14 &                 & 0.146& 0.065 & 0.066 \\ 
ESO540-032   
        & Y & N    & 0.03 & 0.05 & -1.42 & 0.032$\pm$0.003 & 0.146& 0.055 & 0.049 \\ 
        & Y & Y    & 0.02 & 0.03 & -1.47 &                 & 0.166& 0.058 & 0.045 \\ 
        & N & N    & 0.03 & 0.07 & -1.56 &                 & 0.162& 0.043 & 0.042 \\ 
        & N & Y    & 0.02 & 0.03 & -1.48 &                 & 0.162& 0.044 & 0.046 \\ 
M81F6D1      
        & Y & N    & 0.01 & 0.04 & -1.12 & 0.026$\pm$0.003 & 0.133& 0.024 & 0.026 \\ 
        & Y & Y    & 0.01 & 0.02 & -1.25 &                 & 0.145& 0.033 & 0.030 \\ 
        & N & N    & 0.02 & 0.06 & -1.17 &                 & 0.137& 0.030 & 0.028 \\ 
        & N & Y    & 0.02 & 0.01 & -1.20 &                 & 0.149& 0.025 & 0.026 \\ 
M81K61     
        & Y & N    & 0.01 & 0.04 & -1.12 & 0.032$\pm$0.002 & 0.112& 0.020 & 0.021 \\ 
        & Y & Y    & 0.01 & 0.02 & -1.25 &                 & 0.122& 0.034 & 0.035 \\ 
        & N & N    & 0.02 & 0.06 & -1.17 &                 & 0.135& 0.039 & 0.043 \\ 
        & N & Y    & 0.02 & 0.01 & -1.20 &                 & 0.147& 0.044 & 0.047 \\ 
IKN     
        & Y & N    & 0.02 & 0.03 & -1.11 & 0.023$\pm$0.002 & 0.116& 0.029 & 0.031 \\ 
        & Y & Y    & 0.02 & 0.05 & -1.49 &                 & 0.124& 0.028 & 0.031 \\ 
        & N & N    & 0.03 & 0.00 & -1.17 &                 & 0.160& 0.046 & 0.040 \\ 
        & N & Y    & 0.04 & 0.00 & -1.55 &                 & 0.158& 0.033 & 0.033 \\ 
DDO78      
        & Y & N    & 0.01 & 0.14 & -1.12 & 0.041$\pm$0.002 & 0.127& 0.069 & 0.071 \\ 
        & Y & Y    & 0.01 & 0.11 & -1.10 &                 & 0.156& 0.089 & 0.088 \\ 
        & N & N    & 0.01 & 0.16 & -1.17 &                 & 0.150& 0.081 & 0.079 \\ 
        & N & Y    & 0.01 & 0.15 & -1.09 &                 & 0.164& 0.095 & 0.102 \\ 
SCL-DE1      
        & Y & N    & 0.10 & 0.27 & -0.79 & 0.050$\pm$0.005 & 0.314& 0.269 & 0.251 \\ 
        & Y & Y    & 0.06 & 0.11 & -1.57 &                 & 0.358& 0.274 & 0.275 \\ 
        & N & N    & 0.06 & 0.38 & -0.98 &                 & 0.136& 0.061 & 0.052 \\ 
        & N & Y    & 0.08 & 0.05 & -1.63 &                 & 0.212& 0.105 & 0.099 \\ 
DDO71       
        & Y & N    & 0.01 & 0.12 & -1.05 & 0.039$\pm$0.002 & 0.121& 0.021 & 0.023 \\ 
        & Y & Y    & 0.02 & 0.08 & -1.19 &                 & 0.115& 0.029 & 0.030 \\ 
        & N & N    & 0.02 & 0.11 & -1.06 &                 & 0.141& 0.040 & 0.046 \\ 
        & N & Y    & 0.01 & 0.07 & -1.17 &                 & 0.135& 0.038 & 0.043 \\ 
M81F12D1   
        & Y & N    & 0.01 & 0.08 & -1.17 & 0.032$\pm$0.002 & 0.127& 0.020 & 0.018 \\ 
        & Y & Y    & 0.01 & 0.03 & -1.29 &                 & 0.125& 0.023 & 0.024 \\ 
        & N & N    & 0.01 & 0.05 & -1.15 &                 & 0.142& 0.034 & 0.032 \\ 
        & N & Y    & 0.01 & 0.02 & -1.19 &                 & 0.130& 0.023 & 0.024 \\ 
M81K64   
        & Y & N    & 0.01 & 0.08 & -1.25 & 0.042$\pm$0.003 & 0.134& 0.022 & 0.023 \\ 
        & Y & Y    & 0.01 & 0.08 & -1.32 &                 & 0.138& 0.030 & 0.025 \\ 
        & N & N    & 0.01 & 0.07 & -1.26 &                 & 0.143& 0.038 & 0.033 \\ 
        & N & Y    & 0.01 & 0.06 & -1.30 &                 & 0.146& 0.035 & 0.039 \\ 
\multicolumn{10}{c}{Results for mid-to-lowest density rings of galaxies} \\
ESO540-032   
        & N & N    & 0.03 & 0.06 & -1.49 & 0.028$\pm$0.004 & 0.162 & 0.059 & 0.058\\ 
DDO113       
        & N & N    & 0.02 & 0.13 & -1.45 & 0.037$\pm$0.005 & 0.153 & 0.071 & 0.052\\ 
M81K61     
        & N & N    & 0.01 & 0.11 & -1.21 & 0.029$\pm$0.003 & 0.143 & 0.043 & 0.042\\ 
DDO78      
        & N & N    & 0.02 & 0.12 & -1.24 & 0.038$\pm$0.003 & 0.169 & 0.070 & 0.076\\ 
\enddata
\label{tab_ratio}
\end{deluxetable*}